\documentclass{aa}  

\usepackage{graphicx}
\usepackage{txfonts}
\usepackage{colortbl}
\usepackage{hyperref}
\hypersetup{colorlinks=true,linkcolor=[rgb]{1.,0.2,0.2},citecolor=[rgb]{0.1,0.1,1.},filecolor=[rgb]{0.7,0.2,0.2},urlcolor=[rgb]{0.7,0.2,0.2}}
\usepackage{amsmath}
\usepackage{amssymb}
\usepackage{natbib}

\def \LGalaxies{\texttt{L-Galaxies}\,}

\usepackage{ulem}
\def \mbin{MBHBs}
\def \msun{\,\rm M_\odot}
\def \mBHS{MBHs}
\usepackage{color}
\definecolor{myorange}{rgb}{0.8, 0.3, 0.0}

\definecolor{mygreen}{rgb}{0.0, 0.398, 0.0}

\def\msun{\,\rm{M_\odot}}
\begin{document}

   \title{Properties and merger signatures of galaxies hosting LISA coalescing massive black hole binaries}
   \author{David Izquierdo-Villalba$^{*1,2}$ \and Monica Colpi$^{1,2}$ \and Marta Volonteri$^{3}$ \and Daniele Spinoso$^{4}$ \and \\
    Silvia Bonoli$^{5,6}$ \and Alberto Sesana$^{1,2}$  }
   \institute{$^{1}$ Dipartimento di Fisica ``G. Occhialini'', Universit\`{a} degli Studi di Milano-Bicocca, Piazza della Scienza 3, I-20126 Milano, Italy\\
    $^{2}$ INFN, Sezione di Milano-Bicocca, Piazza della Scienza 3, 20126 Milano, Italy\\
    $^{3}$ Institut d’Astrophysique de Paris, Sorbonne Université, CNRS, UMR 7095, 98 bis bd Arago, 75014 Paris, France\\
    $^{4}$ Department of Astronomy, MongManWai Building, Tsinghua University, Beijing 100084, China\\
    $^{5}$ Donostia International Physics Centre (DIPC), Paseo Manuel de Lardizabal 4, 20018 Donostia-San Sebastian, Spain\\
    $^{6}$ IKERBASQUE, Basque Foundation for Science, E-48013, Bilbao, Spain\\ \\
   \email{david.izquierdovillalba@unimib.it}}

   \date{Received; accepted}

% \abstract{}{}{}{}{} 
% 5 {} token are mandatory
 
  \abstract
   {
   The gravitational wave (GW) antenna LISA will detect the signal from coalescing massive black hole binaries (MBHBs) of $\rm 10^4\,{-}\,10^7\, \msun{}$, providing clues on their formation and growth along cosmic history. Some of these events will be localized with a precision of several to less than a deg$^2$, enabling the possible identification of their host galaxy. This work explores the properties of the host galaxies of LISA MBHBs below $z\,{\lesssim}\,3$. We generate a simulated lightcone by using the semi-analytical model \LGalaxies{} applied on the merger trees of the high-resolution N-body cosmological simulation \texttt{Millennium-II}. The model shows that LISA MBHBs are expected to be found in optically dim ($r\,{>}\,20$), star-forming ($\rm sSFR\,{>}\,10^{-10}\, \rm yr^{-1}$), gas-rich ($f_{\rm gas}\,{>}\,0.6$) and disc-dominated ($\rm B/T\,{<}\,0.7$) \textit{low-mass galaxies} of stellar masses $10^8\,{-}\,10^9 \msun{}$. However, these properties are indistinguishable from those of galaxies harboring single massive black holes with comparable mass, making difficult the selection of LISA hosts among the whole population of low-mass galaxies.  Motivated by this, we explore the possibility of using merger signatures to select LISA hosts. We find that $40\,{-}\,80$\% of the galaxies housing LISA MBHBs display merger features related to the interaction which brought the secondary MBH to the galaxy. Despite this, around $60\%$ of dwarf galaxies placed in the surroundings of the LISA hosts will show such kind of features as well, challenging the unequivocal detection of LISA hosts through the search of merger signatures. Consequently, the detection of an  electromagnetic transient associated with the MBHB merger will be vital to pinpoint the star-forming dwarf galaxy where these binary systems evolve and coalesce.
   }

   \keywords{Methods: numerical --- quasars: supermassive black holes -- Gravitational waves -- Galaxies: interactions }
   \titlerunning{Hosts of LISA massive binaries}
    \authorrunning{Izquierdo-Villalba et al}
   \maketitle
%
%-------------------------------------------------------------------

\section{Introduction}

During the last few decades, astronomical observations have proven the existence of massive black holes (MBHs) above $10^5\,\msun{}$ lurking at the center of most of the galaxies \citep[see e.g.][]{Sargent1978,Tonry1984,Dressler1988,Kormendy1995,Magorrian1998,Greene2020}. Furthermore, the masses of these objects show correlations with the properties of their host galaxies, pointing to a co-evolution between MBHs and galaxies \citep{Graham2001,HaringANDRix2004,Kormendy2013,Savorgnan2016,Capuzzo2017}. Our current hierarchical structure formation paradigm establishes that the assembly of galaxies takes place mainly through a series of mergers and accumulation of intergalactic gas funneled into dark matter filaments \citep[]{WhiteandRees1978,WhiteFrenk1991,Kauffmann1999}. Thus, the presence of MBHs in most of the merging galaxies hints at the existence of massive black hole binary systems \citep[MBHBs,][]{Begelman1980}. To date, many different processes have been proposed as plausible mechanisms for the formation and evolution of MBHBs. Dynamical friction with background stars, the interaction with massive gas clumps or torques exerted by galactic bars seems to regulate the evolution of MBHs at galactic scales ($\rm {\sim}\,kpc$, \citealt{Milosavljevic2001,Yu2002,Mayer2007,Fiacconi2013,Bortolas2020,Bortolas2022,Kunyang2022}. At smaller distances ($\rm {<}\,pc$) the interaction with individual stars, circumbinary gaseous discs, or emission of gravitational waves (GWs) tighten the MBHB and drive the two MBHs to the final coalescence  \citep{Quinlan1997,Sesana2006,Vasiliev2014,Sesana2015,Escala2004,Escala2005,Dotti2007,Cuadra2009,Bonetti2020,Franchini2021,Franchini2022}.\\

According to General Relativity, coalescing MBHBs are among the loudest sources of low-frequency GWs. The future space-based  mission LISA (Laser Interferometer Space Antenna, \citealt{LISA2017}) will tackle the search for GW signals  at $\rm 0.1\,{-}\,100 \, mHz$  from coalescing MBHBs of $10^4\,{-}\,10^7\, \msun{}$. The population of MBHBs detected by LISA will help in sorting out several open questions.  For instance, the discovery of $10^4\,{-}\,10^7\, \msun{}$ binary systems at $z\,{<}\,9$  will help to understand which mechanisms trigger the growth of low-mass MBHs \citep{Lupi2016,Pezzulli2017,Trinca2022,Spinoso2022,Sassano2023}. Besides, the detection of MBHs in the first stages of their evolution will shed light on their still unknown origin \citep{Loeb1994,Schneider2002,Yoshida2003,Koushiappas2004,Agarwal2012,Valiante2017,Visbal2018,MayerBonoli2019,Lupi2021,Volonteri2021,Spinoso2022}.\\% It will be possible to disentangle whether MBHs were born with very low mass ($10^2-10^3 \, \rm M_{\odot}$) as a byproduct of the evolution of the first stars of the Universe or rather with a larger mass ($10^4-10^6 \, \rm M_{\odot}$) produced by the direct collapse of giant gas clouds  \citep{Loeb1994,Schneider2002,Yoshida2003,Koushiappas2004,Agarwal2012,Valiante2017,Visbal2018,MayerBonoli2019,Lupi2021,Volonteri2021,Spinoso2022}. \mv{I would like to see a more modern approach to "seeding", I feel that this dichotomy is an old view that does not take into account recent developments.} \monica{I agree with Marta: can we  just say ... unknown origins and  cite few reviews cutting the text on the dichotomy between low and mass seeds? or after origins keep all the refs and full stop.}\\

Another important scientific case of LISA is the possibility of using MBHB mergers as \textit{bright standard sirens}. The simultaneous observation of LISA sources in the GW and electromagnetic (EM) band will allow the constrain of the luminosity distance-redshift relation and provide an independent measure of the Hubble expansion parameter \citep{Petiteau2011,Tamanini2016}. %\monica{\sout{On the one hand, the luminosity distance of the source can be directly measured from the GW signal by using waveform modeling. On the other hand, the redshift must be inferred with the identification of the galaxy hosting the MBHB, a search that is completely ruled by the sky-localization capabilities of LISA }} \mv{the redshift must be measured with emission lines from an active galactic nucleus powered by the MBH itself or from the star-forming galaxy. } 
In particular, the search for the EM counterpart of the merging MBHB is crucially dependent on LISA's sky-localization capabilities. The latest estimates suggest that the sky area constrained by LISA can range from several hundreds of $\rm deg^2$ to fractions of $\rm deg^2$, depending on the intrinsic properties of the binary, its luminosity distance and the detection time prior to the final coalescence \citep{Mangiagli2020,Marsat2021,Piro2022}. The recent work of \cite{Lops2022} proved that the unequivocal identification of the galaxy housing LISA MBHB will be challenging \citep[see also][]{Kocsis2006}. Specifically, by using simulated data, the authors showed that the LISA sky-localization area is expected to be crowded with as many as $10^5$ potential host candidates, especially for the fields of $z\,{>}\,1$ MBHBs. \cite{Lops2022} highlighted that a pre-selection of X-ray AGNs would reduce this number by one order of magnitude. Still, the large number of galaxies lying within the LISA sky-area will hamper the possibility of using LISA MBHBs as standard sirens. Therefore, further theoretical studies are required to provide hints about specific galaxy properties that can be used to unequivocally identify the galaxy housing LISA systems. Several attempts have been done in this direction. For instance, by using the cosmological hydrodynamical simulation \texttt{Illustris}, \cite{DeGraf2021} investigated the mergers of MBHs and their connection with the morphologies of
the galaxies in which they are found. Assuming an instantaneous MBH coalescence right after the galaxy merger, it was shown that the hosts of LISA-like MBHBs will exhibit short-lived merger morphologies (${\sim}\,500\,\rm Myr$). In fact, the incorporation of more realistic dynamics of MBHB will imply that the host galaxy will have enough time to relax prior to the emission of GW and thus, blur even more the connection between coalescing MBHs and post-merger galaxies. Similar results were reported by \cite{Volonteri2020} by analyzing the \texttt{NewHorizon} simulation. On top of this, the authors showed that the long time delays between the galaxy and MBHB merger will cause that disturbed features present in the galaxies hosting GW sources will not correlate at all with the merger which led to the MBHB formation.\\

Under these premises, in this work, we explore the properties of the galaxies hosting $z\,{<}\,3$ LISA MBHBs and study from a statistical point of view if signatures of galaxy mergers can be used as tracers for seeking the hosts of merging MBHBs. To this end, we use the \LGalaxies{} semi-analytical model \citep[SAM,][]{Henriques2015,IzquierdoVillalba2020,IzquierdoVillalba2021,Spinoso2022} applied on top of the high-resolution dark matter (DM) merger trees of the \texttt{Millennium-II} simulation \citep{Boylan-Kolchin2009} to produce a synthetic lightcone of ${\sim}\,1000\,\rm deg^2$ reaching $z\,{\sim}\,3.5$. Thanks to the physics included in \LGalaxies{} the resulting lightcone is detailed enough to account for the cosmological evolution of galaxies and of their single and binary MBHs. We highlight that throughout the whole paper, the LISA MBHB population will be chosen according to the total mass of the binary, its mass ratio, and its time to coalescence. Specifically, we will impose the total mass $\rm 10^{4-7}\, \msun{}$, mass ratio ${\geq}\,0.1$ and the merging time ${<}\,1\, \rm Myr$ limits to ensure that the selected MBHBs will emit eventually GW signals inside the LISA frequency band.\\

The paper is organized as follows: In Section~\ref{sec:SAM_DESCRIPTIONS} we describe the main characteristics of the \texttt{Millennium-II} dark matter simulation and we summarize the physics implemented in \LGalaxies{} to tackle the assembly and evolution of galaxies, MBHs, and MBHBs. In Section~\ref{sec:PropertiesLISAMBHBs} we present the abundance and properties of MBHBs potentially detectable by LISA interferometer. In Section~\ref{sec:HostProperties} we summarize the properties of the galaxies hosting LISA MBHBs and compare them with the ones of galaxies housing single MBHs with the same mass as the binary systems. In Section~\ref{sec:MergerSignatures} we discuss the possibility of using merger signatures as a potential distinctive feature of LISA hosts, i.e the galaxies where detectable LISA MBHBs are placed. In Section~\ref{sec:Caveats} we list several caveats that should be taken into account when interpreting the results. Finally, in Section~\ref{sec:Conclusions} we summarize our main findings. A Lambda Cold Dark Matter $(\Lambda$CDM) cosmology with parameters $\Omega_{\rm m} \,{=}\,0.315$, $\Omega_{\rm \Lambda}\,{=}\,0.685$, $\Omega_{\rm b}\,{=}\,0.045$, $\sigma_{8}\,{=}\,0.9$ and $\rm H_0\,{=}\,67.3\, \rm km\,s^{-1}\,Mpc^{-1}$ is adopted throughout the paper \citep{PlanckCollaboration2014}.

\section{A LIGHTCONE FOR THE STUDY OF THE LISA MASSIVE BLACK HOLE BINARIES} \label{sec:SAM_DESCRIPTIONS}

In this section, we describe the dark matter simulation and galaxy formation model used to generate a lightcone specifically designed to study potential LISA MBHB sources. We use the so-called \LGalaxies{} semi-analytical model (SAM), a state-of-the-art model set to reproduce many different observational constraints such as the stellar mass function, the cosmic star formation rate density evolution, galaxy colors, and the fraction of passive galaxies (we refer to \citealt{Guo2011} and \citealt{Henriques2015} for further details). Among all the versions of the model, we use the one presented in \cite{Henriques2015} with the modifications of \cite{IzquierdoVillalba2019,IzquierdoVillalba2020,IzquierdoVillalba2021}. These changes were included to improve the predictions for galaxy morphology, extend the physics of MBHs and introduce the formation and evolution of \mbin{}. In the following, we summarize the main features of the model, and we refer the reader to the papers cited above for a detailed description of the baryonic physics included.

\subsection{The underlying dark matter population: \texttt{Millennium-II}} \label{sec:DarkMatter}

\LGalaxies{} is a semi-analytical model which self-consistently couples different astrophysical processes with the dark matter (DM) merger trees of N-body simulations. In particular, our SAM can be run on top of the merger-trees of the \texttt{Millennium} suite of simulations: \texttt{Millennium-I} \citep[MS,][]{Springel2005}, \texttt{Millennium-II} \citep[MSII,][]{Boylan-Kolchin2006} and \texttt{Millennium-XXL} \citep[MXXL,][]{Angulo2012}. The different box sizes and DM mass resolution of the \texttt{Millennium} suite offer the possibility of exploring different baryonic processes over a wide range of scales and environments.\\

Among all these \texttt{Millennium} simulations, we use the MSII, whose mass resolution allows the study of \mBHS{} and \mbin{} hosted in galaxies with stellar mass as low as $\rm {\sim}\,10^7\,M_{\odot}$. In brief, the MSII follows the cosmological evolution of $2160^3$ DM particles with mass $6.885 \times 10^6\, \mathrm {M_{\odot}}/h$ within a periodic comoving box of $100\,{\rm Mpc}/h$ on a side. The simulation was stored at 68 different epochs or \textit{snapshots}, to which the \texttt{SUBFIND} algorithm was applied to detect all the DM halos whose minimum halo mass corresponds to 7 times the particle mass (${\sim}\,10^7\,\msun{}/\mathit{h}$). After that, the \texttt{L-HALOTREE} code arranged these structures according to their evolutionary path in the so-called \textit{merger trees} \citep{Springel2001,Springel2005}. The time resolution offered by the finite number of the MSII outputs causes inconveniences in tracing accurately the baryonic physics involved in galaxy evolution. To overcome this, the  SAM does an internal time interpolation between two consecutive snapshots with approximately $\rm {\sim}\,5\,{-}\,20 \, Myr$ of time resolution, depending on redshift. Finally, \LGalaxies{} re-scales the original cosmology of MSII (WMAP1 \& 2dFGRS ‘concordance’ $\Lambda$CDM framework, \citealt{Spergel2003}) to the cosmology of Planck first-year data release \citep{PlanckCollaboration2014} by using the \cite{AnguloandWhite2010} methodology. This re-scaling modifies by a factor of 0.96 and 1.12 the MSII box size and particle mass, respectively. Taking into account this, the merger trees of MSII enables to trace the cosmological assembly of galaxies placed in halos of $5.7\,{\times}\,10^{7}\,{-}\, 3 \,{\times}\,10^{14}\, \msun$. %\dstext{I would bet that these rescaling factors and the cosmology parameters you mentioned in the introduction are referred to Planck18 cosmology, but I may be very wrong}.

\subsection{The assembly of the galaxy population: \texttt{L-Galaxies} formation and evolution model} \label{sec:Baryons_LGal}

%Following the framework of \cite{WhiteFrenk1991}, \LGalaxies{} assumes that the birth of a galaxy starts at the centre of every newly formed DM halo through the infall of baryonic matter. During this process, a fraction of the infalling material (taken to be proportional to the baryon fraction) is shock-heated and forms a diffuse, spherical, and quasi-static hot gas atmosphere with an extension equal to the halo \textit{virial radius} ($R_{200c}$).

Following the framework of \cite{WhiteFrenk1991}, \LGalaxies{} assumes that the birth of a galaxy takes place at the centre of every newly formed DM halo. As soon as a DM halo collapses, a fraction of baryonic matter (proportional to the baryon fraction) is trapped and collapses with it. During this process, the material is shock-heated and forms a diffuse, spherical, and quasi-static hot gas atmosphere with an extension equal to the halo \textit{virial radius} ($R_{\rm vir}$). Part of this hot gas is allowed to cool down and migrate towards the DM halo centre \citep{WhiteandRees1978}. The rate at which this process takes place is determined by cooling functions \citep{SutherlandDopita1993} and the amount of hot gas enclosed within the halo \textit{cooling radius} ($r_{\rm cool}$), defined as the radius at which the cooling time matches the halo dynamical time. At high-$z$ and in low-mass DM halos, the hot gas can cool rapidly (\textit{rapid infall}, $r_{ \rm cool}\,{>}\,R_{\rm vir}$) causing the migration of the whole mass towards the DM halo centre at essentially the free-fall rate. On the other hand, a slow \textit{cooling flow regime} ($r_{\rm cool}\,{<}\,R_{\rm vir}$) takes place at low-$z$ and in massive halos. In these cases, only a fraction of the hot gas is allowed to condensate through cooling flows. After any of these condensation processes, the cold gas settles in a disc with a specific angular momentum inherited from the host DM halo. This newly formed disc is assumed to be distributed with an exponential profile, whose extension is determined according to the evolution of the gas angular momentum \citep[see][]{Guo2011}.\\

Based on the observational results of \cite{Kennicutt1998}, our SAM assumes that star formation (SF) processes take place as soon as the surface density of the cold gas exceeds a critical value. When this occurs, the galaxy begins (or continues) the assembly of its stellar disc on a time scale given by the cold gas disc dynamical time. As a consequence of star formation processes, massive and short-lived stars explode as supernovae (SNe) injecting energy and metals into the cold gas disc (\textit{SNe feedback}). This injection causes the re-heating of a fraction of cold gas and it may additionally expel a fraction of the hot gas beyond the halo virial radius. At later times, this ejected gas can be reincorporated, initiating new star formation events. In addition to SNe feedback, \LGalaxies{} introduces the so-called \textit{radio-mode feedback} as an additional process to regulate the assembly of the stellar component in massive galaxies. This mechanism is activated by the gas accretion onto the MBH from the hot gas atmosphere around the galaxy. The result of this accretion is the release of kinetic energy, whose injection into the surrounding medium can reduce or even suppress the cooling of gas. \LGalaxies{} computes the extent of the stellar disc by following the evolution of its specific angular momentum (modified by star formation events and galaxy mergers) and assuming an exponential profile \citep[see][]{Guo2011}. Finally, our SAM models large-scale effects (also known as \textit{environmental processes}) such as ram pressure stripping or galaxy tidal disruption. These processes occur when the halo hosting the galaxy falls into a larger system. As a consequence, the galaxy can lose its entire hot gas atmosphere and can be deprived of its cold gas and stellar component through tidal forces.

\subsubsection{The assembly of bulges} \label{sec:bulges}

The continuous assembly of the stellar disc causes some galaxies to undergo disc instability (DI) processes. These events refer to the mechanism by which the stellar disc becomes massive enough to suffer non-axisymmetric instabilities. The eventual result is the formation of a central ellipsoidal component via buckling of the nuclear stellar
orbits \citep{MoMaoWhite1997}. The occurrence of DIs is modeled according to the \cite{Efstathio1982} analytic stability test (tested against cosmological simulations, see \citealt{Izquierdo-Villalb_DI_2022}). When the criterion is satisfied, our SAM triggers the formation (or growth) of a bulge by transferring from the disc the minimum stellar mass needed to make it marginally stable again (i.e. $\rm \Delta M_{\rm stars}^{DI}$). The effective radius of the bulge after any DI events is determined by assuming that the transferred stellar mass comes from the innermost part of the disc \citep[see][for further information]{Guo2011}.\\

The hierarchical growth of the DM halos also shapes galaxy properties. The interaction between galaxies is ruled by the merger of the parent DM halos. As soon as two DM halos merge, their galaxies do it as well on a time scale given by the dynamical friction presented in \cite{BinneyTremine1987}. According to the \textit{baryonic} (stars plus gas) merger ratio of the two interacting galaxies ($\rm m_R\,{\leq}\,1$), \LGalaxies{} differentiates between \textit{major} ($\rm m_R\,{>}\,0.2$) and \textit{minor} interactions ($\rm m_R\,{<}\,0.2$). Major mergers completely destroy the discs of the two galaxies, giving rise to a pure spheroidal remnant that undergoes a \textit{collisional starburst}. Conversely, during minor interactions, the disc of the larger galaxy survives and experiences a burst of star formation, while its bulge integrates the entire stellar mass of the satellite. After any type of merger, \LGalaxies{} determines the effective radius of the remnant bulge by assuming the conservation of the binding and orbital energy. In addition to these two merger treatments, the model used in this work includes the prescription of \textit{smooth accretion} presented by \cite{IzquierdoVillalba2019} in order to deal with the physics of extremely-minor mergers. Indeed, \cite{IzquierdoVillalba2019} showed that the inclusion of these processes is important in order to recover the observed morphology of dwarf galaxies ($\rm M_{stellar}\,{\leq}\,10^9 \, M_{\odot}$) in \LGalaxies{} when this one is run on top of the MSII simulation. Specifically,  \textit{smooth accretions} take place in satellite galaxies with low binding energy. As a result, the stellar component of the satellite (i.e the bulge plus disc) gets diluted inside the disc of the central galaxy before being able to reach the nucleus. Consequently, the central galaxy \textit{loses the possibility} of forming (growing) the bulge component.\\

Based on the ratio between the bulge and total stellar component (known as the bulge-to-total ratio, $\rm B/T$), \LGalaxies{} divides the galaxy population into several morphological types:

\begin{enumerate}
    \item[a)] \textbf{Ellipticals}: Galaxies with $\rm B/T \,{>}\,0.7$.
    \item[b)] \textbf{Discs} or \textbf{Spirals}: Galaxies with $\rm B/T\,{<}\,0.7$. The bulge component of these systems is split into three types: 
    \begin{itemize}
    \item \textit{Pseoudobulges}: Galaxies with $\rm 0.01\,{<}\,B/T \,{<}\,0.7$ and more than $2/3$ of the bulge mass has been accumulated through disc instabilities.
    \item \textit{Classical bulges}: Galaxies with  $\rm 0.01\,{<}\,B/T \,{<}\,0.7$ and less than $2/3$ of the bulge mass has been accumulated through disc instabilities. Thus, (minor/major) mergers are the main mechanisms that brought mass into the bulge.
    \item \textit{Extreme late-type}: Bulgeless galaxies or galaxies with $\rm B/T\,{<}\,0.01$ (regardless of the process that gave rise to the bulge component).
    \end{itemize}    
\end{enumerate}

%\begin{figure}
%\centering  
%\includegraphics[width=1.0\columnwidth]{Figures/StellarMassFunction_MSII.pdf}
%\caption[]{Stellar mass function at $z\,{=}\,0$ when the model is run on top of the %\texttt{Millennium-II} merger trees.}
%\label{fig:ocupation_MBHs}
%\end{figure}

\subsection{The population of massive black holes: seeds, growth and spin} \label{sec:MBH_Model}
Thanks to the modifications of \cite{IzquierdoVillalba2020} \cite{IzquierdoVillalba2021} and \cite{Spinoso2022} \LGalaxies{} is able to track self-consistently the formation, growth, and spin evolution of MBHs. In the following subsections, we summarize the main physics included in our SAM to model the evolution of massive black holes.

\subsubsection{The seeding of massive black holes cosmic dawn}

\begin{figure}
\centering  
\includegraphics[width=1.0\columnwidth]{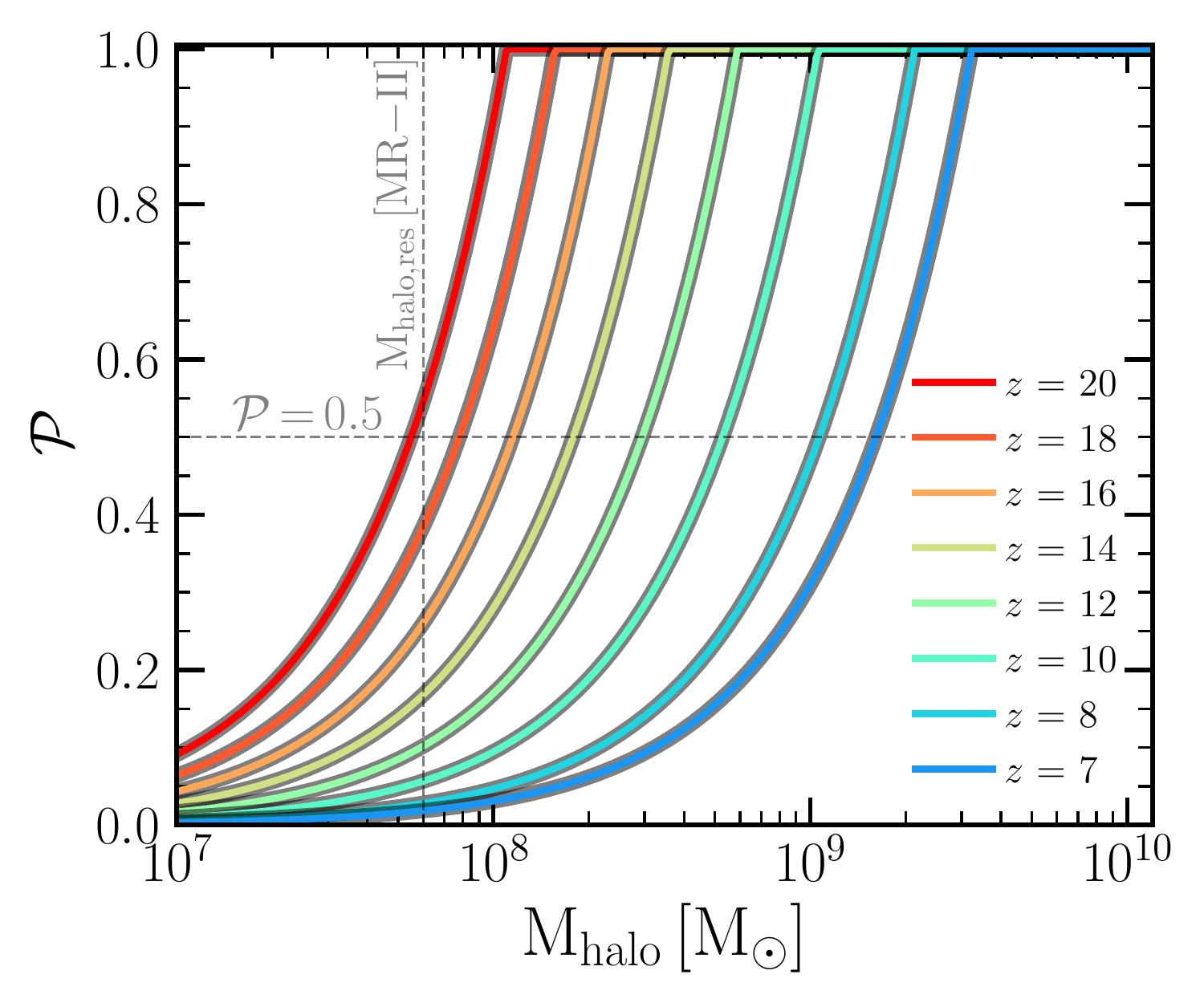}
\caption[]{Probability $\mathcal{P}$ that a given DM halo of mass $\rm M_{halo}$ would be seeded by a MBH. Each color corresponds to a different redshift. Vertical and horizontal grey dashed lines highlight the halo mass resolution of MSII and the value $\mathcal{P}\,{=}\,0.5$, respectively.}
\label{fig:seeding_Probability}
\end{figure}

The formation of massive black hole seeds in \LGalaxies{} has been extensively explored in  \cite{Spinoso2022}. In particular, the genesis of light seeds \citep[PopIII remnants,][]{bromm_larson2004} was accounted for by using a sub-grid approach, while the formation of massive seeds (i.e. intermediate-mass and heavy BH-seeds) was addressed by taking into account the spatial variations of the IGM metallicity and the UV-background produced by star formation events \citep[see][for a recent review]{inayoshi_visbal_haiman2020}. With these models applied on \LGalaxies{} and MSII, the authors showed that the formation of BHs is strongly inhibited at $z\,{\lesssim}\,6\,{-}\,7$ due to the progress of IGM chemical enrichment. In addition, the occupation fraction of newly-formed BH-seeds showed a dependence with the halo mass and redshift, being almost null in halos of ${<}\,10^8\, \msun{}$ at $z\,{<}\,9$. Following these results, in this work we include a simple empirical BH-seeding model where MBHs only form within newly-resolved galaxies at $z\,{\geq}\,7$ with a probability, $\mathcal{P}$, given by:

\begin{figure}
\centering  
\includegraphics[width=1.0\columnwidth]{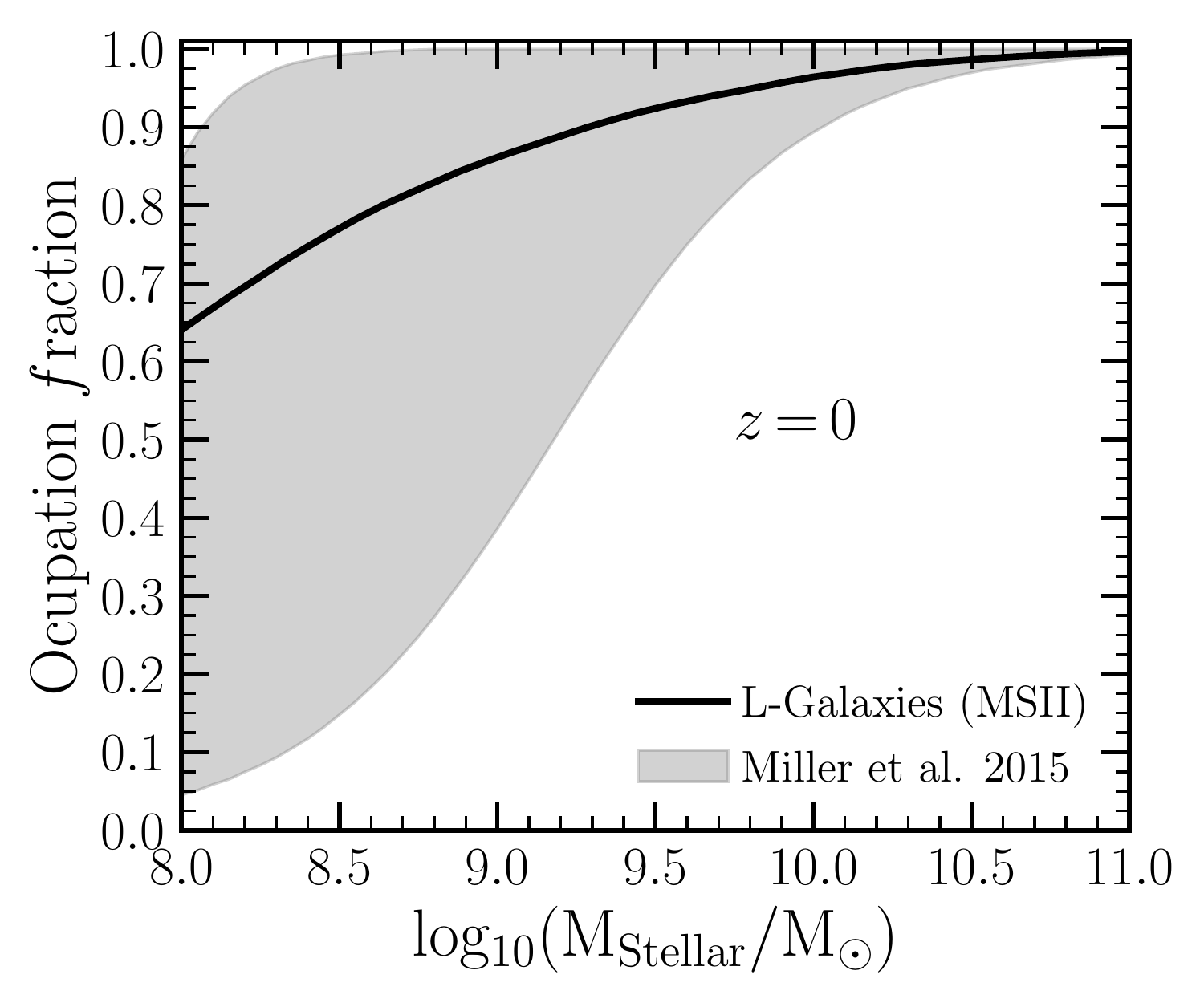}
\caption[]{Massive black hole occupation fraction at $z\,{=}\,0$ as a function of stellar
mass ($\rm M_{Stellar}$). The shaded grey region corresponds to the constraints presented in  the observational work of \protect{\cite{Miller2015}}.}
\label{fig:ocupation_MBHs}
\end{figure}

\begin{equation} \label{eq:ProbabilitySeeding}
\mathcal{P}\,{=}\, \mathcal{A} (1+\mathit{z})^{\gamma} \left( \frac{\rm M_{halo}}{\rm  \, M_{halo}^{th}}\right)
\end{equation}
   %\begin{equation} \label{eq:ProbabilitySeeding}
    % \mathcal{P} = \left\{
	%       \begin{array}{ll}
	%       \mathcal{A} (1+\mathit{z})^{\gamma} \left( \frac{\rm M_{halo}}{\rm  \, M_{halo}^{th}}\right)
	%	 \; \; \;\; \; \;  \; \rm \;\; at \;  \mathit{z}\,{\geq}\,7 \\ \\
	%	 \rm 0 \;\;\;\;\;\;\;\;\;\;\;\;\;\;\;\;\;\;\;\;\;\; \;\;\;\; \;\;\;\; \;\;\;\;\; 
     %         at \;  \mathit{z}\,{<}\,7  , \\ 
	 %      \end{array}
	 %    \right.
   %\end{equation}

\noindent where $\mathcal{A}\,{=}\,0.015$, $\gamma\,{=}\,7/2$ and $\rm M_{halo}^{th}\,{=}\,7\,{\times}\,10^{10}\, \msun{}$ \citep[see also Eq.~9 of][]{Spinoso2022}. To guide the reader, in Fig.~\ref{fig:seeding_Probability} we show how $\mathcal{P}$ varies with redshift and halo mass, $\rm M_{halo}$. As shown, at very high-$z$ the seeding process occurs  mainly in low-mass halos. As the redshift decreases the seeding events shift towards higher mass halos. This evolution is assumed to rise as a combination of an early formation of MBH after the explosion of PopIII stars and a later creation of MBHs via stellar runaway mergers and a direct collapse of pristine gas clouds. Based on Eq.~\ref{eq:ProbabilitySeeding}, every time that a $z\,{\geq}\,7$  galaxy is formed, we compute the value of $\mathcal{P}$ and draw a random value $\mathcal{R}\,{\in}\,[0\,{-}\,1]$. If $\mathcal{R}\,{>}\,\mathcal{P}$ a MBH is placed at the center of the galaxy whose mass and spin is randomly extracted from $10^2\,{-}\,10^4\, \msun{}$ and $0\,{-}\,0.998$, respectively \citep[see Figure 3 of][for further information about the seed mass choice]{Spinoso2022}. To show that this chosen seeding procedure retrieves a reasonable MBH population, in Fig.~\ref{fig:ocupation_MBHs} we present the predicted occupation fraction at $z\,{=}\,0$. As shown, the occupation fraction decreases towards low stellar masses, consistently with the constraints of \cite{Miller2015}. While at $\rm M_{stellar} \,{>}\,10^{10}\, \msun{}$ almost all galaxies host a nuclear MBH, at $\rm M_{stellar} \,{<}\,10^{9}\, \msun{}$ less than 80\% have one. We stress that in future works, we plan to include the full physical seeding model developed by \cite{Spinoso2022}.

\begin{figure*}
\centering  
\includegraphics[width=1.0\columnwidth]{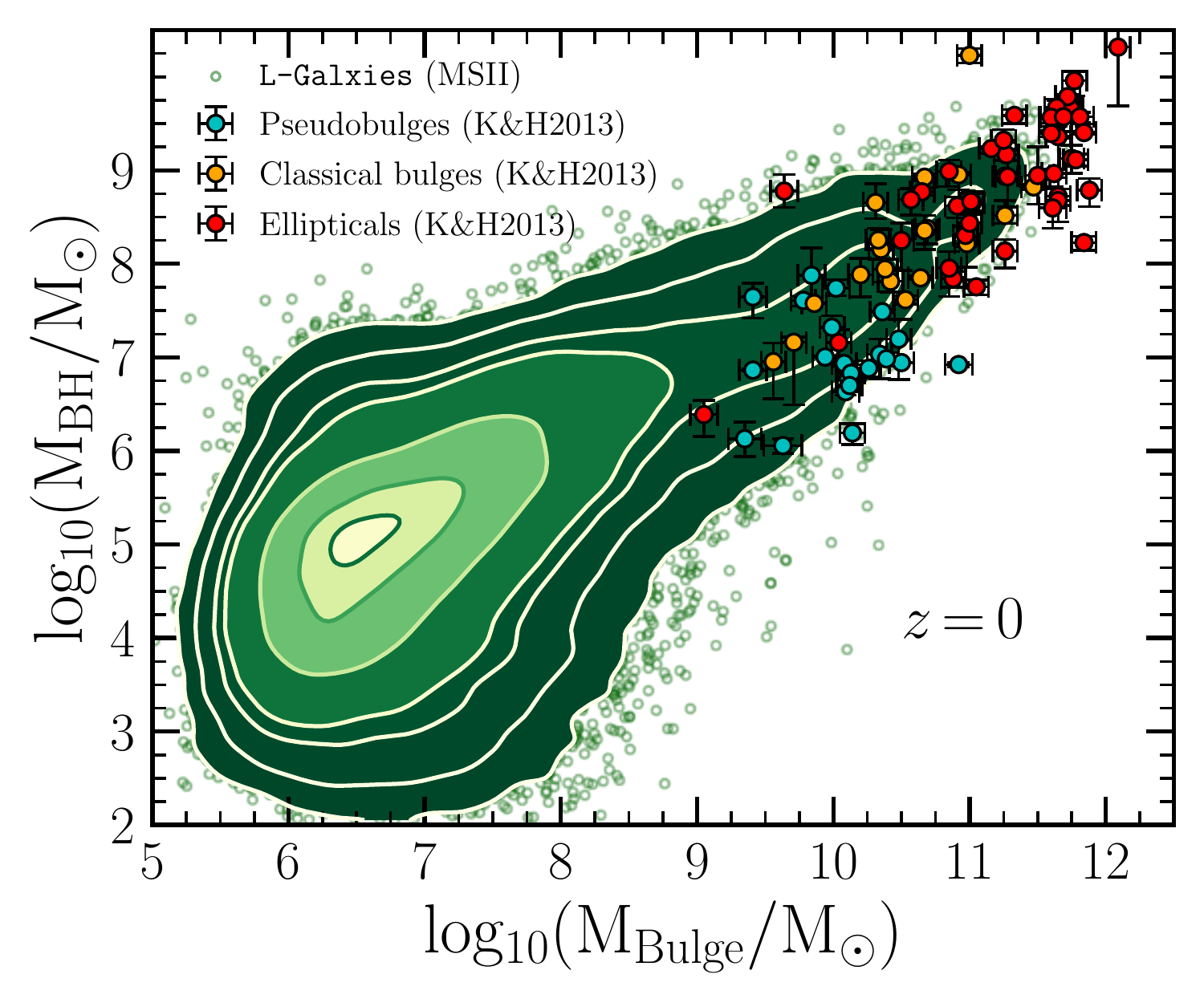}
\includegraphics[width=1.0\columnwidth]{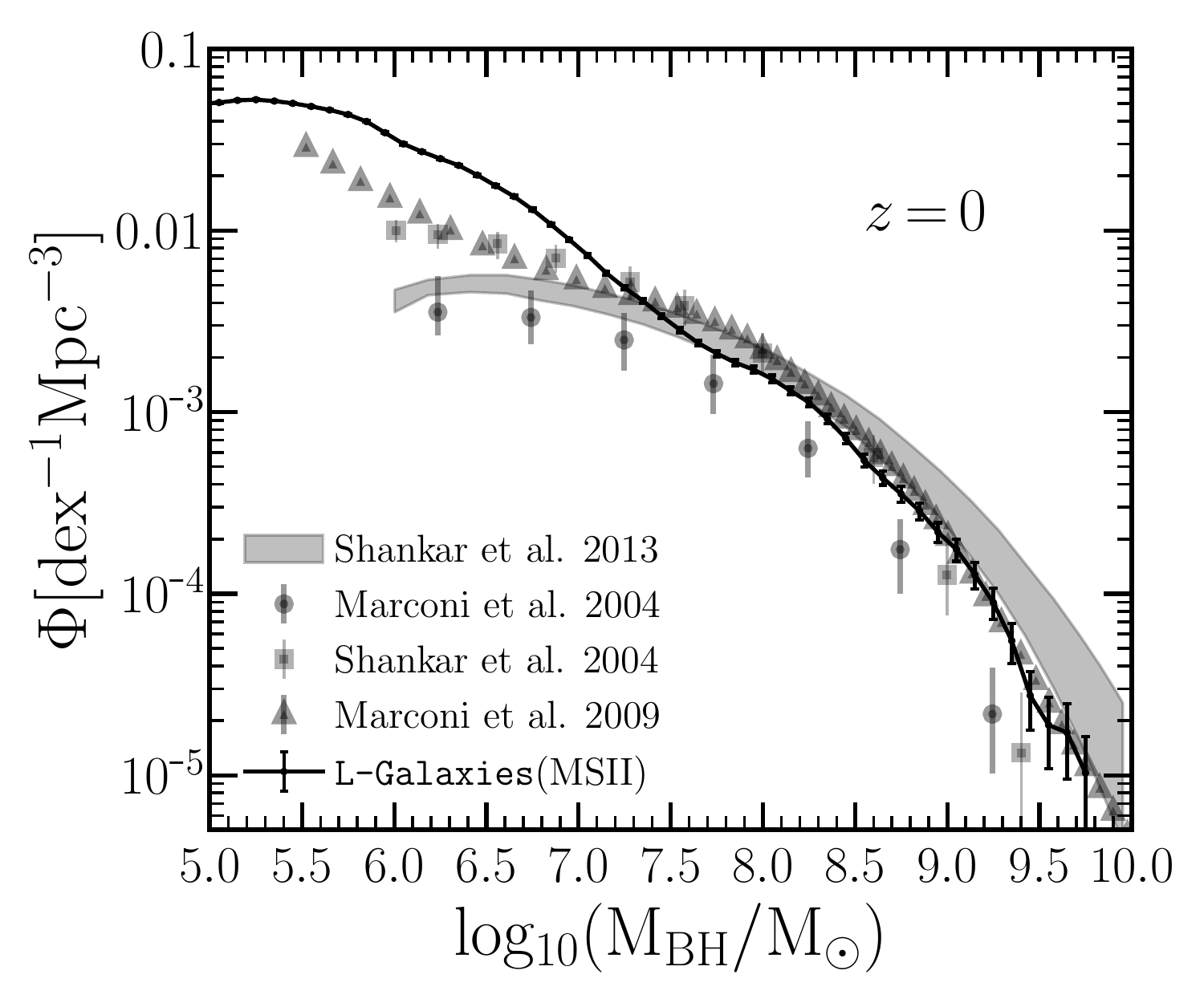}
\caption[]{\textbf{Left panel}: 2D histogram of the bulge-black hole correlation at $z\,{=}\,0$. The predictions are compared with the observations of {\protect \cite{Kormendy2013} (K\&H2013)} where blue, orange, and red points correspond to pseudobulges, classical bulges, and ellipticals, respectively. \textbf{Right panel}: Black hole mass function at $z\,{=}\,0$ when the model is run on top of the \texttt{Millennium-II} merger trees (black lines). Error bars correspond to the Poissonian error. For comparison, we have added the observational constraints of {\protect \cite{Marconi2004,Shankar2004,Shankar2009,Shankar2013}}.}
\label{fig:BHMF_Bulge_MBH}
\end{figure*}

%Each halo is seeded with a MBH of $\rm 10^{4} M_{\odot}$\footnote{\textbf{ The large seed mass chosen in this work is motivated by the dark matter resolution of the \texttt{Millennium} simulation. The vast majority of the newly formed halos have masses $\rm {\sim}\,10^{10}\, M_{\odot}$. This implies that the SAM can not access the assembly of the galaxy and its central MBH before their parent dark matter halo is resolved. Therefore, to account for this unresolved evolution we decided to place a relatively massive MBH seed in each galaxy.}} and random spin in the range $0\,{<}\,\chi\,{<}\,0.998$ (further improvements on the seeding paradigm will be done by including the model of \citealt{Spinoso2022}). 

\subsubsection{The growth of massive black holes} \label{sec:GrwothNuclear}

In the semi-analytical model, MBHs can grow via three different channels: \textit{cold gas accretion}, \textit{hot gas accretion}, and \textit{mergers} with other MBHs. In the following lines, we summarize the main assumptions related to gas consumption processes:

\begin{itemize}

\item \textit{Hot gas accretion}: This channel of growth is triggered by a continuous gas accretion from the hot gas atmosphere that surrounds the galaxy hosting the MBH \citep{Croton2006a}. The rate of accretion is usually orders-of-magnitude below the Eddington limit and is linked with the so-called radio mode feedback which injects energy into the hot atmosphere, halting the cooling gas inflows which supply  gas to the galaxy. In the model, the accretion due to hot gas is determined as \citep{Henriques2015}:
\begin{equation}\label{eq:Radio_mode}
\dot{\rm M}_{\rm  BH} \rm \,{=}\, \mathit{k}_{AGN} \left( \frac{M_{hot}}{10^{11}M_{\odot}} \right) \left( \frac{M_{BH}}{10^{8}M_{\odot}}\right),
\end{equation}
where $\rm M_{hot}$ the total mass of hot gas surrounding the galaxy and $\rm \mathit{k}_{AGN}$ is a free parameter set to $\rm 9\,{\times}\,10^{-5} M_{\odot}/yr$ to reproduce the turnover at the massive end of the galaxy stellar mass function.\\

\item \textit{Cold gas accretion}: It is the main channel driving the black hole growth and is triggered by both \textit{galaxy mergers/smooth accretion} and \textit{disc instability} events. In particular, after a galaxy merger or smooth accretion, the nuclear MBH can accumulate a fraction of the galaxy cold gas given by:
\begin{equation}\label{eq:QuasarMode_Merger}
\rm   \Delta {M}_{BH}^{gas} \,{=}\,\mathit{f}_{BH}^{merger} (1+\mathit{z}_{merger})^{5/2} \frac{m_{R}}{1 + (V_{BH}/V_{200})^2}\, M_{\rm gas},
\end{equation}
where $z_{\rm merger}$ is the redshift of the galaxy merger,  $\rm M_{\rm gas}$ the cold gas mass of the galaxy, $\rm V_{200}$ the virial velocity of the host DM halo and $\rm V_{BH}$, $\rm \, \mathit{f}_{BH}^{\rm merger}$ two adjustable parameters set to $\rm 280 \, km/s$ and $0.014$, respectively. On the other hand, after a disc instability, the black hole accretes an amount of cold gas proportional to the mass of stars that has triggered the stellar instability, $\rm \Delta M_{\rm stars}^{DI}$:
\begin{equation}\label{eq:QuasarMode_DI}
\rm    \Delta {M}_{BH}^{gas} \,{=}\, \mathit{f}_{BH}^{DI} (1+\mathit{z}_{DI})^{5/2} \frac{\Delta M_{stars}^{DI}}{{1 + (V_{BH}/V_{200})^2}},
\end{equation}
where $\rm \mathit{z}_{DI}$ is the redshift in which the disc instability occurs, and $\rm \mathit{f}_{BH}^{DI}$ is a free parameter that takes into account the gas accretion efficiency, set to $0.0014$. All these adjustable parameters  have been tuned to give the best agreement between the observations and model predictions for the $z\,{=}\,0$ black hole-bulge correlation and the BH mass function for $\rm M_{BH}\,{>}\,10^6\, \rm M_{\odot}$. \\

After a galaxy merger or a disc instability, the cold gas available for accretion (see Eq.~\ref{eq:QuasarMode_Merger} and Eq.~\ref{eq:QuasarMode_DI}) is assumed to settle in a reservoir around the black hole (with total mass $\rm M_{Res}$), which is progressively consumed according to a two phases model extensively used in \cite{IzquierdoVillalba2020} and \cite{IzquierdoVillalba2021}. The first phase corresponds to an Eddington-limited growth, which lasts until the MBH consumes a faction $\mathcal{F}$ of the available gas reservoir. The free parameter $\mathcal{F}$ is set to 0.7 in order to match the faint end of the low-$z$ AGN LFs \citep[see also][]{Marulli2008,Bonoli2009}. Once this phase ends, the BH enters in a self-regulated or quiescent growth regime characterized by progressively smaller accretion rates. To show that the model of MBHs explained in this section gives rise to a population in good agreement with the observations, in Fig.~\ref{fig:BHMF_Bulge_MBH} we present the black hole mass function (BHMF) and the bulge-MBH mass correlation in the local Universe. As we can see, the scaling relation raised in our SAM is consistent with the results reported by \cite{Kormendy2013}. On the other side, the BHMF is in good agreement with the observations at $\rm M_{BH}\,{>}\,10^7\,\msun{}$. For lower masses, \LGalaxies{} applied in the MSII merger trees finds a steeper increase than the one seen by observations. Interestingly, this rise is also seen in cosmological hydrodynamical simulations such as \texttt{EAGLE} \citep{Rosas-Guevara2016} or \texttt{TNG} \citep{Habouzit2021}.\\

The large number density of low-mass black holes reported in Fig.\ref{fig:BHMF_Bulge_MBH} has an impact on the evolution of active MBHs. Fig.~\ref{fig:LFs_MS_MSII} shows the evolution of the bolometric luminosity function (LF) predicted by \LGalaxies{}. As shown, our SAM applied on the \texttt{Millennium-II} merger trees is able to reproduce the observed trends of bright objects ($\rm L_{bol}\,{\geq}\,10^{46} \, erg/s$). However, it overpredicts the faint end of the LFs \citep[see similar trends in][]{Sijacki2015,Griffin2018,Weinberger2018,Marshall2020,Trinca2022}. Interestingly, this excess is not seen in runs with the \texttt{Millennium} simulation, pointing out that the evolution of the faint end of the LFm and thus the low-mass MBH population, is strongly affected by the resolution of the underlying dark matter simulation. The nature of the inconsistency between simulated and observed LFs is still an open issue. From one side, it is  challenging to cover wide sky areas with large depths, thus a robust sampling of high-$z$ faint AGNs is still missing \citep{Siana2008,Masters2012,McGreer2013,Niida2016,Akiyama2018}. On the other hand, some studies have explored different ways to suppress the large excess of faint AGNs seen in most of the SAMs and hydro-dynamical simulations \citep[see e.g][]{Hirschmann2014,Griffin2018,Habouzit2022}. For instance, varying the efficiency of MBH seed formation or the ability of newly born MBH to accrete matter have been postulated as plausible mechanisms \cite[see e.g][]{Degraf2010,Fanidakis2012,DeGraf2020,Spinoso2022,Trinca2022}. Despite that, no clear answer has been proposed yet, and further investigations are needed. 

%\mv{I think that the AGN LF at $z=2-3$ should be shown, or say in the text how well it is reproduced at the redshifts of interest ($z<3$).}

\begin{figure}
\centering  
\includegraphics[width=1.0\columnwidth]{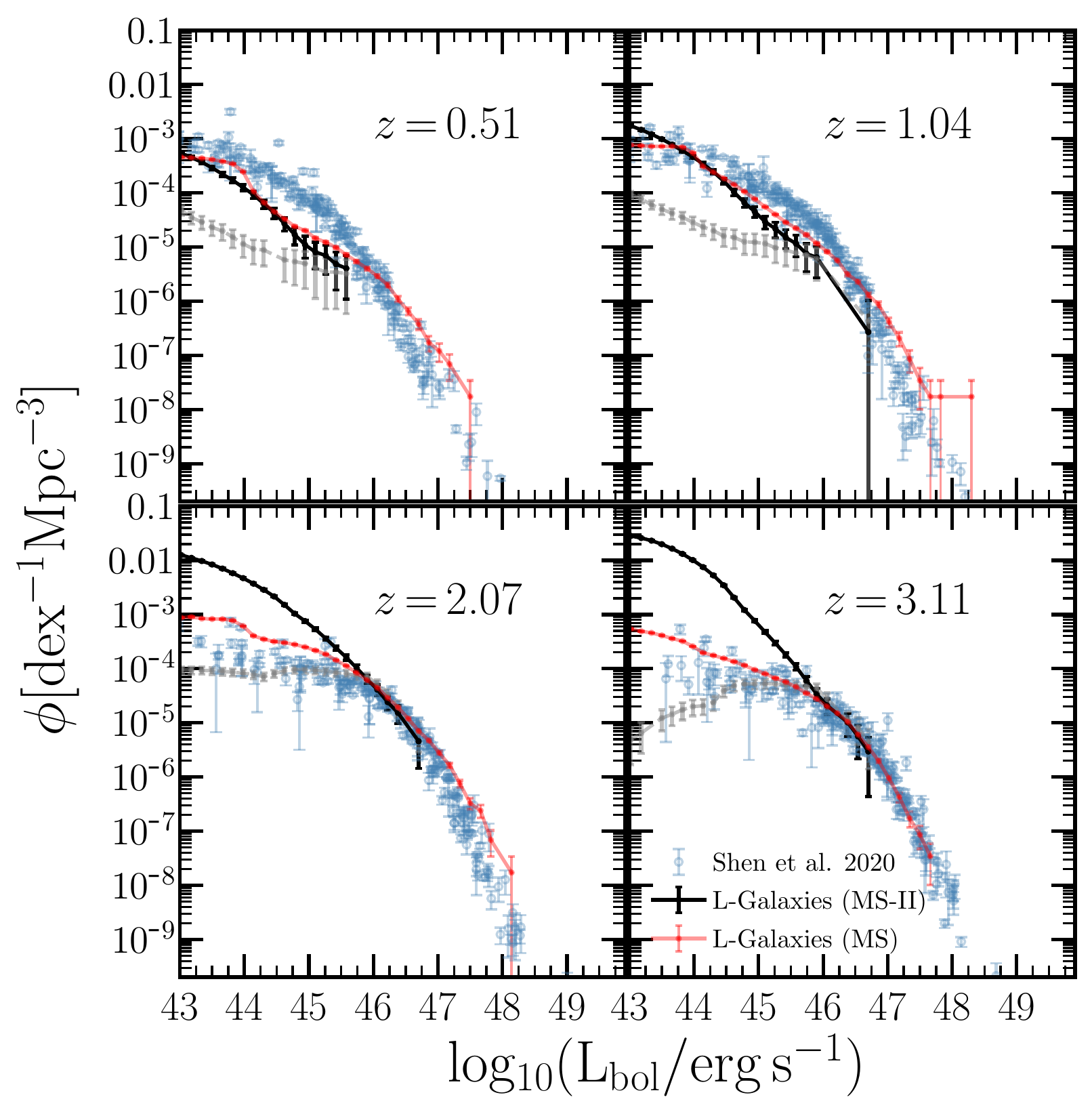}
\caption[]{Quasar bolometric luminosity functions ($\rm L_{bol}$) at $z{\sim}\,0.5,1.0,2.0$, and $3.0$ predicted by \LGalaxies{} when it is used MS (red) and MSII (black) DM merger trees. For comparison, the grey dashed lines represent the luminosity functions of \LGalaxies{} applied on the MSII when a cut of black hole mass $\rm {\geq}\,10^{7}\, M_{\odot}$ is performed. For comparison, these predictions are compared with the data presented in {\protect \cite{Shen2020}} (blue circles). In all the plots, error bars correspond to the Poissonian error (large error bars correspond to the cases where only one object is found).}
\label{fig:LFs_MS_MSII}
\end{figure}

\end{itemize}

\subsection{The population of massive black holes binaries} \label{sec:MBHBHs_Model}

%\mv{Divided the text into more paragraphs, I find it difficult to read very long blocks of text, I hope it's okay}
Besides including a comprehensive model for MBH growth, \LGalaxies{} deals with the dynamical evolution and growth of massive black hole binaries \citep[see][]{IzquierdoVillalba2021}. In the next sections, we summarize key aspects of the model included in \LGalaxies{}.

\subsubsection{The dynamical evolution}

The evolution of MBHBs inside \LGalaxies{} is divided into three different stages \citep{Begelman1980}: \textit{pairing, hardening} and \textit{gravitational wave} phase. %\dstext{I am really ignorant on this topic but what about changing ``gravitational wave phase'' to ``final inspiral phase''? I am saying this just to avoid additional mentions to ``gravitational wave'' or ``GW'' in the text}

\begin{itemize}

    \item \textit{Pairing phase}: After a galaxy merger, the MBH hosted by the satellite galaxy experiences dynamical friction that leads to its pairing with the nuclear MBH of the primary galaxy. This process causes the satellite MBH (typically deposited at a few kpc) to sink toward the galactic center of its new host. The time needed for the satellite MBH to reach the nuclear part of the galaxy, $t_{\rm dyn}^{\rm BH}$, is determined by using \citep{BinneyTremaine2008}:
\begin{equation}  \label{eq:DynamicalFriction}
    t_{\rm dyn}^{\rm BH} \,{=} \, 19 \, f_s \, f(\varepsilon)  \left( \frac{r_0}{5 \, \rm kpc} \right)^2 \left( \frac{\sigma}{200 \,\rm km/s}\right) \left( \frac{10^8 \, \rm M_{ \odot}}{\rm M_{BH}} \right) \, \frac{1}{\Lambda}\, \rm [Gyr] , 
\end{equation}
where $f(\varepsilon)\,{=}\,\varepsilon^{0.78}$ is a function with depends on the orbital circularity of the MBH \citep[$\varepsilon$,][]{Lacey1993}, $r_0$ is the initial position of the black hole deposited by the satellite galaxy after the merger, $\sigma$ is the velocity dispersion of the remnant galaxy, $\rm M_{BH}$ is the mass of the satellite black hole and $\rm \Lambda\,{=}\,\ln(1 + M_{stellar}/M_{BH})$ is the Coulomb logarithm  \citep{MoWhite2010}. The computation of all these quantities can be found in \cite{IzquierdoVillalba2021}.\\

The variable $f_s$ takes into account the stochastic insparalling of MBHs seen in simulations of clumpy (gas-rich) and barred galaxies. For instance, the simulations of gas-rich galaxies of \cite{Tamburello2017} showed that the interaction between MBHs and massive clumps typically lags the pairing phase of MBHs. At the other extreme, we find the results of \cite{Lupi2015}, who, by simulating the late stages of a gas-rich galaxy merger, found that the gravitational torques after the interaction can be very efficient in forming massive gas clumps that substantially perturb the orbits of the infalling MBHs. These perturbations result in impulsive kicks that lead to the formation of a gravitational bound MBHB in ${\sim}\,10\, \rm Myr$ from the start of the merger. Concerning galactic structures, the works of \cite{Bortolas2020,Bortolas2022} showed that bar structures induce an erratic motion in pairing MBHs, causing either a delay or a boost in the inspiral.\\

Taking into account these studies, we assume that $f_s$ is  set to $1$ when the galaxy is gas poor ($f_{\rm gas}\,{<}\,0.5$) or it does not display a pseudobulge structure (i.e bar related morphology). For the other cases, $f_s$ is a random value extracted from a log-normal distribution whose free parameters are a median of $0.2$ and a variance of $0.6$. The choice of these values is motivated by the shape of the resulting log-normal distribution, which peaks at ${\sim}\,1$ and features a positive skewness (i.e a long tail towards values ${>}\,1$). We stress that during this pairing phase, the two MBHs do not form a bound system. Instead, they can be considered as dual MBHs or dual AGNs in case both of them undergo an active phase.\\

\item \textit{Hardening \& gravitational phase}: Once the dynamical friction phase ends,  the satellite MBH reaches the galactic nucleus of the new galaxy and it binds with the central MBH ($\rm {\sim}\,pc$ separation) forming a massive black hole binary. Hereafter, we will refer to the most massive black hole in the system as \textit{primary black hole} (with mass $\rm M_{BH,1}$), whereas the less massive one is tagged as \textit{secondary black hole} (with mass $\rm M_{BH,2}$). The total mass of the binary and its mass ratio will be denoted as $\rm M_{Bin} \,{=}\, M_{BH,1}\,{+}\,M_{BH,2}$ and $q\,{=}\,\rm M_{BH,2}/M_{BH,1}$, respectively. Regarding the initial properties of the binary orbit, the code assumes that the eccentricity of the binary, $e_0$, starts with a random value between $[0\,{-}\,1]$ while the initial separation, $a_0$, is set to the scale in which $\rm M_{Bulge}({<}\mathit{a_0})\,{=}\,2\,M_{BH,2}$, where $\rm M_{Bulge}({<}\mathit{a_0})$ corresponds to the mass in stars of the hosting bulge within $a_0$. To determine $a_0$, \LGalaxies{} assumes that the bulge mass density profile follows a Sérsic model \citep{Sersic1968} with an index extracted randomly according to the observed distribution of $z\,{=}\,0$ pseudobulges, classical bulges, and ellipticals \citep{Gadotti2009}. We stress that no assumptions are needed to determine the normalization and scale radius of the Sérsic profile since \LGalaxies{} computes self-consistently the redshift evolution of the bulge mass and effective radius (see \citealt{Guo2011,IzquierdoVillalba2019}).\\ %Based on that profile, $a_0$ can be determined by solving \citep{Terzic2005}:
%\begin{equation}\label{eq:a0}
%    \gamma \left(n(3-p),b\frac{a_0}{R_e}\right) \,{=}\,\frac{\rm M_{BH,2}}{2\pi \,\rho_0 \, R_e^3\, n\, b^{n(p-3)}},
%\end{equation}
%where $R_e$ is the bulge effective radius, $\rho_0$ is the central bulge density, $n$ its is the Sérsic index, $\gamma$ is the incomplete gamma function, $p$ and $b$ are two different quantities that depend on the Sérsic index of the bulge:  $p \,{=} \, 1 - 0.6097 n^{-1} + 0.05563 n^{-2}$ and $b\,{=}\, 2n - 0.33 + 0.009876n^{-1}$ \citep{Marquez2000}. As described in Section~\ref{sec:Baryons_LGal},  \LGalaxies{} computes self-consistently the redshift evolution of the bulge mass, effective radius, and type (classical bulge, elliptical, and pseudobulge, see \citealt{Guo2011,IzquierdoVillalba2019}). Therefore, no assumptions are needed to determine the values of $\rho_{0}$ and $R_e$. The main limitation concerns the Sérsic index since \LGalaxies{} does not provide such information. To attach a Sérsic value to each galaxy, we follow the methodology presented in \cite{IzquierdoVillalba2021}. In brief, based on the observational data of \cite{Gadotti2009} it is computed and fitted the observed Sérsic index distribution of $z\,{=}\,0$ pseudobulges, classical bulges, and ellipticals. Based on these distributions and the bulge type, a random Sérsic index is assigned to each simulated bulge.\\

Once the two MBHs bind at the galactic center, the separation ($a_{\rm BH}$) and eccentricity ($e_{\rm BH}$) of the binary system are evolved depending on the environment in which the binary is embedded. If the gas reservoir around the binary ($\rm M_{Res}$) is larger than its total ($\rm M_{Bin}$), the evolution of the system is driven by the interaction with a  circumbinary gaseous disc and then GWs emission \citep{Dotti2015}. Otherwise, the system evolves thanks owing to the interaction with single stars embedded in a Sérsic profile and the emission of GWs \citep{Quinlan1997,Sesana2015}. We refer the reader to \cite{IzquierdoVillalba2021} for a full description of the equations used to evolve $a_{\rm BH}$ and $e_{\rm BH}$. %Therefore, the evolution of $a_{\rm BH}$ and $e_{\rm BH}$ can be expressed as:

\end{itemize}

\subsubsection{Triple interactions}

In some cases, the lifetime of a MBHB can be long enough that a third MBH can reach the galaxy nucleus and interact with the binary  system. In this scenario, multiple outcomes are allowed. To deal with these triple MBH interactions, \LGalaxies{} uses the tabulated values of \cite{Bonetti2018ModelGrid}. Based on the mass of the intruder MBH and the mass ratio of the MBHBs, the model determines if the triple interaction leads to the prompt merger of the MBHB or causes the ejection of the lightest MBH from the system. In case the latter scenario takes place, the separation of the leftover MBHB is computed following \cite{Volonteri2003} and the resulting $e_{\rm BH}$ is selected as a random value between $[0\,{-}\,1]$.% For reference, the typical result of the majority of these interactions is that the primary MBH does not change while the new secondary MBH is the most massive object between the secondary MBH and the MBH that finished its dynamical friction phase. 

\subsubsection{The growth of massive black hole binaries} \label{sec:GrowthBinaries}

%\begin{figure*}
%\centering  
%\includegraphics[width=1.0\columnwidth]{Figures/Mass_ratio_LISA_binaries.pdf}
%\includegraphics[width=1.0\columnwidth]{Figures/spin_LISA_binaries.pdf}
%\caption[]{\textbf{Left panels}: Median binary mass ratio as a function of redshift for LISA sources. Grey areas represent the percentile $\rm 68^{th}\,{-}32^{nd}$. The horizontal dashed line corresponds to $q\,{=}\,0.1$, the value above which we consider that a binary is an equal mass system. Each panel correspond to binaries with different total masses: $10^4\,{-}\,10^5\,\msun{}$ (top), $10^5\,{-}\,10^6\,\msun{}$ (middle) $10^6\,{-}\,10^7\,\msun{}$ (bottom). \textbf{Right panels}: The same as in the left panel but for the spin modulus, $a$, of the primary (solid line) and secondary (dashed line) MBH.}
%\label{fig:Binary_Mass_ratio_spin}
%\end{figure*}

We model the growth of massive black hole binaries in a different way than the one of single MBHs. In particular, our SAM assumes that the accretion rates of the two MBHs of the binary are correlated, as proposed by \cite{Duffell2020}: 
\begin{equation} \label{eq:Relation_accretion_hard_binary_blac_hole}
\dot{\rm M}_{\rm BH_1} =  \dot{\rm M}_{\rm BH_2} (0.1+0.9\mathit{q}),
\end{equation} 
where $\dot{\rm M}_{\rm BH_1}$ and  $\dot{\rm M}_{\rm BH_2}$ are respectively the accretion rate of the primary and secondary MBHs. For simplicity, the latter is set to the Eddington limit, as in \cite{IzquierdoVillalba2021}.% independently of the gas content near the MBHB.

\subsubsection{The growth of massive black holes in the pairing phase} \label{sec:GrowthBinariesPairing}

On top of the growth of MBHBs, \LGalaxies{} deals with the gas accretion of MBHs in the dynamical friction phase. For these objects, the gas consumption is modeled in the same way as for nuclear black holes (Section~\ref{sec:GrwothNuclear}). The growth lasts until the MBH consumes the total gas reservoir stored prior to the merger. This reservoir is set as the sum of all the gas that the MBH accumulated before the galaxy merger (i.e, as a consequence of past disc instabilities or mergers) and an extra amount computed at the time of the galaxy merger as Eq.~\ref{eq:QuasarMode_Merger}, where the cold gas in the equation is assumed to be the one of the satellite galaxy. This extra accumulation of gas is motivated by the hydrodynamical simulations of merging galaxies with central MBHs by \cite{Capelo2015}. The authors showed that during the merging process, the secondary galaxy suffers important perturbations during the pericenter passages around the central one. In such cases, the black hole of the secondary galaxy experiences accretion enhancements mainly correlated with the galaxy mass ratio.

\subsection{Lightcone construction}

In this work, we explore the properties of LISA MBHBs and their host galaxies by making use of a simulated \textit{lightcone}, i.e a mock Universe in which only galaxies whose light has just enough time to reach the observer are included.\\

The main limitation to creating a lightcone by using \LGalaxies{} and MSII is the small box-side length of the latter ($L\,{\sim}\,100\, \mathrm{Mpc}/h$) which is insufficient to represent the Universe beyond redshift $0.025$. To overcome this, we employ the methodology presented in \cite{IzquierdoVillalba2019LC}. In brief, to reach a desired redshift depth the procedure exploits the periodic boundaries of the MSII and replicates its simulated box (i.e fundamental box) a number of $\mathcal{N}$ times in each Cartesian coordinate. Once the replication is made, the methodology establishes the location and line-of-sight (LOS) of the observer. Specifically, the observer is placed at the origin of the first replica while the LOS is chosen according to \cite{KitzbichlerWhite2007} to minimize the structure replication inside the lightcone. As shown by \cite{KitzbichlerWhite2007} selecting a LOS given by $\hat{u}\,{=}\,(n,m,nm)/|(n,m,nm)|$ (where $n$ and $m$ are two different integers with no common factor) implies that the observer will pass through the first periodic image at a distance given by $mnL$. Furthermore, no point of the fundamental box is imaged more than once within $mnL$ when a rectangular footprint of size $1/m^2n \,{\times}\, 1/n^2m$ (radians) is selected. Once the position and LOS of the observer are set, the methodology presented in \cite{IzquierdoVillalba2019LC} places galaxies inside the lightcone by determining the moment at which they (and their corresponding single and binary MBHs) cross the observer past lightcone. To this end, the galaxy merger trees provided by \LGalaxies{} were used given that they accurately follow in time the cosmological evolution of individual galaxies between the DM snapshots with a fine time step resolution (see Section~\ref{sec:DarkMatter}).\\%\dstext{I find the wording of this last sentence slightly confusing/redundant: L-Galaxies does not provide merger-trees itself and the typical timestep length was already introduced. It's not clear to me what message does this sentence need to deliver.} \davcoment{L-Galaxies is generating the galaxy merger trees, right? based, of course, on the halo merger tree.}\\

Since we are interested in $z\,{<}\,3$ galaxies, we have set $\mathcal{N}\,{=}\,48$, $n\,{=}\,5$ and $m\,{=}\,9$. This selection implies a LOS of $\rm (RA,DEC) \,{=}\, (77.1,60.95) \, \rm deg$. We highlight that with this set up no structure repetition would be allowed  up to $z\,{\sim}\,3$  for a FOV of $0.14\,{\times}\,0.25 \, \rm deg^2$. Since in this work we require a large area to have enough statistics we allow some structure repetition and we set the lightcone footprint as a rectangular shape with extension $\rm (\delta RA,\delta DEC) \,{=}\, (45.6,22.5) \, \rm deg$ (corresponding to $1027\, \rm deg^2$).

\section{From gravitationally bound systems to potential LISA sources}\label{sec:PropertiesLISAMBHBs}

In this section, we explore the abundance of MBHBs at different masses and distances. Specifically, we will define LISA MBHBs (or LISA systems) as those binaries whose total mass is $\rm 10^4\,{\leq} \, M_{Bin} \,{\leq}\,10^7\, \msun{}$, $q\,{\geq}\,0.1$\footnote{We have checked that the results presented in this work mildly change when no mass ratio cut is imposed.} and time to coalescence ${<}\,1\, \rm Myr$ (i.e., GW dominated phase). These cuts will ensure that the selected population will emit GWs at $0.1\,{-}\,100\, \rm  mHz$.\\

\begin{figure}
\centering  
\includegraphics[width=1.0\columnwidth]{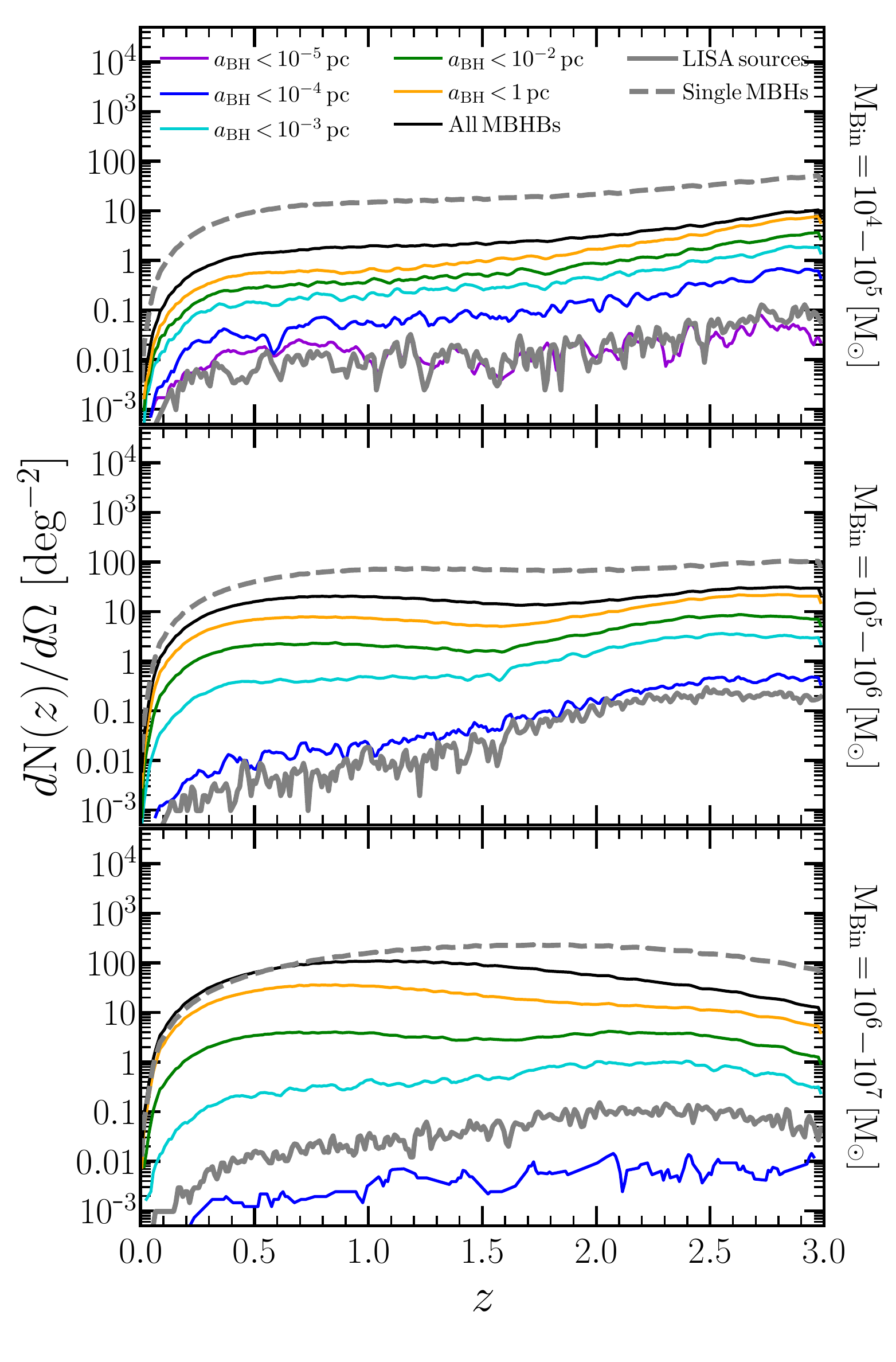}
\caption[]{Number of hard MBHBs (grey solid) and single MBHs (grey dashed) per $\rm deg^2$ ($d\mathrm{N}(z)/d\Omega$, black) as a function of redshift. Different colors represent the same but when the population of MBHBs is divided by semi-major axis ($a_{\rm BH}$). Each panel corresponds to binaries with different total masses: $10^4\,{-}\,10^5\,\msun{}$ (top), $10^5\,{-}\,10^6\,\msun{}$ (middle) $10^6\,{-}\,10^7\,\msun{}$ (bottom). Overall, the figure shows that LISA MBHBs represent $0.1\,{-}\,1$\% of the MBHs between $\rm 10^4\,{-}\,10^7\, \msun$.}
\label{fig:Ndensity}
\end{figure}

In Fig.~\ref{fig:Ndensity} we present the number of MBHBs per $\rm deg^2$ as a function of redshift. The population of MBHBs has been divided into three different mass bins. %For convenience, we have compared the selected binaries with the population of single MBHs with the same mass. 
The lightest systems, binaries with $\rm M_{Bin}\,{=}\,10^4\,{-}\,10^5\,\msun{}$ are less numerous towards low-$z$. For instance, at $z\,{\sim}\,3$ the number of hard binaries can reach up to $10 \, \rm deg^{-2}$ while at $z\,{\sim}\,0.5$ it drops down to $1 \, \rm deg^{-2}$. A similar trend is shown by single MBHs with the same mass but their abundances can be up to ${\sim}\,5$ times higher, regardless of redshift. When the MBHB sample is divided into bins according to the semi-major axis, the large majority of the systems are at $a_{\rm BHB}\, {<}\,1 \, \rm pc$ with very few of them (a number 100 times smaller) at ${<}\,10^{-4} \, \rm pc$. Specifically, the population of LISA MBHBs coincides with binaries at ${<}\,10^{-5} \, \rm pc$ whose number density does not overpass ${\sim}\,10^{-2}\, \rm deg^{-2}$ and represent only 1\% of the whole population of $10^4\,{-}\,10^5\,\msun{}$ hard binaries.\\

MBHBs with a total mass between $\rm 10^5\,{-}\,10^6\,\msun{}$ can be up to 3 times more abundant than the ones in the previous mass range, with values reaching ${\sim}\,10\,{-}\,30 \, \rm deg^{-2}$, regardless of redshift. Despite these large numbers, single MBHs with the same mass are a factor $4$ more numerous, with $\rm{\sim}\,100$ objects per $\rm deg^{2}$. Concerning the MBHB separation, the bulk of the population displays a semi-major axis of $0.1\,{<}\,a_{\rm BHB}\,{<}\,1\, \rm pc$. LISA systems are a much tighter sample with separations ${<}\,10^{-4}\, \rm pc$ and an abundance that decreases towards low-$z$. For instance, at $z\,{\sim}\,3$ the number of objects per $\rm deg^2$ reaches 0.5 while at $z\,{\sim}\,0.5$ it drops down to 0.01. Finally, MBHBs of $10^6\,{-}\,10^7\,\msun{}$ display a different trend with respect to the two previous mass bins. The number of objects in the sky rises from $z\,{\sim}\,3$ ($10 \, \rm deg^{-2}$) down to $z\,{\sim}\,0.5$ ($100 \, \rm deg^{-2}$), redshift at which the abundance of MBHBs reaches its maximum. Interestingly, single and binary MBHBs show the same abundance at $z\,{<}\,1$, pointing out that half of the low-$z$ MBH population of $\rm 10^6\,{-}\,10^7\,\msun{}$ reside in binary systems. Finally, at these masses, LISA MBHBs have separations of $a_{\rm BHB}\,{<}\,10^{-3}\, \rm pc$ with an abundance of ${\sim}\,0.01 \rm \, deg^{-2}$, independently of redshift.\\ %\mv{As a curiosity, In 2010MNRAS.404.2143V assuming rapid binary evolution we were finding smaller binary fractions by a factor about 10.}\\

All together, the numbers shown in this section point out that ${\sim}\,20\%$ of low-mass MBHs are expected to be in relatively wide binaries ($\rm {\sim}\,pc$). Furthermore, this fraction can rise up to ${\sim}\,50\%$ for MBHs of $\rm 10^{6}\,{-}\,10^{7}\, \msun$ at low-$z$. These predictions are relatively larger than the ones presented in other works. For instance, the study of \cite{Volonteri2003} showed that ${\sim}\,5\,{-}\,10\%$ of $z\,{\sim}\,0$ halos host an MBHB\footnote{Notice that \cite{Volonteri2003} reported that these fractions are not constant in time but they rise towards low redshifts and halo mass.}. Among these, ${\sim}\,60\%$ have separation ${>}\,0.1 \, \rm kpc$ and only ${\sim}\,10\%$ feature an advanced hardening stage ($a_{\rm BH}\,{<}\,10 \, \rm pc$). Regardless of these differences (raised most likely by the different approaches and assumptions) previous and current studies highlight that MBHBs at parsecs scales located in low-mass galaxies have a non-negligible contribution to the MBH population. Despite this, the faint nature and the small separation of these MBHBs will challenge their discovery. Detecting and characterizing the presence of Doppler-shifting in broad AGN emission lines can be a good avenue to detect (sub)parsec MBHBs with current spectroscopy facilities such as SDSS \citep[see e.g.][]{Bogdanovic2009,Tsalmantza2011,Montuori2011, Eracleous2012,Shen2013}. On the other hand, upcoming deep surveys like LSST \citep{Ivezic2019} or Athena \citep{Nandra2013} will help in building a complete census of active low-mass MBHs at cosmological distances. Furthermore, the possibility of getting periodic lightcurves from the data of these observatories will open a new path to identify and characterize potential low-mass MBHB candidates \citep[see e.g.][]{Valtonen2008,Graham2015,Charisi2016,Liu2016,Liu2019,Liao2021,Witt2021}.\\

%\mv{I'd conlcude the para with a short summary, saying that overall about 1 in 5 MBHs are expected to be in wide binaries, at $\sim pc$ separation, and for massive binaries at low redshift half of MBHs are in binaries. And something about perspectives for detecting such binaries in current and future surveys}

\section{The hosts of LISA massive black hole binaries} \label{sec:HostProperties}

\begin{figure}
\centering  
\includegraphics[width=1.0\columnwidth]{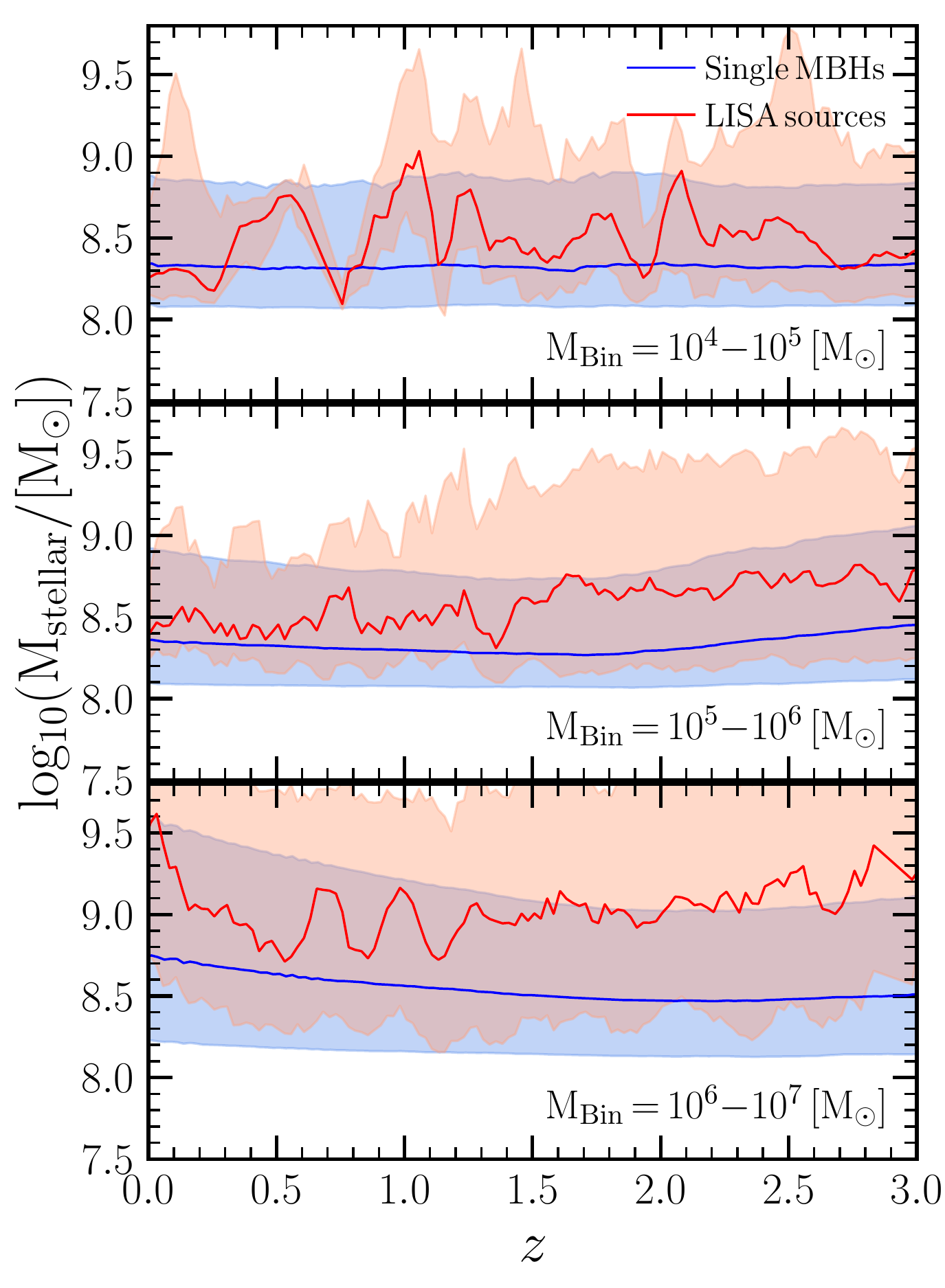}
\caption[]{The median stellar mass of galaxies hosting LISA sources (red) and single MBHs (blue) at different redshifts. Red and blue areas represent the percentile $\rm 16^{th}\,{-}\,84^{th}$. Different panels corresponds to MBHs and MBHBs with masses: $10^4\,{-}\,10^5\,\msun{}$ (top), $10^5\,{-}\,10^6\,\msun{}$ (middle) $10^6\,{-}\,10^7\,\msun{}$ (bottom).}
\label{fig:StellarMass}
\end{figure}

%In this section, we explore the properties of the galaxies in which MBHBs potentially detected by LISA reside. 
%The detection of GWs coming from $10^4\,{-}10^7\, \msun{}$ merging MBHBs will provide a new window to unveil the assembly of MBHs. Furthermore, combining the detection of GW signals emitted from these light MBHBs with the identification of their host galaxies will open the possibility of using MBHBs as standard sirens for cosmology. In this section, we study the properties of the galaxies housing $z\,{<}\,3$ LISA MBHBs to guide their future search.\\

The detection of GWs coming from $10^4\,{-}10^7\, \msun{}$ merging MBHBs, together with the identification of their host galaxies will open a new window to study the population of MBHs and constrain our standard cosmological model. In this section, we aim at guiding the future search of $z\,{<}\,3$ LISA hosts by determining their masses and properties. To this end, in Fig.~\ref{fig:StellarMass} we present the median stellar mass of the galaxies where LISA MBHBs are placed. As shown, these GW sources inhabit low-mass galaxies (i.e \textit{dwarf range}) of $\rm M_{stellar} \,{\sim}\,10^{8-9}\,\msun{}$, expected values according to the MBH-galaxy mass relation (see the left panel of Fig.~\ref{fig:BHMF_Bulge_MBH}). Interestingly, no redshift evolution is seen in the typical mass of the galaxy hosting LISA systems.  %Whereas no significant redshift variations are seen in the hosts of $\rm 10^4\,{-}\,10^5\msun{}$ and $10^5\,{-}\,10^6\msun{}$ MBHBs, the hosts of $\rm 10^6\,{-}\,10^7\msun{}$ show a mild redshift evolution. At $z\,{\sim}\,3$ these galaxies have $\rm M_{stellar}\,{\sim}\,10^{8.5}\, \msun{}$ while at $z\,{\sim}\,0$ the hosts display up to $\rm M_{stellar}\,{\sim}\,10^{9}\, \msun{}$ % \dstext{Extremely minor detail: are these the same, exact galaxies (i.e. did you follow their merger trees)? The wording ``\textit{their masses rise up to...''} seems to suggest that they are. In case they are not, I would slightly rephrase this sentence}. 
In the same figure, we have included the median stellar mass of galaxies harboring single MBHs with the same mass as LISA binaries. This comparison enables us to determine if galaxies housing LISA systems represent untypical hosts of low-mass MBHs. As shown, at fixed MBH mass, LISA systems are placed in slightly more massive galaxies than single MBHs. However, these differences are relatively small with values ${<}\,0.1\,{-}\,0.5\, \rm dex$, depending on the specific mass of the binary. This trend is also seen in recent cosmological hydrodynamical simulations. For instance, the work of \cite{DongPez2023a} showed that when post-processing dynamical delays between galaxies and MBHs merger are taken into account in the \texttt{OBELISK} simulation, merging MBHs tend to be placed slightly above the $\rm M_{BH}\,{-}\,M_{stellar}$ scaling relation than singles MBHs. Despite this, the difference is small enough that they agree within the scatter of the global relation.\\

\begin{figure}
\centering  
\includegraphics[width=1.0\columnwidth]{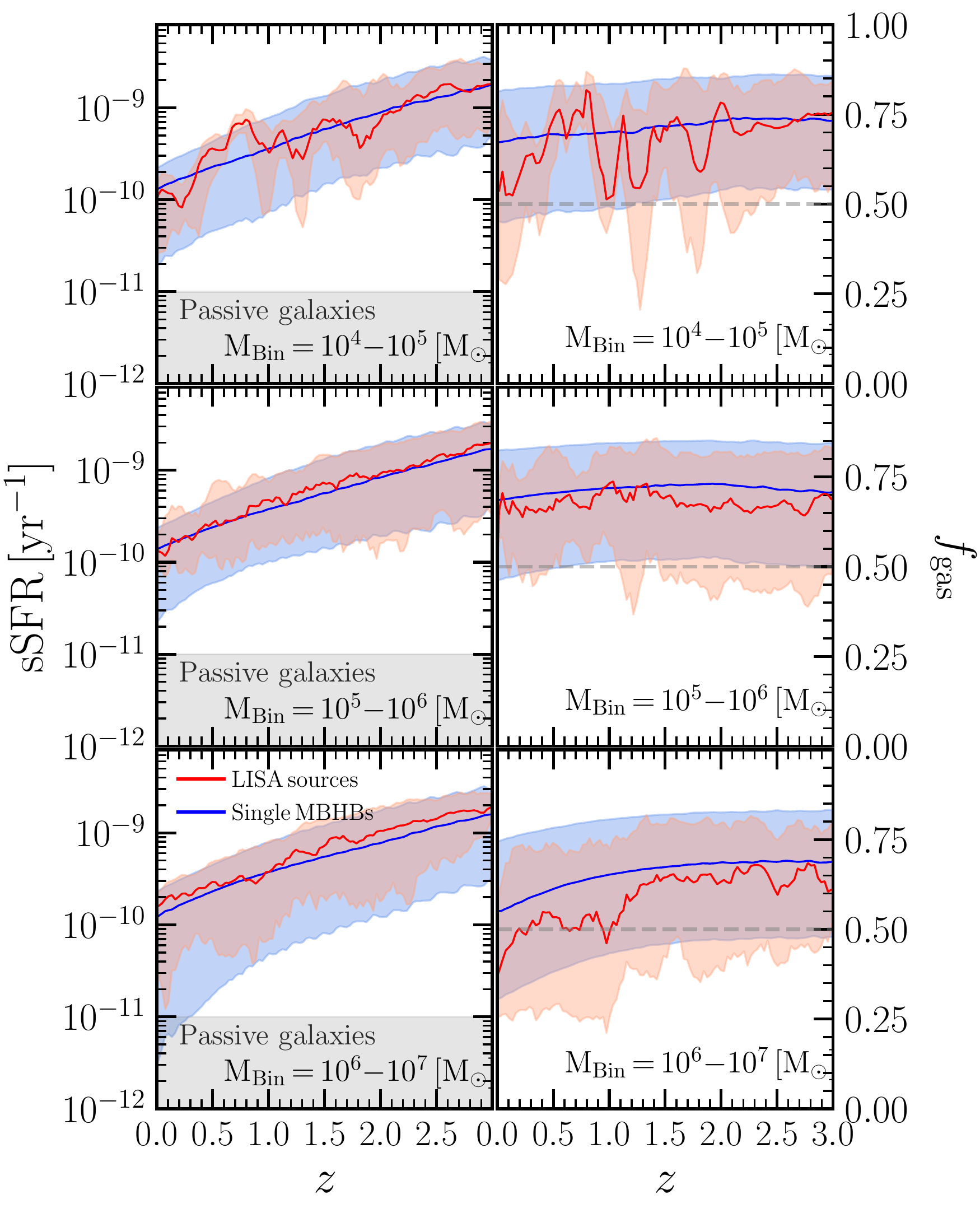}
\caption[]{\textbf{Left panel}: The median specific star formation rate (sSFR) of galaxies hosting LISA sources (red) and single MBHs (blue) at different redshifts. The grey area represents the region where galaxies are considered passive galaxies ($\rm sSFR\,{<}\,10^{-11}\, yr^{-1}$). \textbf{Right panel}: The same as the left panel but for the gas fraction ($f_{\rm gas}$). The grey dotted line highlights the value $f_{\rm gas}\,{=}\,0.5$. In all the panels, red and blue areas represent the  $\rm 16^{th}\,{-}\,84^{th}$ percentile. The results shown in this figure highlight that the galaxies hosting LISA MBHBs are gas-rich and star-forming.}
\label{fig:ssFR_gas_Fraction}
\end{figure}

The results presented above evince that LISA MBHBs will be hosted by dwarf galaxies. Given that these represent the most abundant population of galaxies in the Universe, it is fundamental to determine if the LISA hosts display any distinctive property allowing for their unequivocal identification. Motivated by this, in Fig.~\ref{fig:ssFR_gas_Fraction} we explore the specific star formation rate ($\rm sSFR\,{=}\, SFR/M_{Stellar}$) of the LISA MBHBs hosts. As expected, their values decrease towards low-$z$ but they are systematically larger than $\rm 10^{-10} \, yr^{-1}$, compatibly with a population of star-forming galaxies. %As we can see, all the galaxies show a decreasing sSFR toward low-$z$. Regardless of this decrease, all the hosts display $\rm sSFR\,{>}\,10^{-10} \, yr^{-1}$, compatible with a population of active galaxies. 
For comparison, Fig.~\ref{fig:ssFR_gas_Fraction} displays the sSFR evolution of galaxies hosting single MBHs with the same mass as LISA binaries. As shown, their values and trends are indistinguishable from the ones featured by LISA hosts \citep[see similar results presented in][but for higher redshifts]{DongPez2023a}. Therefore, the stellar activity of dwarf galaxies will not be a good discriminant to unequivocally pinpoint the galaxies where LISA MBHBs reside.\\

On top of the specific star formation, in the right panels of Fig.~\ref{fig:ssFR_gas_Fraction} we show the gas fraction of the LISA hosts, $f_{\rm gas}$, defined as $\rm M_{cold}/(M_{cold} + M_{stellar})$. As shown, LISA binaries are placed in galaxies with large content of gas (${>}\,60\%$) regardless of redshift. Similar trends have been reported by \cite{Li2022} which, by analyzing the outputs of the \texttt{Illustris-TNG} hydrodynamical simulation \citep{NelsonTNGDataReleas2019}, showed that most of the detected LISA MBHBs would be located in gas-rich galaxies with gas fractions in the range of $0.6\,{-}\,0.9$. %\mv{Say something about what Li et al. 2022ApJ...933..104L were finding in Illustris TNG?}. 
In line with what was seen before, the hosts of $\rm 10^4\,{-}\,10^6\msun{}$ MBHBs do not show important redshift variations in $f_{\rm gas}$. Conversely, the galaxies of $\rm 10^6\,{-}\,10^7\msun{}$ MBHBs tend to be less gas-rich towards low-$z$, probably a consequence of their larger stellar mass content (the growth of the stellar component implies a significant depletion of the gas reservoir). Following the comparison done with the stellar mass, Fig.~\ref{fig:StellarMass} includes the gas fraction of galaxies harboring single MBHs with the same mass as our selected LISA systems. Despite the absence of evident differences, the hosts of single MBHs display slightly larger $f_{\rm gas}$. This is most likely due to their smaller stellar mass, suggesting that a smaller amount of their gas was converted into stars.\\

\begin{figure}
\centering  
\includegraphics[width=1.0\columnwidth]{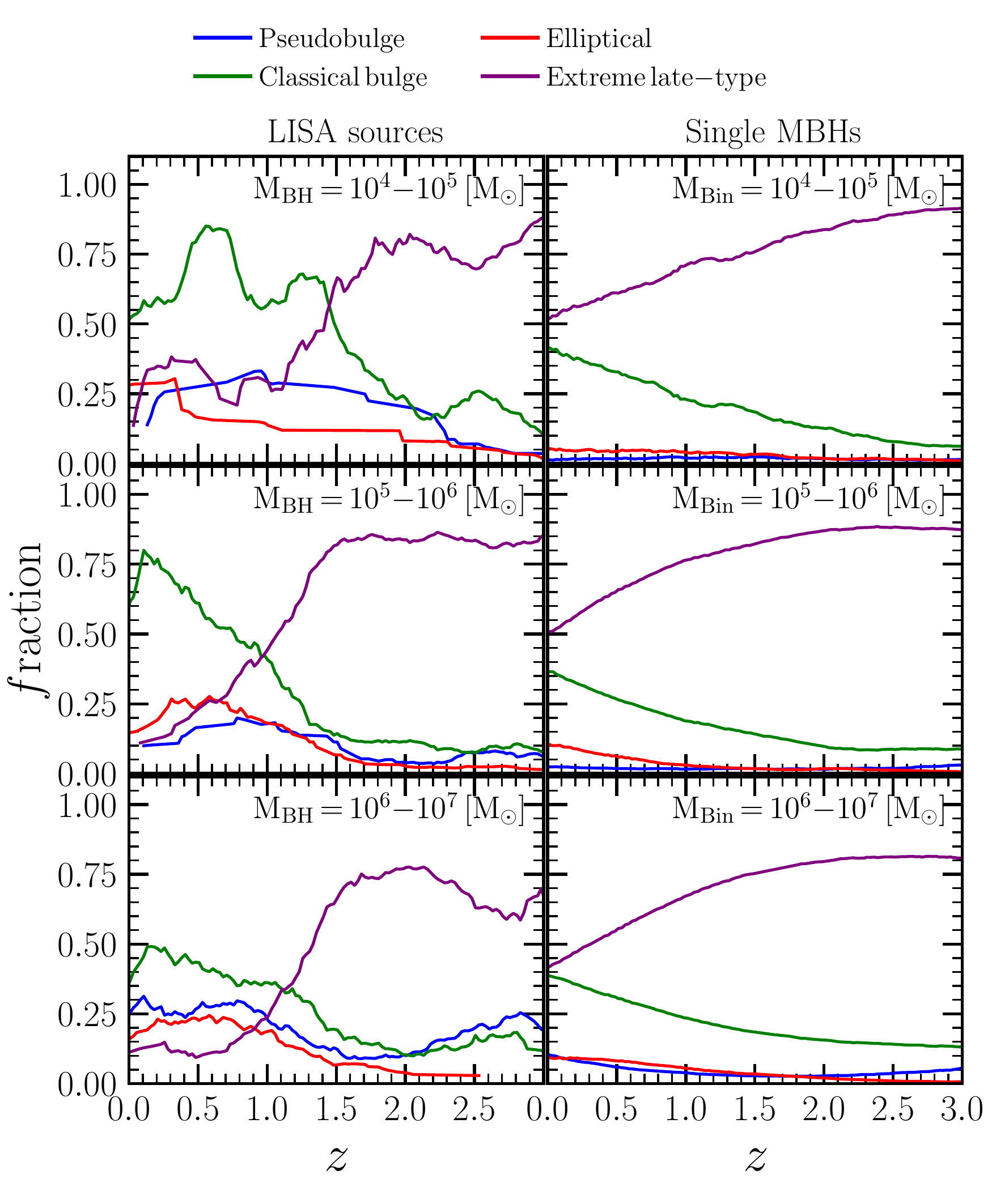}
\caption[]{The fraction LISA MBHB sources (left panels) and single MBHs (right panels) hosted in elliptical (red), classical bulge (green), pseudobulge (blue), and extreme-late type (purple) galaxies at different redshifts. Top, middle, and bottom panels correspond to different mass bins: $10^4\,{-}\,10^5\,\msun{}$, $10^5\,{-}\,10^6\,\msun{}$ and $10^6\,{-}\,10^7\,\msun{}$, respectively. Note that the MBHB hosts have noisier distributions since their number density is up to $2\,{-}\,3$ orders of magnitude smaller than single MBHs (see Fig.~\ref{fig:Ndensity}). In brief, galaxies hosting LISA systems display a disc-dominated morphology whose bulge component is more pronounced toward lower redshifts.}% and in less massive binaries. For the more massive MBHBs, their galaxies show a classical bulge or a tiny small bulge. }}
\label{fig:Morphology}
\end{figure}

\begin{figure}
\centering  
\includegraphics[width=1.0\columnwidth]{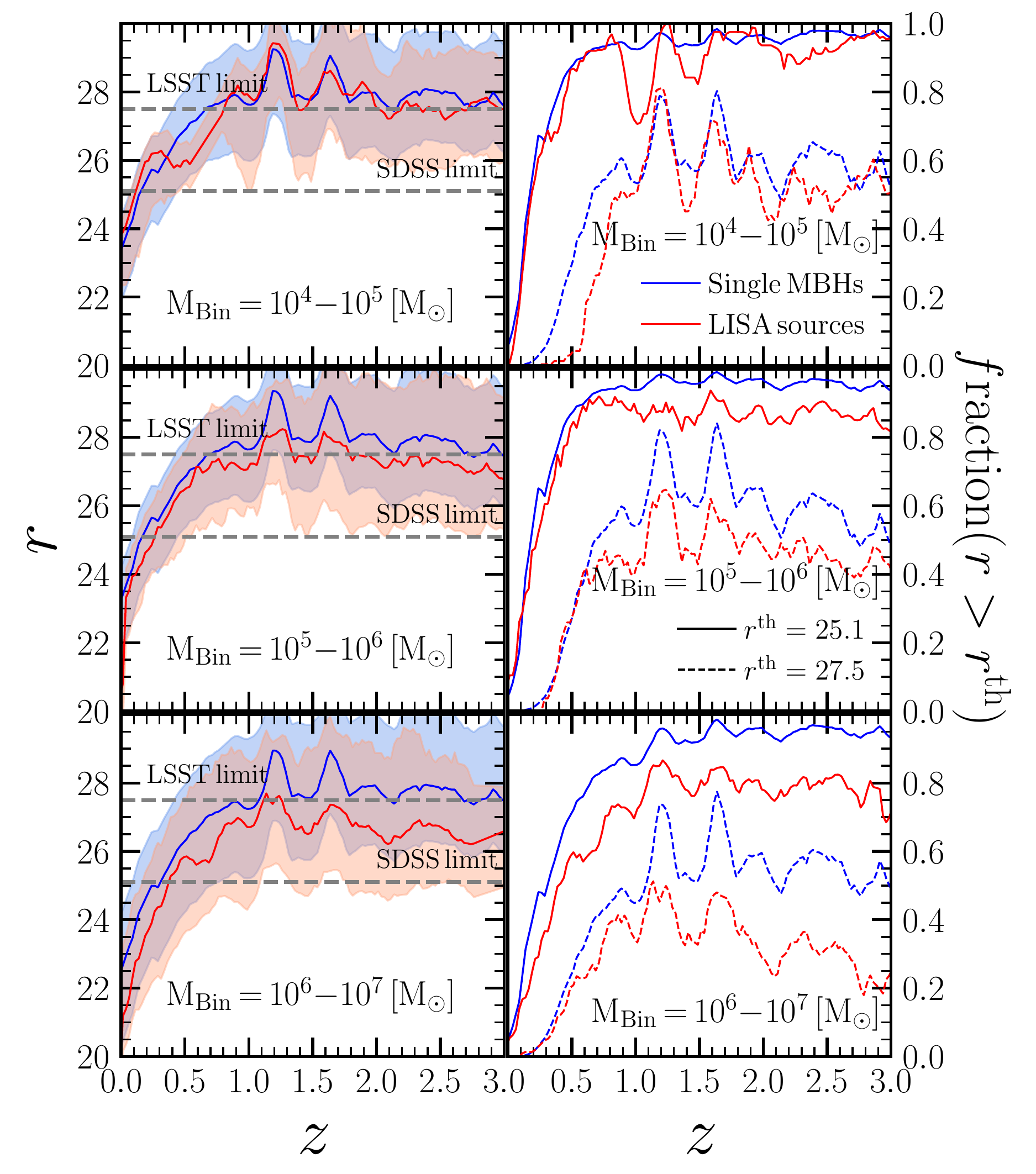}
\caption[]{\textbf{Left panel}: Median $r-$band magnitude of the galaxies hosting LISA MBHBs (red) and single MBHs (blue). Shaded areas represent the percentile $\rm 16^{th}\,{-}\,84^{th}$. Dashed grey lines corresponds to the $r$ band limiting magnitude of SDSS ($r\,{=}\,25.1$) and LSST ($r\,{=}\,27.5$). \textbf{Right panel}: Fraction of galaxies hosting LISA MBHBs (red) and single MBHs (blue) whose $r-$band magnitude is larger than $r^{\rm th}\,{=}$ 25.1 (solid line) and 27.5 (dashed line). Top, middle, and bottom panels correspond to different mass bins: $10^4\,{-}\,10^5\,\msun{}$, $10^5\,{-}\,10^6\,\msun{}$ and $10^6\,{-}\,10^7\,\msun{}$, respectively. Overall, the figure shows that LISA MBHBs are hosted in dim galaxies. In fact, half of them are fainter than the detection limit of current photometric surveys.}
\label{fig:Magnitude}
\end{figure}

\begin{figure*}
\centering  
\includegraphics[width=1.\columnwidth]{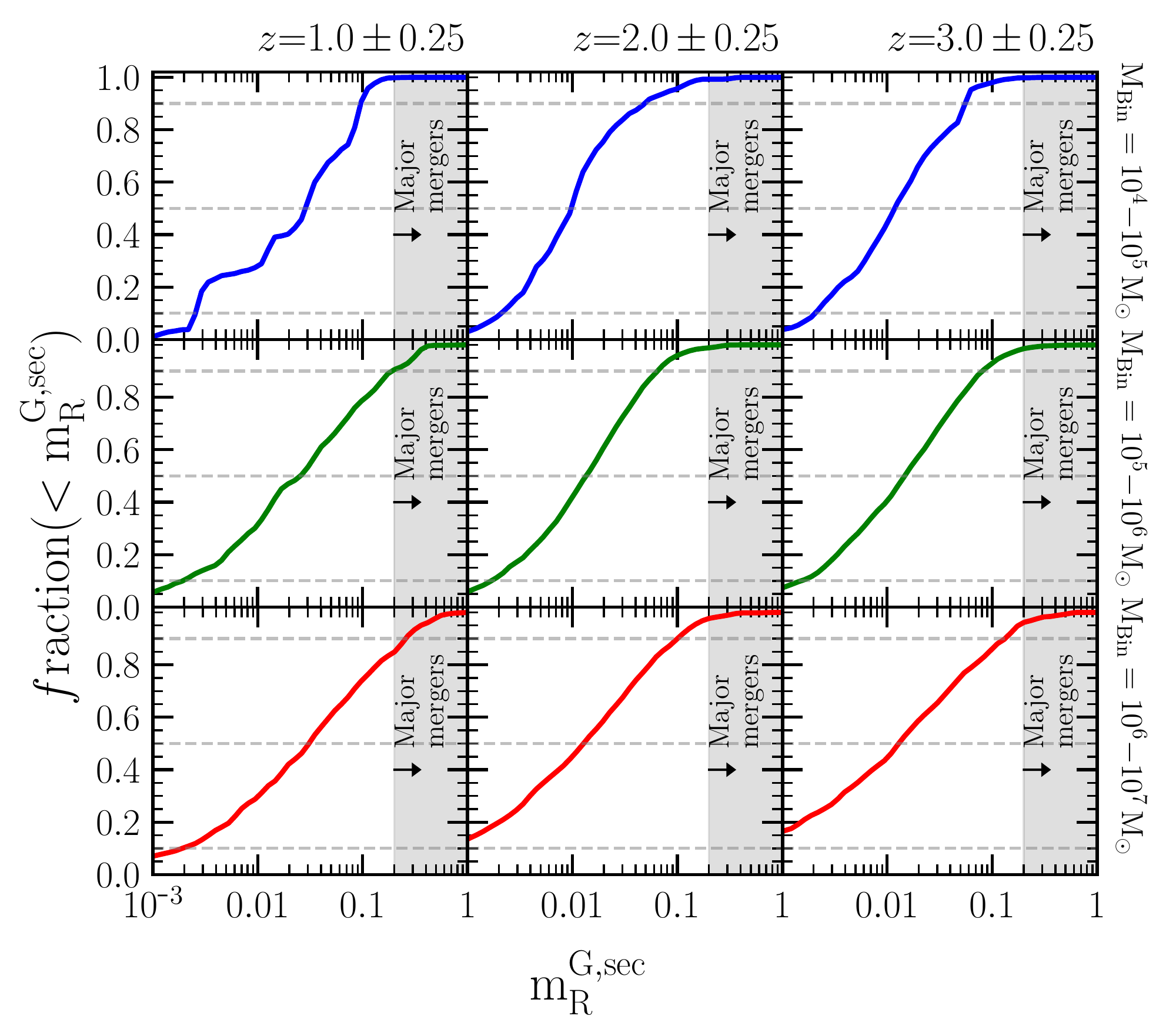}
\includegraphics[width=1.\columnwidth]{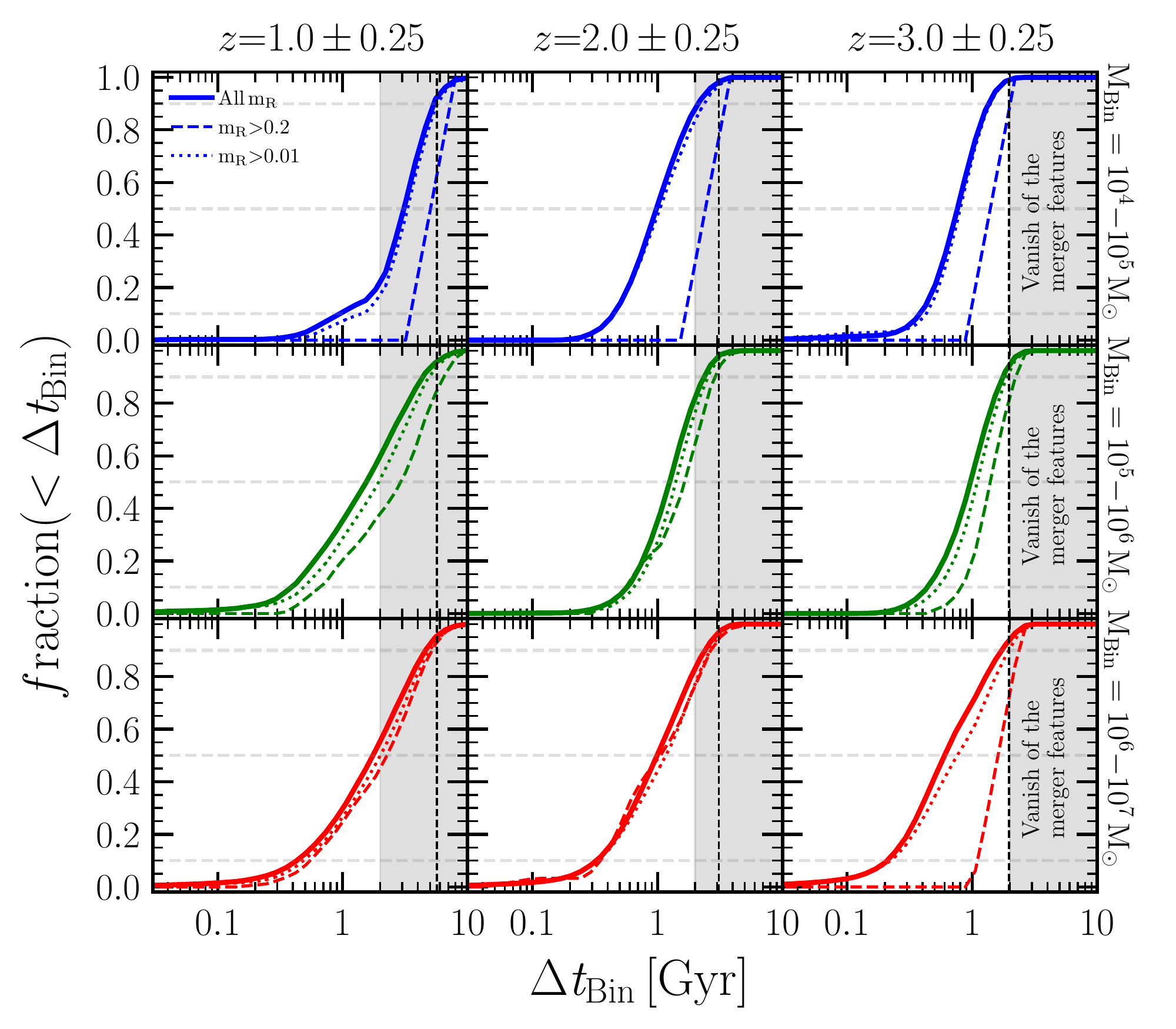}
\caption[]{\textbf{Left panel}: Cumulative distribution of the baryonic mass ratio ($\rm m_R^{G,sec}$) of the two galaxies involved in the interaction which brought the secondary MBH to the primary galaxy. Different column corresponds to a different redshift bin: $z\,{=}\,1\,{\pm}\,0.25$  (left), $z\,{=}\,2\,{\pm}\,0.25$ (middle) and $z\,{=}\,3\,{\pm}\,0.25$ (right). The horizontal dashed grey lines highlight the values 0.1, 0.5, and 0.9. The shaded region corresponds to the merger ratios associated with major mergers in \LGalaxies{}. \textbf{Right panel}: The same as the left panel but for the values of  the delay time, $\Delta t_{\rm Bin}$, between the galaxy-galaxy merger and that of the binary. In this case, the shaded area represents the region where the merger signatures have vanished at the time of LISA detection. Dashed and dotted lines represent the same but for galaxy interactions with $\rm m_{R}\,{>}\,0.2$ and $0.01$, respectively. In all panels, vertical dashed lines correspond to the Hubble time at $z\,{=}\,1$ (left panel), $z\,{=}\,2$ (middle panel) and $z\,{=}\,3$ (right panel). In general, the distributions shown in these figures show that more than 50\% of the LISA MBHBs result from galaxy mergers with a baryonic mass ratio larger than 0.01, with a minor contribution to major mergers. We stress that these baryonic ratios do not imply that the binary display the same mass ratios at inspiral and coalescence. This is due to the gas accretion of MBHBs along the galaxy evolution.}
\label{fig:MergerRatioBroughtSecondary_And_deltaTime}
\end{figure*}    
%In addition to the intrinsic galaxy properties such as star formation rate or gas fraction, it is also interesting to explore if the galaxy morphology could be a guide to select, among all the low-mass galaxies, the ones harboring LISA systems. To this end, in the right panel of Fig.~\ref{fig:Morphology} we show the fraction 

In addition to the intrinsic galaxy properties such as star formation rate or gas fraction, in Fig.~\ref{fig:Morphology} we explore if the galaxy morphology could be a guide to select low-mass galaxies harboring LISA systems. Specifically, Fig.~\ref{fig:Morphology} shows the fraction of LISA MBHBs placed in ellipticals and spiral galaxies (divided into pseudobulges, classical bulges, and extreme late-type, see Section~\ref{sec:bulges}). As shown, at $z\,{>}\,1.5$  ${\sim}\,75\%$ of MBHBs with $\rm M_{Bin}\,{=}\,10^4\,{-}\,10^7 \, \msun{}$ are hosted in extreme late-type systems. This trend is the result of the merger history of these high-$z$ galaxies, ruled by interactions with small baryonic merger ratios (either extreme minor mergers or smooth accretion) whose capability of building bulges is negligible. Interestingly, at these high-$z$ elliptical and pseudobulge morphologies have a marginal contribution at any mass, representing less than $20\%$ of the total population.

At $z\,{<}\,1.5$ the trends for $\rm 10^4\,{-} \, 10^7 \, \msun{}$ and $\rm 10^5\,{-}\, 10^6 \, \msun{}$ change and ${\sim}\,75\%$ of these MBHBs inhabit disc dominated galaxies with a classical bulge. For systems with $\rm 10^6\,{-}\,10^7\, \msun{}$, classical bulges dominate as well, being the typical bulge structure for 50\% of the population. However, pseudobulges and elliptical structures have large relevance too, representing the other ${\sim}\,50\%$ of the population. As a reference, in the left panels of Fig.~\ref{fig:Morphology} we show the morphological properties of the galaxies hosting single MBHs with the same mass as the LISA MBHBs. Interestingly, the morphology of these galaxies does not share the redshift evolution seen in the LISA hosts. At any redshift and mass, disc-dominated galaxies with an extreme late-type morphology dominate ($40\,{-}\,75\%$) the hosts of low-mass (${<}\,10^7\, \msun{}$) single MBHs. This small difference seen between the morphology of normal dwarf galaxies and LISA hosts could help in the identification of the galaxies where $z\,{<}\,1$ MBHBs of $\rm 10^4\,{-}\,10^6\, \msun{}$ are placed: low-mass galaxies with a more predominant bulge component than the average population are more likely to harbor a LISA system.\\

Finally, we explore the optical counterpart of the galaxies where LISA sources reside. Our aim is to determine the possibility of detecting the LISA host in case no AGN counterpart associated with the MBHB is found, i.e the binary is in an inactive phase. To this end, in Fig.~\ref{fig:Magnitude} we present the apparent magnitude of the LISA hosts in the optical $r$ band\footnote{\LGalaxies{} computes the photometry of each simulated galaxy (in a given set of filters) by using on-the-fly stellar population synthesis models combined with a dust-reddening \cite[see][for further information]{Henriques2015}.}. We stress that similar behaviors are found in the other optical bands such as $g$ or $i$. As expected, high-$z$ galaxies are dimmer than low-$z$ ones. For instance, $z\,{>}\,2$ galaxies have $r\,{\sim}\,27$ while galaxies at $z\,{<}\,0.5$ display $r\,{<}\,25$. Besides,  Fig.~\ref{fig:Magnitude} shows that the magnitude of LISA MBHB hosts does not increase (i.e become dimmer) at $z\,{>}\,1$ but it remains constant. This is the effect of the fast rise of the galaxy star formation towards high-$z$ (see Fig.~\ref{fig:ssFR_gas_Fraction}) which is able to compensate for the effect of the luminosity distance in making sources dimmer. In the same figure,  we have included the median $r$ band magnitude of galaxies harboring single MBHs with the same mass as LISA binaries. As shown, no significant differences are seen between these two samples except for the case of $\rm 10^6\,{-}\,10^7 \, \msun$, in which the hosts of LISA systems are slightly brighter than the ones of single MBHs. This deviation is driven by the stellar mass of the galaxy hosts, which tend to be slightly larger for the case of MBHBs (see Fig.~\ref{fig:StellarMass}).  \\

To guide the reader about the observability of LISA MBHBs host galaxies, in Fig.~\ref{fig:Magnitude} we have highlighted the $r$ band detection limits of SDSS ($r\,{=}\,25.1$) and LSST ($r\,{=}\,27.5$). As shown, SDSS will be only able to detect the optical emission of $z\,{<}\,0.5$ LISA hosts. On the other hand, LSST can extend this detection up to $z\,{\sim}\,2\,{-}\,2.5$. Consequently, these results highlight that without any AGN emission raised by the MBHB, the optical identification of LISA hosts will be only feasible at $z\,{\leq}\,1$, while at higher redshift the galaxies are only borderline detectable. To show this, in the right panel of Fig.~\ref{fig:Magnitude} it is presented the redshift evolution of the fraction of galaxies below the detection limit of SDSS and LSST (i.e not detectable). As shown, at $z\,{>}\,1$ around the 80\% and 50\% of the galaxies will not be found out by the optical surveys SDSS and LSST, respectively.\\%\mv{The magnitude remains rather flat at $z>1$, why? Since the luminosity distance increases the magnitude should became fainter at fixed galaxy luminosity: is the intrinsic luminosity higher because galaxies are more star-forming at the same galaxy mass (cf. Fig. 6)? Can you show me the same plot but with the absolute R magnitude? In any case I'd change "the optical identification of LISA hosts will be only feasible at $z\,{\leq}\,2$." to "the optical identification of LISA hosts will be only feasible at $z\,{\leq}\,1$, while at higher redshift the galaxies are only borderline detectable". I think that a figure with the fraction of galaxies above $r=25$ and $r=27$ vs redshift would be a nice addition}\\

Overall, the results presented in this section indicate that LISA MBHBs will be placed in dim galaxies with small stellar content, large gas fraction, and an active star formation history. However, they will not be peculiar systems since little differences are found with respect to the galaxies housing single MBHs with the same mass as LISA binaries. In this way, the unequivocal identification of LISA hosts through standard galaxy properties will be challenging. Motivated by this, in the following section we explore the presence of merger signatures as a tracer that can be used to select LISA hosts among the population of dwarf galaxies.

\section{Merger signatures: An important feature to detect the hosts of LISA MBHBs?} \label{sec:MergerSignatures}

According to our current paradigm, galaxy mergers are essential requisites for the creation of MBHBs \citep{Begelman1980}. Based on this, galaxies hosting LISA systems might display visible merger signatures raised by the galaxy interaction which led to the MBHB formation. In this section we explore the feasibility of pinpointing LISA MBHB hosts through the identification of merger signatures in dwarf galaxies. We stress that the merger ratios reported in this work correspond to the baryonic ones, i.e accounting for the stellar and gas component. 

%According to our current paradigm, the merger of two galaxies marks the starting point of the MBHB life \citep{Begelman1980}. Right after the galaxy merger, the dynamical friction caused by background stars and gas causes the central MBHs of the two galaxies to be dragged toward the center of the remnant galaxy, eventually forming a gravitational bound system. Consequently, galaxy mergers are essential requisites for the creation of MBHBs. Based on this, galaxies harboring MBHBs must have undergone at least one galaxy merger, leading to the formation of possible visible merger features. In this section, we explore the feasibility of pinpointing LISA MBHB hosts through the identification of these merger signatures. We highlight that all the merger ratios reported in this section correspond to the baryonic ones, i.e accounting for the stellar and gas component. 

\subsection{Merger signatures in the hosts of LISA MBHBs}

The left panel of Fig.~\ref{fig:MergerRatioBroughtSecondary_And_deltaTime} presents the baryonic mass ratio of the galaxy interaction which deposited the secondary MBH to the galaxy, hereafter $\rm m_R^{G,sec}$. For simplicity we have only presented the LISA MBHBs at three different redshifts bins: $z\,{=}\,1\,{\pm}\,0.25$, $2\,{\pm}\,0.25$ and $3\,{\pm}\,0.25$. As shown, the 50\% of the mergers which brought the secondary MBH to the galaxy display $\rm 0.002\,{<}\,m_R^{G,sec}\,{<}\,0.02$, regardless of redshift and binary mass. On the other hand, major mergers ($\rm m_R^{G,sec}\,{>}\,0.2$) contribute for 10\% of the cases.\\ %Therefore, our galaxy formation model suggests that mergers leading to the formation of LISA sources display typically intermediate to small baryonic merger ratios.\\
%The moment at which these mergers %responsible for the MBHB formation 
%took place is another important quantity to take into account. 

Besides merger ratios, another important quantity to take into account is the moment at which these mergers took place. This quantity can provide precious information about still visible signatures related to the galaxy interaction such as stellar tidal tails, bridges, streams, and shell structures  \citep{Toomre1972,Gerber1994,Lotz2004}. For instance, by analyzing hydrodynamical cosmological simulations \cite{Mancillas2019} found out that major and intermediate merger events (i.e stellar merger ratios ${\geq}\,0.1$) leave post-merger signatures with a survival timescale of $0.7\,{-}\,4$ Gyr. While tidal tails remain visible between $0.7\,{-}\,1\, \rm Gyr$, shells, and streams could last up to $3\,{-}\,4\,\rm Gy$ and $1.5\,{-}\,3\, \rm Gyr$ respectively. Taking into account these results, hereafter we will assume that merger signatures last on average $2\, \rm Gyr$, independently of the high redshift explored here. We highlight that this duration of merger signature should be considered as an \textit{upper limit} for high-redshift galaxies. Since the galaxy dynamical time goes as $(1\,{+}\,z)^{-3/2}$ \citep{MoMaoWhite1997}, distortion/non-asymmetric features raised during mergers would prevail up to 8 times less at $z\,{=}\,3$ than at $z\,{=}\,0$ \citep[see further discussion in][]{Volonteri2021}. %\mv{I'd note that is likely an upper limit for high-z galaxies: features should disappear in a few dynamical times and the dynamical time goes as $(1+z)^{-3/2}$, so it's 8 times shorter at $z=3$ compared to $z=0$}. 
On top of this, we stress that we do not take into account the surface brightness of these features and the results presented in this section should be considered as \textit{upper limits}, i.e the merger signature(s) could be present in the galaxy but its associated surface brightness could be low enough that will hamper its detection.\\

\begin{figure}
\centering  
\includegraphics[width=1.\columnwidth]{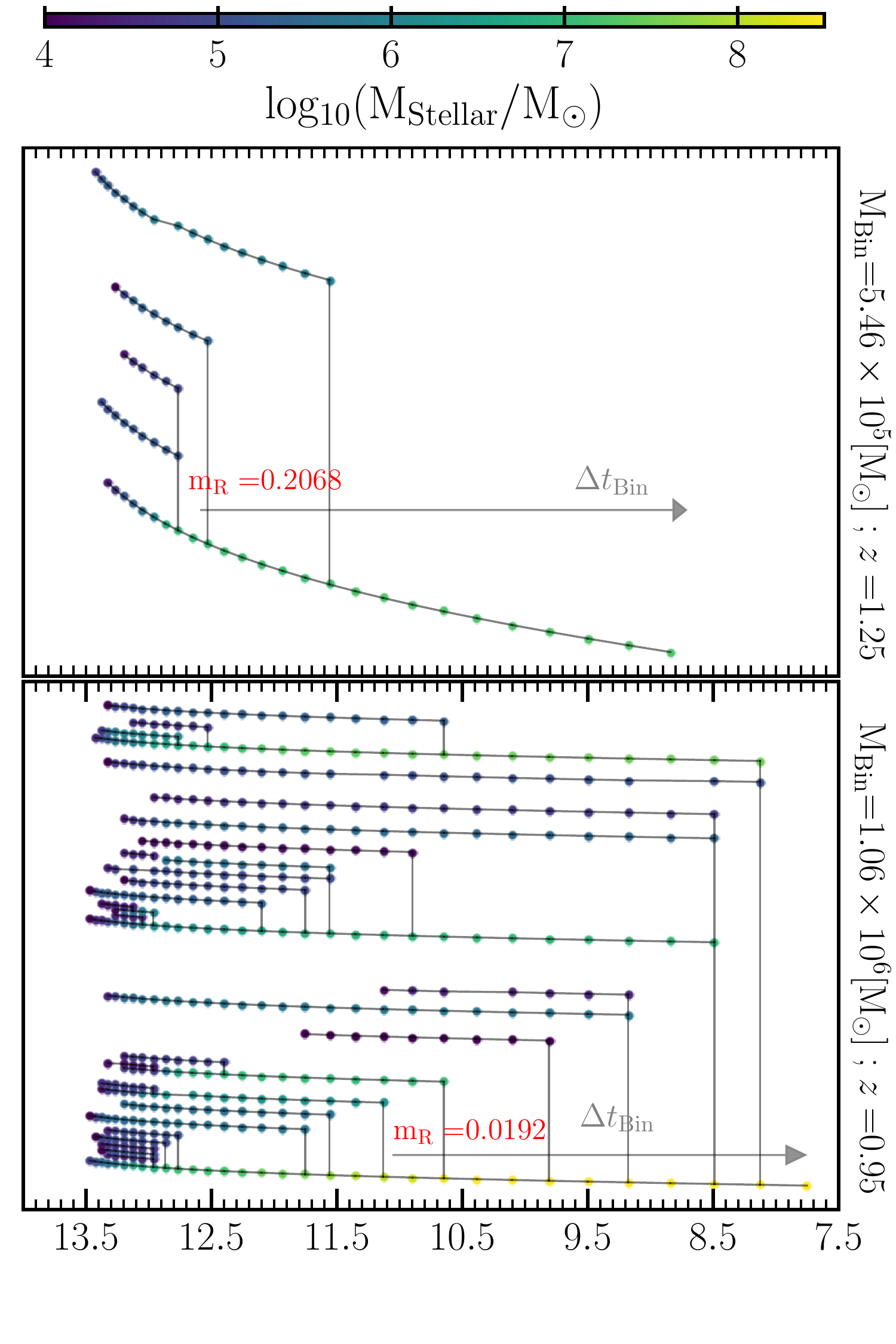}
\caption[]{Galaxy merger tree of the LISA host at $z\,{\sim}\,1.0$. The time evolution is represented as a function of the lookback time ($t_{\rm lookback}$). For reference, $t_{\rm lookback}\,{=}\,8.5\,(11.5)\, \rm Gyr$ corresponds to $z\,{\sim}\,1\, (z\,{\sim}\,13)$. Vertical black lines connect the secondary branches with the main ones (always at the bottom). Thus, the linking points between branches correspond to galaxy mergers. In red we have highlighted the baryonic mass ratio of the merging galaxies that caused the binary formation %\monica{Later on you mention "baryonic mass ratio", are you mentioning only the baryonic one or the dark matter mass ratio? I got confused}. 
The color of each dot encodes the stellar mass of the galaxy. Finally, the extension of the grey arrow corresponds to the length of $\Delta t_{\rm Bin}$. The two galaxy merger trees reported in this figure show that multiple episodes of galaxy mergers can happen before and after the formation of the LISA source.}
\label{fig:MergerTree_Host_LISA}
\end{figure}

\begin{figure}
\centering  
\includegraphics[width=1.\columnwidth]{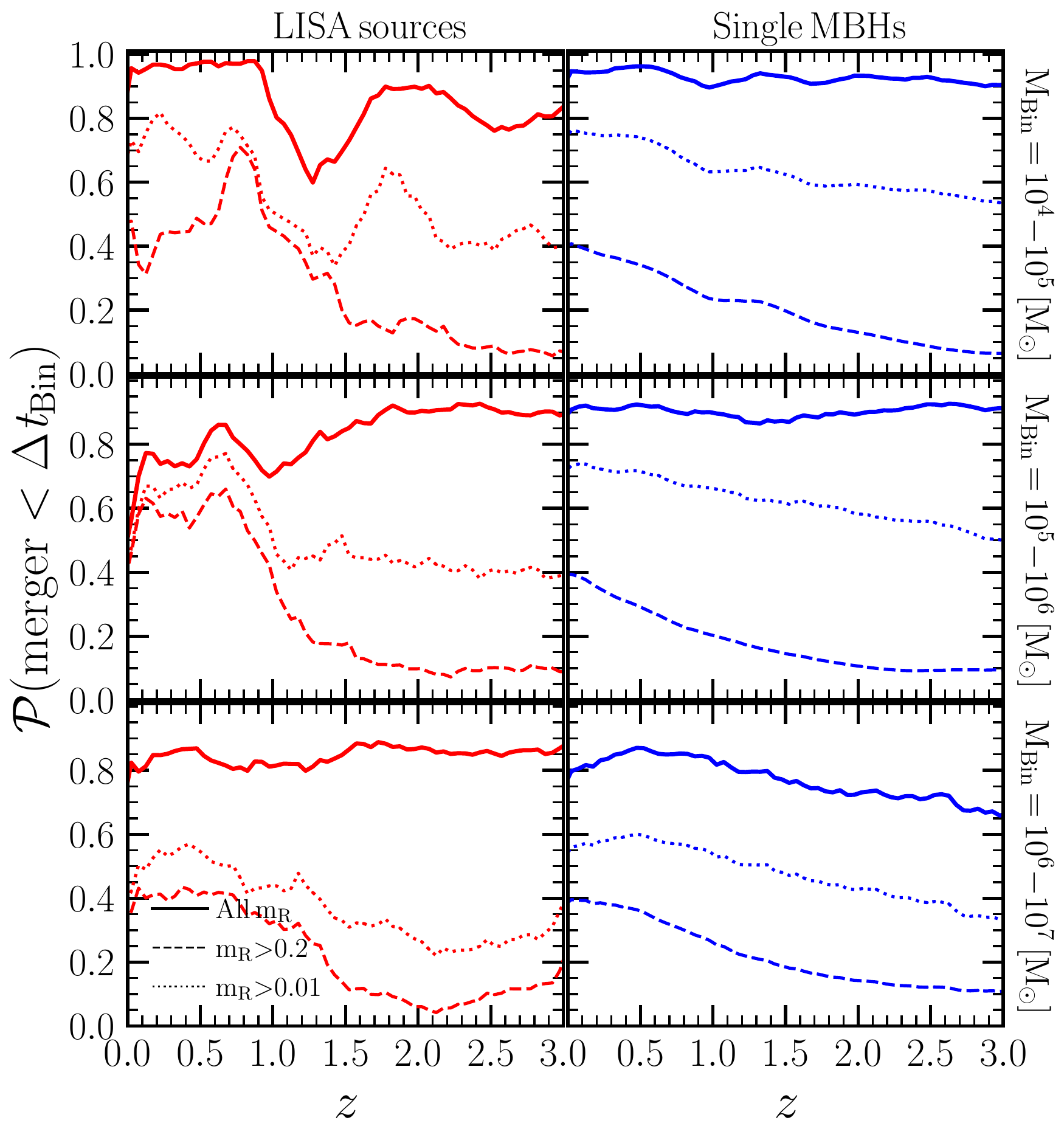}
\caption[]{Probability, $\mathcal{P}$, that a galaxy hosting a LISA MBHB (left) and a single MBHs (right) would undergo a galaxy merger within $\Delta t_{\rm Bin}$. Solid, dashed, and dotted lines correspond to any baryonic mass ratio ($\rm m_R$), $\rm m_R\,{>}\,0.2$ and $\rm m_R\,{>}\,0.01$, respectively. Top, middle and bottom panels display the results for $\rm M_{Bin}\, {=}\, 10^4\,{-}\,10^5\, \msun{}$, $10^5\,{-}\,10^6\, \msun{}$ and $10^6\,{-}\,10^7\, \msun{}$, respectively. The large probability values shown in this figure show that galaxies hosting the LISA source display merger signatures uncorrelated to the one that led to the MBHB formation and coalescence.}
\label{fig:UncorrelatedMergers}
\end{figure}

%In the right panel of Fig.~\ref{fig:MergerRatioBroughtSecondary_And_deltaTime} we explore the presence of merger signatures related to the galaxy interaction which brought the secondary MBH to the galaxy. To this end, we present the cumulative distribution function of $\Delta t_{\rm Bin}$, defined as:

To explore the presence of merger signatures related to the galaxy interaction which brought the secondary MBH to the galaxy, in Fig.~\ref{fig:MergerRatioBroughtSecondary_And_deltaTime} we present the cumulative distribution of $\Delta t_{\rm Bin}$, defined as:
\begin{equation} \label{eq:Delta_T_binary}
 \Delta t_{\rm Bin}\,{=}\,t_{\rm now} \,{-}\, t_{\rm G,sec}   
\end{equation}
where $t_{\rm now}$ is the lookback time associated with the redshift at which the binary is detected, and $t_{\rm G,sec}$ is the lookback time at which the galaxy merger leading to the binary formation took place. To guide the reader, the smaller the value of $\Delta t_{\rm Bin}$, the smaller the time elapsed between the observation (i.e. GW detection) and the galaxy merger involved in the formation of the binary. As shown, the large majority of LISA hosts at $z\,{\sim}\,3$ display $ \Delta t_{\rm Bin}\,{<}\,1\,\rm Gyr$, pointing out that they would still display merger signatures at the moment of the MBHB detection. Galaxies at $z\,{\sim}\,3$ hosting $10^6\,{-}\,10^7\,\msun{}$ are the ones that have the smallest $\Delta t_{\rm Bin}$ values,  with 50\% of them having $\Delta t_{\rm Bin}\,{<}\,0.8\, \rm Gyr$. These small $\Delta t_{\rm Bin}$ values at $z\,{\sim}\,3$ are caused by the fact that MBHB evolution (pairing and hardening) is faster for dense high-$z$ galaxies with relatively massive MBHs.\\  %\mv{Perhaps give the reason for this? I expect this is because binary evolution is faster for dense high-z galaxies with relatively massive MBHs.}

When the distributions are divided by merger ratios, the galaxy interactions associated with major mergers ($\rm m_R\,{>}\,0.2$) are skewed towards smaller $\Delta t_{\rm Bin}$ values, irrespective of the mass bin studied. This points out that, as expected, pairing and hardening evolution is faster in major galaxy mergers.  %\mv{Why? This seems to imply that dual and binary evolution is longer for major mergers, which is not what I'd have expected} 
Despite some displacement can be also seen for intermediate-merger ratios ($\rm m_R\,{>}\,0.01$) in $\rm M_{Bin}\,{>}\,10^{5}\,{-}\,10^{7}\, \msun$, they are systematically smaller than in the case of major interactions. %\mv{I'm confused by this sentence: I see similar displacements for the two mass ratio bins only for the most massive binaries at $z=1$ and for the intermediate mass bin at almost all redshifts} 
According to these distributions, $70\,{-}\,60\%$ of the galaxies hosting $z\,{\sim}\,3$ LISA sources would display a merger signature related to the major/intermediate interaction that led the MBHB formation. These described trends are kept when lower redshifts are studied. However, less number of galaxies housing LISA sources would display the presence of merger signatures. For instance, at $z\,{\sim}\,2$ ($z\,{\sim}\,1$) only 50\%  ($40\,{-}\,30\%$) of the LISA sources associated with major/intermediate galaxy mergers display signs that reveal the interaction, regardless of the MBHB mass. In particular, among all the galaxies harboring LISA systems, the ones with $\rm M_{Bin}\,{=}\,10^4\,{-}\,10^5\,\msun{}$ at $z\,{\sim}\,1$ are the ones in which the presence of merger signatures related to the formation of the MBHB is the smallest, with ${<}\,10\%$ of the cases.\\

\begin{figure*}
\centering  
\includegraphics[width=1.\columnwidth]{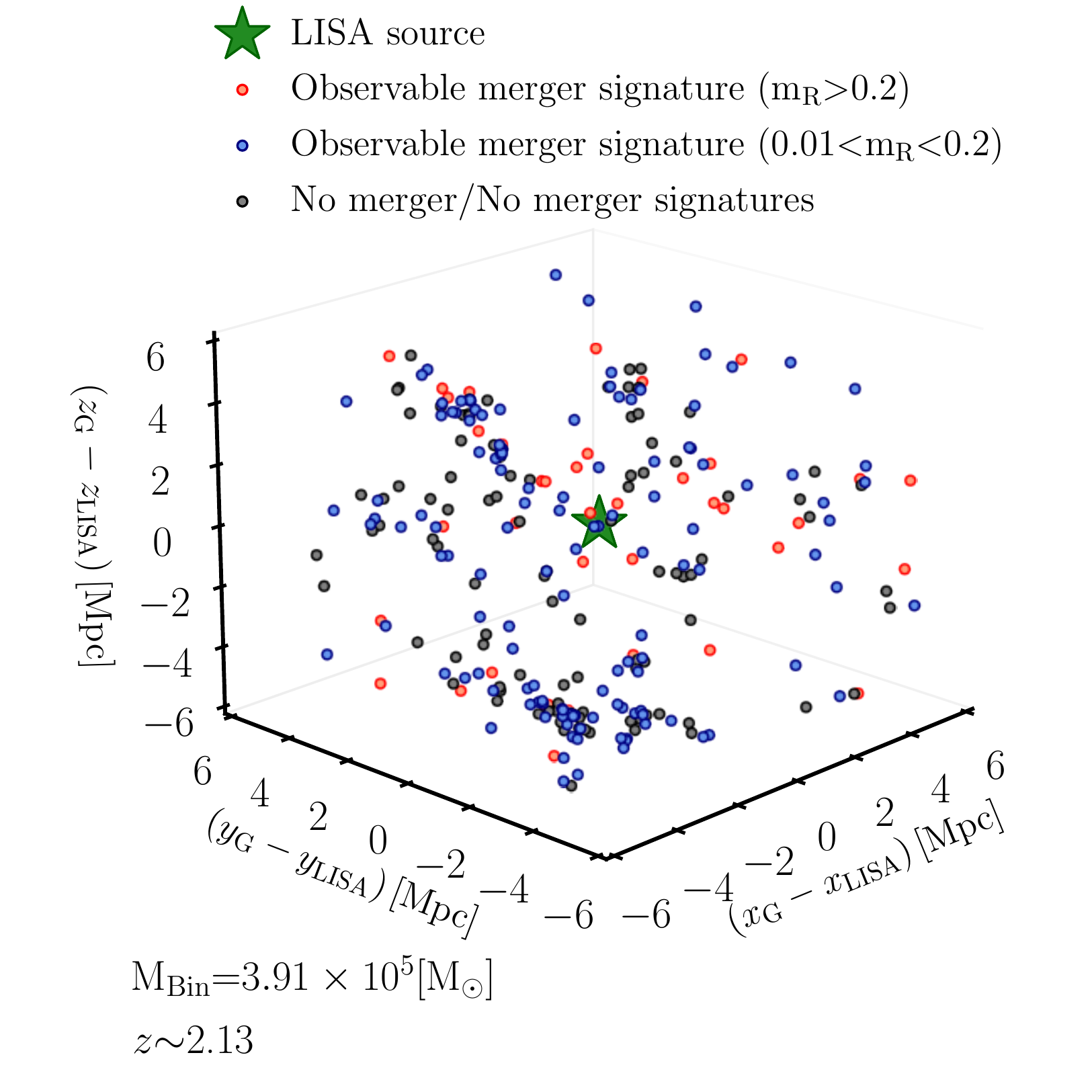}
\includegraphics[width=1.\columnwidth]{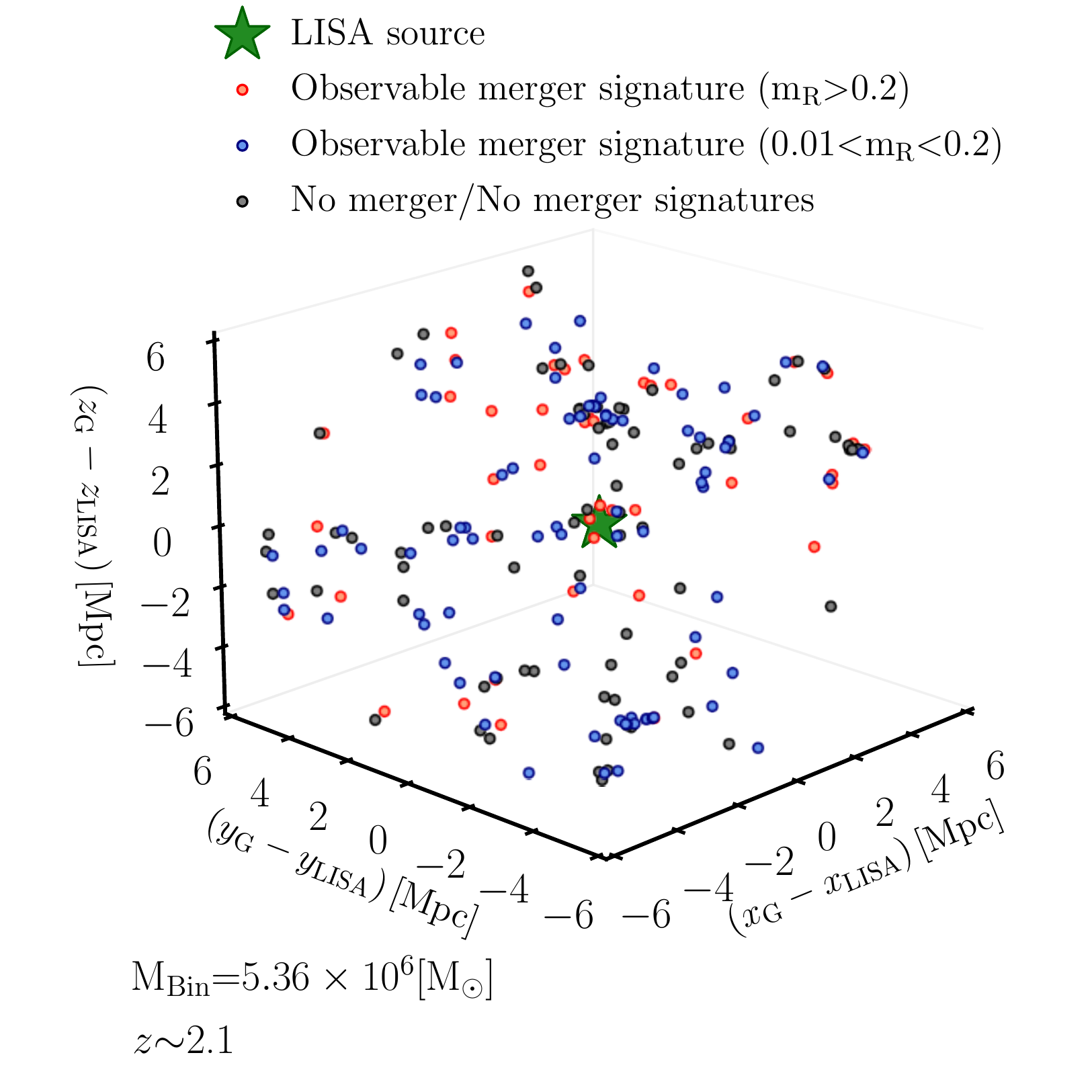}
\caption[]{Sky map of a LISA source at $z\,{\sim}\,2$ in the simulated universe.  While left panel displays the case of $\rm M_{Bin}\,{=}\,10^5\,{-}\,10^6\,\msun{}$ LISA MBHB, right corresponds to a MBHB of $10^6\,{-}\,10^7\,\msun{}$, right. Each dot corresponds to a galaxy of mass $\rm 10^8\,{<}\,M_{Stellar}\,{<}10^9\,\msun{}$ whose distance to the LISA source is smaller than $8 \,\rm Mpc$. Black dots represent galaxies that never experienced a merger or a merger that happened more than $2\, \rm Gyr$ ago. Red dots correspond to galaxies that experienced at least one major merger in the last $2\,\rm Gyr$. Blue dots display the galaxies that experience a merger with mass ratio  $\rm 0.01\,{<}\,m_R\,{<}\,0.2$ in the last $2\,\rm Gyr$. In brief, the galaxy environment around a LISA source is rich in low-mass galaxies with and without merger signatures, hampering the unambiguous identification of the host.}
\label{fig:Enviroment}
\end{figure*}

The analysis of the assembly history of LISA hosts highlights the fact that at the moment of the GW detection, some galaxies harboring LISA sources would still display merger signatures related to the interaction which brought the secondary MBH to the galaxy. Despite this, the lifetime of a galaxy is rather complicated and multiple mergers can happen between the observation time and the merger responsible for the binary formation \citep[see][]{Volonteri2020}. To illustrate this, in Fig.~\ref{fig:MergerTree_Host_LISA} we present the merger tree of two random galaxies hosting $\rm 10^5\,{-}\,10^6\, \msun{}$ and $\rm 10^6\,{-}\,10^7\, \msun{}$ LISA MBHBs at $z\,{\sim}\,1$. As we can see, the galaxy interaction responsible for the MBHB formation took place at $z\,{\sim}\,12\,{-}\,11$ but the galaxy underwent subsequent galaxy mergers until $z\,{\sim}\,1$. Hence, the LISA hosts might show merger signs not correlated with the ones raised by the galaxy interaction leading to the MBHB formation. This can cause confusion in case the merger features are used to infer the mass of the satellite galaxy involved in the interaction or the time at which that merger took place and, thus, determine the binary lifetime.\\

To explore the level of confusion, in Fig.~\ref{fig:UncorrelatedMergers} we present the probability, $\mathcal{P}$, that a LISA MBHB host experienced a further merger within $\Delta t_{\rm Bin}$. As shown, regardless of the MBHB mass, LISA hosts have $\mathcal{P}$ values ${>}\,80\%$. This probability drops when different galaxy merger ratios are taken into account. Concerning major mergers ($\rm m_R\,{>}\,0.2$), the hosts of $\rm 10^4\,{-}\,10^7\msun{}$ MBHBs placed at $z\,{>}\,1.5$ display a ${\sim}\,10\%$ probability of undergoing one of these events within $\Delta t_{\rm Bin}$. This value rises up to $40\,{-}\,50\%$ when lower redshifts are considered. % Hosts of $10^6\,{-}\,10^7\msun{}$ MBHBs show similar behavior, but the redshift evolution is less pronounced having a probability of ${\sim}\,60\%$ at $z\,{<}\,1$ and ${\sim}\,40\%$ at higher redshifts.\\ 
When accounting for intermediate merger ratios ($\rm m_R\,{>}\,0.01$), the results show a similar trend to the one seen in the case of major mergers. However, some differences are seen for the hosts of $\rm 10^4\,{-}\,10^5\msun{}$ and $\rm 10^5\,{-}\,10^6\msun{}$  MBHBs where $\mathcal{P}$ can be a factor ${\sim}\,3$ larger at high-$z$, reaching values of $40\%$. Finally, in the right panel of Fig.~\ref{fig:UncorrelatedMergers} we present the probability $\mathcal{P}$ for systems harboring single MBHs with the same mass as our selected LISA MBHBs. In these cases, it is not possible to set the value of $t_{\rm G,sec}$ (see Eq.~\ref{eq:Delta_T_binary}). As a reference, we have chosen the median $t_{\rm G,sec}$ computed from all the LISA MBHBs placed at the explored redshift. The figure shows small differences for major merger events, where single MBHs at $z\,{<}\,1$ display values of $\mathcal{P}$ around ${\sim}\,10\%$ smaller. Despite that, the trends and values of $\mathcal{P}$ are shared between the LISA and single MBH hosts.\\

The analysis performed in this section about the merger history of LISA hosts reveals that merger features should be taken with caution. The galaxies where LISA MBHBs are placed will display often merger signs that will be uncorrelated from the galaxy interaction that led to the LISA MBHB formation. Besides, dwarf galaxies not hosting MBHBs can also display similar merger features.

\subsection{Merger signatures in the fields of LISA MBHBs}

\begin{figure}
\centering  
\includegraphics[width=1.\columnwidth]{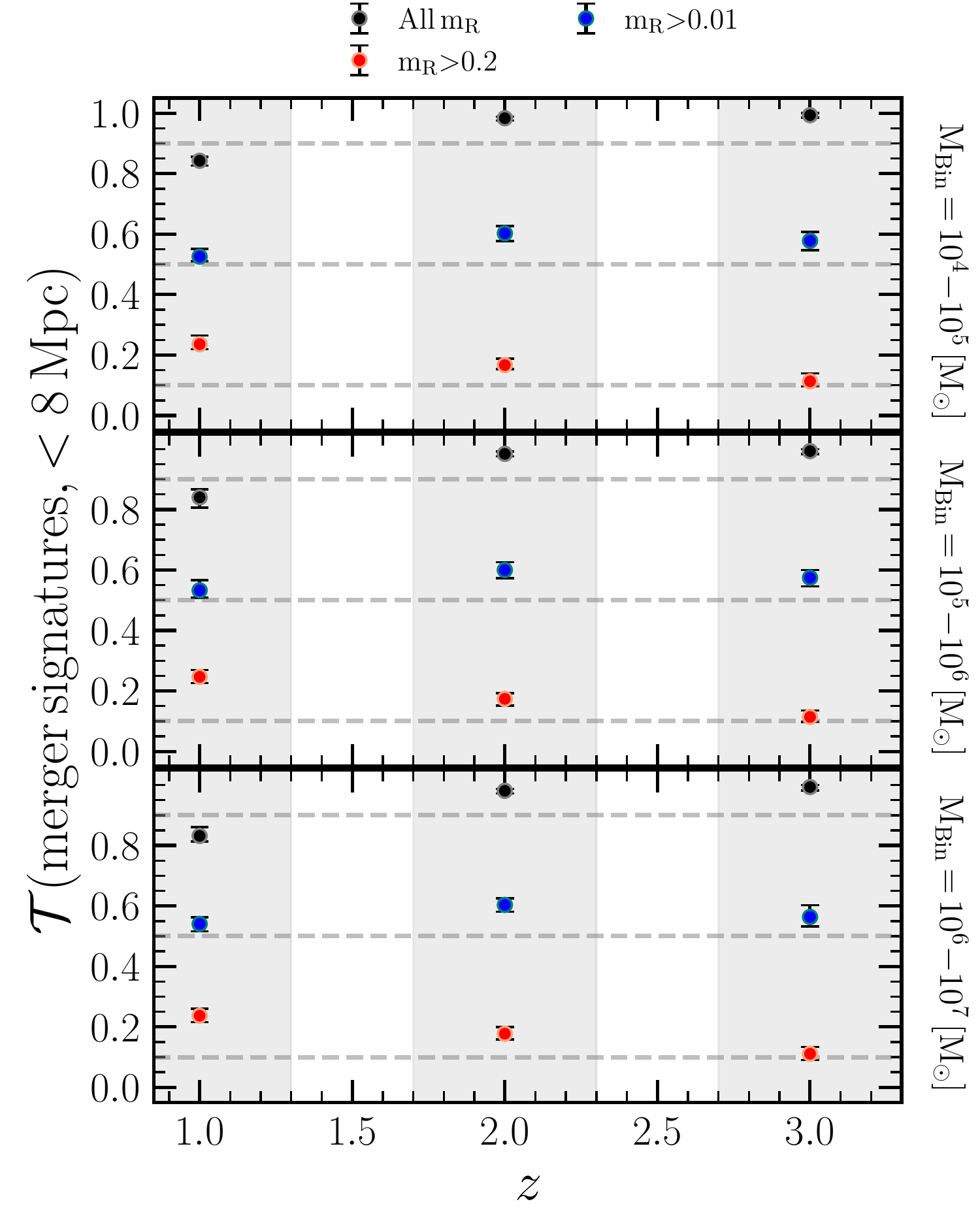}
\caption[]{Probability, $\mathcal{T}$, that the environments of $z\,{=}\,1\,{\pm}\,0.25$, $2\,{\pm}\,0.25$ and $3\,{\pm}\,0.25$ LISA MBHBs have at least one galaxy of $\rm 10^8\,{<}\,M_{stellar}\,{<}\,10^{9}\, \msun{}$ featuring merger signatures. Black, red, and blue dots correspond to the values of $\mathcal{T}$ when the population of $\rm 10^8\,{<}\,M_{stellar}\,{<}\,10^{9}\, \msun{}$ galaxies is divided according to galaxy merger ratios: all ratios, $\rm m_R\,{>}\,0.2$ (major mergers) and $\rm m_R\,{>}\,0.01$ (intermediate mergers), respectively. %The green dots represent the values of $\mathcal{T}$ when considering merger events that had larger baryonic merger ratios than the one which brought the secondary MBH to the galaxy ($\rm m_R^{G,Sec}$). 
The error bars correspond to the $\rm 32^{th}\,{-}\,68^{th}$ percentile. Horizontal dashed lines highlight the values $\mathcal{T}\,{=}\,0.1$, $0.5$ and $0.9$. Shortly, this figure quantifies what shown in Figure \ref{fig:Enviroment}, highlighting the difficulty in identifying the true LISA host.}
\label{fig:Fraction_of_galaxies_with_mergers}
\end{figure}  

%The unequivocal detection of the LISA hosts is ruled by the LISA sky-localization capabilities. 
Recent works have shown that once the GW signal is detected, LISA can localize the MBHB with a sky area that ranges from several hundreds of $\rm deg^2$ to fractions of a $\rm deg^2$, depending on the intrinsic properties of the binary and its redshift. Given these eventually wide areas, the number of potential candidates can be relatively large, hampering the unequivocal detection of the LISA host galaxy. Indeed, using simulated data, \cite{Lops2022} showed that the number of galaxies lying within the LISA sky area can be as large as $10^5$. This implies that the unequivocal detection of LISA hosts would benefit from a pre-selection based on some specific galaxy properties or features. Driven by this, in this section, we explore how feasible it is to identify the LISA host among all the galaxies in its surrounding by using merger features. Hereafter, we will define LISA MBHB environment as the distribution of galaxies within $\rm 8 \, Mpc$ distance. This distance selection has been chosen such as the sky-projected area resembles the sky-localization of LISA when it detects a $z\,{>}\,1$ MBHB at merger time ($\rm {\sim}\,0.1\,{-}\,0.01\,deg^2$, see e.g \citealt{Mangiagli2020,Piro2022}). %\mv{justify the choice of 8 Mpc}. 
Besides this definition, we will only focus on galaxies with a stellar mass of $\rm 10^8\,{<}\,M_{stellar}\,{<}\,10^9\,\msun{}$, i.e systems with stellar masses similar to those of LISA hosts (see Section~\ref{sec:HostProperties}).\\

In Fig.~\ref{fig:Enviroment} we present the environment of two random LISA binaries with mass $10^5\,{-}\,10^6\, \msun{}$ and $10^6\,{-}\,10^7\, \msun{}$ at $z\,{=}\,2$. As we can see, the number of galaxies with similar stellar mass to the LISA hosts is relatively large. When these galaxies are highlighted according to their merger history, a large number of them experienced a galaxy interaction with an intermediate merger ratio ($\rm 0.01\,{<}\,m_R\,{<}\,0.2$) within the last $2 \, \rm Gyr$ (i.e the average lifetime of merger features). On the other hand, the number of galaxies that had a major merger ($\rm m_R\,{>}\,0.2$) within the last $2\, \rm Gyr$ is smaller but not negligible. These plots suggest that the environment of LISA sources would be crowded by dwarf galaxies featuring merger signatures. This would hinder the unequivocal detection of LISA sources by pre-selecting galaxies with recent signs of interaction.\\

To quantify the previous results, in Fig.~\ref{fig:Fraction_of_galaxies_with_mergers} we present the probability $\mathcal{T}$, that a galaxy with $\rm 10^8\,{<}\,M_{stellar}\,{<}\,10^9\,\msun{}$ within the $8\rm \, Mpc$ around the LISA host displays a merger signature\footnote{i.e, the last interaction underwent by the dwarf galaxy took place ${<}\,2\,\rm Gyr$ ago.}. We explored three different redshift bins: $z\,{=}\,1\,{\pm}\,0.25$, $2\,{\pm}\,0.25$ and $3\,{\pm}\,0.25$. As shown, the probability $\mathcal{T}$ displays a redshift evolution, regardless of the considered mass bin. At $z\,{\sim}\,3$ the value of $\mathcal{T}$ reaches 100\% while at $z\,{\sim}\,1$ the probability drops down to ${\sim}\,80\%$. This redshift dependence vanishes when only moderate baryonic merger ratios are considered (i.e. $\rm m_R{>}\,0.01$). Specifically, a rather constant probability of ${\sim}\,60\%$ is reached in these cases. Finally, for dwarf galaxies displaying major mergers ($\rm m_R\,{>}\,0.2$) signatures, the results show an increasing probability towards low-$z$. Whereas at $z\,{\sim}\,3$ galaxies with major merger signatures have ${\sim}\,10\%$ probability of appearing in the LISA host environments, at $z\,{\sim}\,1$ the values of $\mathcal{T}$ increases up to $\,{\sim}\,25\%$.% \mv{I don't understand the very last bit of this sentence... is one of the two $z\,{\sim}\,3$ a typo?}\dstext{I agree with the MV comment: the last sentence is not clear (from ``Whereas...'', in my opinion)}

\section{Caveats} \label{sec:Caveats}

In this section, we discuss some caveats related to the model scheme of MBHs and MBHBs that can cause variations in the results presented in this work. As shown in \cite{IzquierdoVillalba2020}, the physical prescriptions of \LGalaxies{} applied on top of the \texttt{Millennium} merger trees generate a population of active MBHs that resembles the one reported in observations. However, \cite{Spinoso2022} showed that \LGalaxies{} tend to over-predict the growth of high-$z$ ($z\,{>}\,3$) small MBHs when the higher-resolution merger trees of \texttt{Millennium-II} are used and the MBH spin is neglected. In fact, this short-coming is not a unique feature of \LGalaxies{} and other semi-analytical models and hydrodynamical simulations feature it \citep[see e.g][]{Degraf2010,DeGraf2020,Marshall2019,Trinca2022}. This over-prediction could cause the model to generate more frequently over-massive black holes in low-mass galaxies. Consequently, the predictions concerning the stellar content of LISA hosts should be considered as a \textit{lower limit}. In a future paper, (Spinoso et al. in prep.) a further investigation of MBH growth in dwarf galaxies will be performed.\\

Another caveat to take into account is the specific treatment of the hardening phase of MBHBs. Since \LGalaxies{} can not track the stellar distribution inside galactic bulges, it is required the assumption of stellar profiles. Specifically, our SAM uses the Sérsic model which has been shown to be a good profile to represent the bulge population at several cosmological times \citep[see e.g][]{Drory2007,Fisher2008,Gadotti2009,Shibuya2015}. However, as proved by \cite{Biava2019} different assumptions about the stellar mass distributions around the binaries lead to variations in the MBHB lifetimes. For instance, the commonly used Dehnen profile \citep{Dehnen1993} predicts a smaller MBHB lifetime than the Sérsic model used in this work. On a related topic, the \LGalaxies{} model neglects the presence of nuclear stellar clusters (a feature that will be addressed in Polkas et al. in prep and Hoyer et al. in prep). Including the large stellar concentrations of this kind of system would cause a faster hardening phase of the MBHBs, changing the expected distribution of the number density of LISA systems predicted in this work.

%All the caveats presented above highlight the fact that modeling the dynamical evolution and growth of low-mass (single and dual) MBHB is not an easy task and  
%{\color{red} Here I'll add some caveats about the model implementation that can change some results. I should add that the model might overpredict the growth of MBHs in their first stage of growth, with no inclusion of nuclear stellar clusters around the MBHs.} 

\section{Conclusions} \label{sec:Conclusions}
In this paper, we study the properties of galaxies hosting LISA massive black hole binaries with masses between $10^4\,{-}\,10^7\msun$ at $z\,{\leq}\,3$. To this end, we generate a simulated lightcone by using the \LGalaxies{} semi-analytical model applied to the high-resolution \texttt{Millennium-II} merger trees ($5.7\,{\times}\,{10}^7\,{-}\,3\,{\times}\,{10}^{14} \msun$ halo mass range). The version of the SAM used in this work corresponds to the one presented in \cite{IzquierdoVillalba2020,IzquierdoVillalba2021} with an improved prescription for the formation of MBHs (based on \citealt{Spinoso2022}). This \LGalaxies{} variant includes different physical models to tackle self-consistently the growth of MBHs and the dynamical evolution of MBHBs (from the galaxy merger down to the GW inspiral phase). The resulting lightcone contains galaxies, MBHs, and MBHBs of different masses up to $z\,{\sim}\,3$, and features a line of sight $\rm (RA,DEC) \,{=}\, (77.1,60.95) \, \rm deg$ with an angular extension $\rm (\delta RA,\delta DEC) \,{=}\, (45.6,22.5) \, \rm deg$.\\

The main results of this work can be summarized as follows: 

\begin{itemize}

    \item LISA systems represent less than 1\% of the $10^4\,{-}\,10^7\msun$ MBH population, and their abundance is not strongly correlated with the mass of the binary. While at $z\,{\sim}\,3$ LISA systems with a total mass of $10^4\,{-}\,10^7\, \msun$ display abundances of ${\sim}\,0.1 \rm \, deg^{-2}$, at $z\,{\sim}\,0.5$ they have ${\sim}\,0.01 \, \rm deg^{-2}$.\\
    
    %While at $z\,{\sim}\,3$ LISA systems with a total mass of $10^4\,{-}\,10^5\, \msun$ display abundances of ${\sim}\,0.01$ object/$\rm deg^{2}$, systems with $10^6\,{-}\,10^7\, \msun$ reach ${\sim}\,0.1$ objects/$\rm deg^{2}$ at comparable redshifts.\\ %\mv{I'm confused by this bullet point. If I look at fig. 4, I consider "LISA sources" those with a thick solid grey curve. The number of these sources is not strongly dependent on mass or redshift: it's between 0.01 and 0.1 for all redshifts and mass ranges. What am I missing? }\\

    \item LISA MBHBs are hosted in \textit{dwarf galaxies} of $10^8\,{-}\,10^9 \msun{}$ stellar masses at any $z\,{<}\,3$. At fixed total black hole mass, these hosts are ${\sim}\,0.1\,{-}\,0.5\, \rm dex$ more massive than the ones housing single MBHs. However, these differences are not large enough to be easily measured from an observational point of view.\\
    
    \item The galaxies hosting LISA MBHBs are gas-rich (gas fraction ${>}\,0.6$) and star-forming galaxies ($\rm sSFR\,{>}\,10^{-10}\, yr^{-1}$). Galaxies hosting single black holes  of comparable mass share similar properties.\\ 
        
    \item LISA hosts have a disc-dominated morphology whose spheroidal component varies with redshift. While systems at $z\,{>}\,1$ display a tiny or negligible bulge contribution ($\rm B/T\,{<}\,0.01$), at lower redshifts the spheroidal component takes a more important role ($\rm 0.01\,{<}\,B/T\,{<}\,0.7$). Galaxies harboring single MBHs with the same mass as LISA sources are also placed in disc-dominated galaxies. However, their bulge importance does not show any redshift evolution, having a negligible contribution to the galaxy morphology ($\rm B/T\,{<}\,0.01$).\\
    
    %LISA hosts display a disc-dominated morphology with a classical bulge component whose contribution to the total galaxy stellar mass varies from $0.1\,{-}70\%$. Galaxies harboring single MBHs with the same mass as LISA sources are also placed in disc-dominated galaxies with a classical bulge. However, for these galaxies, the bulge contribution is higher, with values larger than 1\% of the total galaxy stellar content.\\

    \item LISA MBHBs are placed in faint galaxies, with a magnitude in the $r$ optical band larger than $20$. Taking into account this and the fact that the MBHB might be in an inactive phase (i.e no AGN emission), current optical facilities will be only able to identify the LISA hosts at $z\,{\leq}\,0.5$. The use of surveys like LSST will extend this search up to $z\,{\sim}\,2$.
    
\end{itemize}  

The small differences found (at a fixed MBH mass) between the hosts of LISA MBHBs and single MBHs underline that the unequivocal identification of LISA hosts through standard galaxy properties will be challenging. Motivated by this, we also explore the presence of merger signatures as a property that can be used to select LISA hosts among the population of dwarf galaxies. The main results on this topic can be summarized as:

\begin{itemize}    

    \item Around 80\% of the secondary MBHs forming LISA MBHBs were deposited in the galaxy after the accretion of a companion galaxy with ${\sim}\,5\,{-}\,100$ times smaller  baryonic mass. These interactions induced in the LISA galaxy host merger signatures which are visible for 80\% of the cases at $z\,{\sim}\,2$. This fraction drops down to 40\% for $z\,{\sim}\,1$ hosts.\\ 

    %\monica{I find difficult to understand this item}  Around 80\% of the LISA systems have secondary MBHs which were deposited in the galaxy after an interaction with a baryonic mass ratio between $0.01\,{-}\,0.2$. Around 80\% of the LISA hosts at $z\,{\sim}\,2$ exhibit visible merger signatures related to these interactions but this fraction drops down to 40\% for $z\,{\sim}\,1$ hosts.\\
    
    \item  Given the long lifetime of LISA MBHBs, their hosts have a high chance of undergoing several galaxy interactions. This implies that LISA hosts will have a large probability of displaying merger signatures uncorrelated with the ones produced by the merger which led to the binary formation. In particular, we find that this probability mildly depends on the binary mass and it rises towards low-$z$, with values reaching ${\sim}\,60\,{-}\,80\%$ at $z\,{<}\,1$ when accounting for mergers with a baryonic mass ratio of ${>}\,0.01$.\\

    %80\% of the LISA hosts at $z\,{\sim}\,2$ exhibit visible merger signatures related to the galaxy interactions which brought the secondary MBH to the galaxy. This fraction drops down to 40\% for $z\,{\sim}\,1$ hosts. However, during the lifetime of LISA MBHBs their hosts undergo several galaxy interactions. The probability of undergoing these kind of events rises towards low-$z$, with values saturating at ${\sim}\,80\,{-}\,90\%$ when accounting for mergers with a baryonic mass ratio of ${>}\,0.01$ at $z\,{<}\,1$. This large probability implies that most of the LISA hosts will display merger signatures uncorrelated with the ones produced by the merger which  brought the secondary MBH to the galaxy.\\
    
    \item The environments of LISA hosts will be populated by a large number of dwarf galaxies displaying signs of recent mergers as well. In fact, ${\sim}\,60\%$ of them will have merger signatures caused by galaxy interactions with intermediate baryonic merger ratio ($>0.01$), i.e similar to the ones displaying LISA hosts. This number drops down to $20\%$ when only major mergers are considered.\\
    
\end{itemize}   

Taking into account all the results summarized above, LISA hosts can be expected to be faint  low-mass galaxies whose intrinsic characteristics such as star-forming activity, or extrinsic properties like merger signatures will not be good tracers for their unequivocal identification. In view of these results, other approaches  for pinpointing LISA hosts should be proposed and studied. For instance, strategies based on the search for EM counterparts associated to faint AGNs with  light curves and spectra characteristic of a merger \citep{Dascoli2018,Yuan2021}, would be an optimal avenue \citep{Mangiagli2022,Lops2022,DongPaez2023}. %\sout{\monica In future works, we will explore the EM emission raised by LISA MBHB, studying if these features will allow the unequivocal identification of the LISA hosts.} \mv{produced instead of raised}

\section*{Acknowledgements}
D.I.V. and A.S. acknowledge the financial support provided under the European Union’s H2020 ERC Consolidator Grant ``Binary Massive Black Hole Astrophysics'' (B Massive, Grant Agreement: 818691). M.C. acknowledges funding from MIUR under the grant PRIN 2017-MB8AEZ and from the INFN TEONGRAV initiative. MV acknowledges funding from the French National Research Agency (grant ANR-21-CE31-0026, project MBH\_waves) and from the Centre National d’Etudes Spatiales. DS acknowledges funding from National Key R\&D Program of China (grant no. 2018YFA0404503), the National Science Foundation of China (grant no. 12073014), the science research grants from the China Manned Space project with no. CMS-CSST2021-A05 and Tsinghua University Initiative Scientific Research Program (no. 20223080023). S.B. acknowledges partial support from the project PGC2018-097585-B-C22. 


\bibliographystyle{aa} 
\bibliography{references}

\begin{thebibliography}{149}
\expandafter\ifx\csname natexlab\endcsname\relax\def\natexlab#1{#1}\fi

\bibitem[{{Agarwal} {et~al.}(2012){Agarwal}, {Khochfar}, {Johnson}, {Neistein},
  {Dalla Vecchia}, \& {Livio}}]{Agarwal2012}
{Agarwal}, B., {Khochfar}, S., {Johnson}, J.~L., {et~al.} 2012, \mnras, 425,
  2854

\bibitem[{{Akiyama} {et~al.}(2018){Akiyama}, {He}, {Ikeda}, {Niida}, {Nagao},
  {Bosch}, {Coupon}, {Enoki}, {Imanishi}, {Kashikawa}, {Kawaguchi}, {Komiyama},
  {Lee}, {Matsuoka}, {Miyazaki}, {Nishizawa}, {Oguri}, {Ono}, {Onoue}, {Ouchi},
  {Schulze}, {Silverman}, {Tanaka}, {Tanaka}, {Terashima}, {Toba}, \&
  {Ueda}}]{Akiyama2018}
{Akiyama}, M., {He}, W., {Ikeda}, H., {et~al.} 2018, \pasj, 70, S34

\bibitem[{{Amaro-Seoane} {et~al.}(2017){Amaro-Seoane}, {Audley}, {Babak},
  {Baker}, {Barausse}, {Bender}, {Berti}, {Binetruy}, {Born}, {Bortoluzzi},
  {Camp}, {Caprini}, {Cardoso}, {Colpi}, {Conklin}, {Cornish}, {Cutler},
  {Danzmann}, {Dolesi}, {Ferraioli}, {Ferroni}, {Fitzsimons}, {Gair}, {Gesa
  Bote}, {Giardini}, {Gibert}, {Grimani}, {Halloin}, {Heinzel}, {Hertog},
  {Hewitson}, {Holley-Bockelmann}, {Hollington}, {Hueller}, {Inchauspe},
  {Jetzer}, {Karnesis}, {Killow}, {Klein}, {Klipstein}, {Korsakova}, {Larson},
  {Livas}, {Lloro}, {Man}, {Mance}, {Martino}, {Mateos}, {McKenzie},
  {McWilliams}, {Miller}, {Mueller}, {Nardini}, {Nelemans}, {Nofrarias},
  {Petiteau}, {Pivato}, {Plagnol}, {Porter}, {Reiche}, {Robertson},
  {Robertson}, {Rossi}, {Russano}, {Schutz}, {Sesana}, {Shoemaker}, {Slutsky},
  {Sopuerta}, {Sumner}, {Tamanini}, {Thorpe}, {Troebs}, {Vallisneri},
  {Vecchio}, {Vetrugno}, {Vitale}, {Volonteri}, {Wanner}, {Ward}, {Wass},
  {Weber}, {Ziemer}, \& {Zweifel}}]{LISA2017}
{Amaro-Seoane}, P., {Audley}, H., {Babak}, S., {et~al.} 2017, arXiv e-prints,
  arXiv:1702.00786

\bibitem[{{Angulo} {et~al.}(2012){Angulo}, {Springel}, {White}, {Cole},
  {Jenkins}, {Baugh}, \& {Frenk}}]{Angulo2012}
{Angulo}, R.~E., {Springel}, V., {White}, S.~D.~M., {et~al.} 2012, \mnras, 425,
  2722

\bibitem[{{Angulo} \& {White}(2010)}]{AnguloandWhite2010}
{Angulo}, R.~E. \& {White}, S.~D.~M. 2010, \mnras, 405, 143

\bibitem[{{Begelman} {et~al.}(1980){Begelman}, {Blandford}, \&
  {Rees}}]{Begelman1980}
{Begelman}, M.~C., {Blandford}, R.~D., \& {Rees}, M.~J. 1980, \nat, 287, 307

\bibitem[{{Biava} {et~al.}(2019){Biava}, {Colpi}, {Capelo}, {Bonetti},
  {Volonteri}, {Tamfal}, {Mayer}, \& {Sesana}}]{Biava2019}
{Biava}, N., {Colpi}, M., {Capelo}, P.~R., {et~al.} 2019, \mnras, 487, 4985

\bibitem[{{Binney} \& {Tremaine}(1987)}]{BinneyTremine1987}
{Binney}, J. \& {Tremaine}, S. 1987, {Galactic dynamics}

\bibitem[{{Binney} \& {Tremaine}(2008)}]{BinneyTremaine2008}
{Binney}, J. \& {Tremaine}, S. 2008, {Galactic Dynamics: Second Edition}

\bibitem[{{Bogdanovi{\'c}} {et~al.}(2009){Bogdanovi{\'c}}, {Eracleous}, \&
  {Sigurdsson}}]{Bogdanovic2009}
{Bogdanovi{\'c}}, T., {Eracleous}, M., \& {Sigurdsson}, S. 2009, \apj, 697, 288

\bibitem[{{Bonetti} {et~al.}(2018){Bonetti}, {Haardt}, {Sesana}, \&
  {Barausse}}]{Bonetti2018ModelGrid}
{Bonetti}, M., {Haardt}, F., {Sesana}, A., \& {Barausse}, E. 2018, \mnras, 477,
  3910

\bibitem[{{Bonetti} {et~al.}(2020){Bonetti}, {Rasskazov}, {Sesana}, {Dotti},
  {Haardt}, {Leigh}, {Arca Sedda}, {Fragione}, \& {Rossi}}]{Bonetti2020}
{Bonetti}, M., {Rasskazov}, A., {Sesana}, A., {et~al.} 2020, \mnras, 493, L114

\bibitem[{{Bonoli} {et~al.}(2009){Bonoli}, {Marulli}, {Springel}, {White},
  {Branchini}, \& {Moscardini}}]{Bonoli2009}
{Bonoli}, S., {Marulli}, F., {Springel}, V., {et~al.} 2009, \mnras, 396, 423

\bibitem[{{Bortolas} {et~al.}(2022){Bortolas}, {Bonetti}, {Dotti}, {Lupi},
  {Capelo}, {Mayer}, \& {Sesana}}]{Bortolas2022}
{Bortolas}, E., {Bonetti}, M., {Dotti}, M., {et~al.} 2022, \mnras, 512, 3365

\bibitem[{{Bortolas} {et~al.}(2020){Bortolas}, {Capelo}, {Zana}, {Mayer},
  {Bonetti}, {Dotti}, {Davies}, \& {Madau}}]{Bortolas2020}
{Bortolas}, E., {Capelo}, P.~R., {Zana}, T., {et~al.} 2020, \mnras, 498, 3601

\bibitem[{{Boylan-Kolchin} {et~al.}(2006){Boylan-Kolchin}, {Ma}, \&
  {Quataert}}]{Boylan-Kolchin2006}
{Boylan-Kolchin}, M., {Ma}, C.-P., \& {Quataert}, E. 2006, \mnras, 369, 1081

\bibitem[{{Boylan-Kolchin} {et~al.}(2009){Boylan-Kolchin}, {Springel}, {White},
  {Jenkins}, \& {Lemson}}]{Boylan-Kolchin2009}
{Boylan-Kolchin}, M., {Springel}, V., {White}, S.~D.~M., {Jenkins}, A., \&
  {Lemson}, G. 2009, \mnras, 398, 1150

\bibitem[{{Bromm} \& {Larson}(2004)}]{bromm_larson2004}
{Bromm}, V. \& {Larson}, R.~B. 2004, \araa, 42, 79

\bibitem[{{Capelo} {et~al.}(2015){Capelo}, {Volonteri}, {Dotti}, {Bellovary},
  {Mayer}, \& {Governato}}]{Capelo2015}
{Capelo}, P.~R., {Volonteri}, M., {Dotti}, M., {et~al.} 2015, \mnras, 447, 2123

\bibitem[{Capuzzo-Dolcetta \& Tosta~e Melo(2017)}]{Capuzzo2017}
Capuzzo-Dolcetta, R. \& Tosta~e Melo, I. 2017, Monthly Notices of the Royal
  Astronomical Society, 472, 4013

\bibitem[{{Charisi} {et~al.}(2016){Charisi}, {Bartos}, {Haiman},
  {Price-Whelan}, {Graham}, {Bellm}, {Laher}, \& {M{\'a}rka}}]{Charisi2016}
{Charisi}, M., {Bartos}, I., {Haiman}, Z., {et~al.} 2016, \mnras, 463, 2145

\bibitem[{{Croton} {et~al.}(2006){Croton}, {Springel}, {White}, {De Lucia},
  {Frenk}, {Gao}, {Jenkins}, {Kauffmann}, {Navarro}, \&
  {Yoshida}}]{Croton2006a}
{Croton}, D.~J., {Springel}, V., {White}, S. D.~M., {et~al.} 2006, \mnras, 365,
  11

\bibitem[{{Cuadra} {et~al.}(2009){Cuadra}, {Armitage}, {Alexander}, \&
  {Begelman}}]{Cuadra2009}
{Cuadra}, J., {Armitage}, P.~J., {Alexander}, R.~D., \& {Begelman}, M.~C. 2009,
  \mnras, 393, 1423

\bibitem[{{d'Ascoli} {et~al.}(2018){d'Ascoli}, {Noble}, {Bowen}, {Campanelli},
  {Krolik}, \& {Mewes}}]{Dascoli2018}
{d'Ascoli}, S., {Noble}, S.~C., {Bowen}, D.~B., {et~al.} 2018, \apj, 865, 140

\bibitem[{{Degraf} {et~al.}(2010){Degraf}, {Di Matteo}, \&
  {Springel}}]{Degraf2010}
{Degraf}, C., {Di Matteo}, T., \& {Springel}, V. 2010, \mnras, 402, 1927

\bibitem[{{DeGraf} \& {Sijacki}(2020)}]{DeGraf2020}
{DeGraf}, C. \& {Sijacki}, D. 2020, \mnras, 491, 4973

\bibitem[{{DeGraf} {et~al.}(2021){DeGraf}, {Sijacki}, {Di Matteo},
  {Holley-Bockelmann}, {Snyder}, \& {Springel}}]{DeGraf2021}
{DeGraf}, C., {Sijacki}, D., {Di Matteo}, T., {et~al.} 2021, \mnras, 503, 3629

\bibitem[{{Dehnen}(1993)}]{Dehnen1993}
{Dehnen}, W. 1993, \mnras, 265, 250

\bibitem[{{Dong-P{\'a}ez} {et~al.}(2023{\natexlab{a}}){Dong-P{\'a}ez},
  {Volonteri}, {Beckmann}, {Dubois}, {Mangiagli}, {Trebitsch}, {Vergani}, \&
  {Webb}}]{DongPaez2023}
{Dong-P{\'a}ez}, C.~A., {Volonteri}, M., {Beckmann}, R.~S., {et~al.}
  2023{\natexlab{a}}, arXiv e-prints, arXiv:2303.09569

\bibitem[{{Dong-P{\'a}ez} {et~al.}(2023{\natexlab{b}}){Dong-P{\'a}ez},
  {Volonteri}, {Beckmann}, {Dubois}, {Trebitsch}, {Mangiagli}, {Vergani}, \&
  {Webb}}]{DongPez2023a}
{Dong-P{\'a}ez}, C.~A., {Volonteri}, M., {Beckmann}, R.~S., {et~al.}
  2023{\natexlab{b}}, arXiv e-prints, arXiv:2303.00766

\bibitem[{{Dotti} {et~al.}(2007){Dotti}, {Colpi}, {Haardt}, \&
  {Mayer}}]{Dotti2007}
{Dotti}, M., {Colpi}, M., {Haardt}, F., \& {Mayer}, L. 2007, \mnras, 379, 956

\bibitem[{{Dotti} {et~al.}(2015){Dotti}, {Merloni}, \& {Montuori}}]{Dotti2015}
{Dotti}, M., {Merloni}, A., \& {Montuori}, C. 2015, \mnras, 448, 3603

\bibitem[{{Dressler} \& {Richstone}(1988)}]{Dressler1988}
{Dressler}, A. \& {Richstone}, D.~O. 1988, \apj, 324, 701

\bibitem[{{Drory} \& {Fisher}(2007)}]{Drory2007}
{Drory}, N. \& {Fisher}, D.~B. 2007, \apj, 664, 640

\bibitem[{{Duffell} {et~al.}(2020){Duffell}, {D'Orazio}, {Derdzinski},
  {Haiman}, {MacFadyen}, {Rosen}, \& {Zrake}}]{Duffell2020}
{Duffell}, P.~C., {D'Orazio}, D., {Derdzinski}, A., {et~al.} 2020, \apj, 901,
  25

\bibitem[{{Efstathiou} {et~al.}(1982){Efstathiou}, {Lake}, \&
  {Negroponte}}]{Efstathio1982}
{Efstathiou}, G., {Lake}, G., \& {Negroponte}, J. 1982, \mnras, 199, 1069

\bibitem[{{Eracleous} {et~al.}(2012){Eracleous}, {Boroson}, {Halpern}, \&
  {Liu}}]{Eracleous2012}
{Eracleous}, M., {Boroson}, T.~A., {Halpern}, J.~P., \& {Liu}, J. 2012, \apjs,
  201, 23

\bibitem[{{Escala} {et~al.}(2004){Escala}, {Larson}, {Coppi}, \&
  {Mardones}}]{Escala2004}
{Escala}, A., {Larson}, R.~B., {Coppi}, P.~S., \& {Mardones}, D. 2004, \apj,
  607, 765

\bibitem[{{Escala} {et~al.}(2005){Escala}, {Larson}, {Coppi}, \&
  {Mardones}}]{Escala2005}
{Escala}, A., {Larson}, R.~B., {Coppi}, P.~S., \& {Mardones}, D. 2005, \apj,
  630, 152

\bibitem[{{Fanidakis} {et~al.}(2012){Fanidakis}, {Baugh}, {Benson}, {Bower},
  {Cole}, {Done}, {Frenk}, {Hickox}, {Lacey}, \& {Del P.
  Lagos}}]{Fanidakis2012}
{Fanidakis}, N., {Baugh}, C.~M., {Benson}, A.~J., {et~al.} 2012, \mnras, 419,
  2797

\bibitem[{{Fiacconi} {et~al.}(2013){Fiacconi}, {Mayer}, {Ro{\v{s}}kar}, \&
  {Colpi}}]{Fiacconi2013}
{Fiacconi}, D., {Mayer}, L., {Ro{\v{s}}kar}, R., \& {Colpi}, M. 2013, \apjl,
  777, L14

\bibitem[{{Fisher} \& {Drory}(2008)}]{Fisher2008}
{Fisher}, D.~B. \& {Drory}, N. 2008, \aj, 136, 773

\bibitem[{{Franchini} {et~al.}(2022){Franchini}, {Lupi}, \&
  {Sesana}}]{Franchini2022}
{Franchini}, A., {Lupi}, A., \& {Sesana}, A. 2022, \apjl, 929, L13

\bibitem[{{Franchini} {et~al.}(2021){Franchini}, {Sesana}, \&
  {Dotti}}]{Franchini2021}
{Franchini}, A., {Sesana}, A., \& {Dotti}, M. 2021, \mnras, 507, 1458

\bibitem[{{Gadotti}(2009)}]{Gadotti2009}
{Gadotti}, D.~A. 2009, \mnras, 393, 1531

\bibitem[{{Gerber} \& {Lamb}(1994)}]{Gerber1994}
{Gerber}, R.~A. \& {Lamb}, S.~A. 1994, \apj, 431, 604

\bibitem[{{Graham} {et~al.}(2001){Graham}, {Erwin}, {Caon}, \&
  {Trujillo}}]{Graham2001}
{Graham}, A.~W., {Erwin}, P., {Caon}, N., \& {Trujillo}, I. 2001, \apjl, 563,
  L11

\bibitem[{{Graham} {et~al.}(2015){Graham}, {Djorgovski}, {Stern}, {Drake},
  {Mahabal}, {Donalek}, {Glikman}, {Larson}, \& {Christensen}}]{Graham2015}
{Graham}, M.~J., {Djorgovski}, S.~G., {Stern}, D., {et~al.} 2015, \mnras, 453,
  1562

\bibitem[{{Greene} {et~al.}(2020){Greene}, {Strader}, \& {Ho}}]{Greene2020}
{Greene}, J.~E., {Strader}, J., \& {Ho}, L.~C. 2020, \araa, 58, 257

\bibitem[{{Griffin} {et~al.}(2019){Griffin}, {Lacey}, {Gonzalez-Perez},
  {Lagos}, {Baugh}, \& {Fanidakis}}]{Griffin2018}
{Griffin}, A.~J., {Lacey}, C.~G., {Gonzalez-Perez}, V., {et~al.} 2019, \mnras,
  487, 198

\bibitem[{{Guo} {et~al.}(2011){Guo}, {White}, {Boylan-Kolchin}, {De Lucia},
  {Kauffmann}, {Lemson}, {Li}, {Springel}, \& {Weinmann}}]{Guo2011}
{Guo}, Q., {White}, S., {Boylan-Kolchin}, M., {et~al.} 2011, \mnras, 413, 101

\bibitem[{{Habouzit} {et~al.}(2021){Habouzit}, {Li}, {Somerville}, {Genel},
  {Pillepich}, {Volonteri}, {Dav{\'e}}, {Rosas-Guevara}, {McAlpine}, {Peirani},
  {Hernquist}, {Angl{\'e}s-Alc{\'a}zar}, {Reines}, {Bower}, {Dubois}, {Nelson},
  {Pichon}, \& {Vogelsberger}}]{Habouzit2021}
{Habouzit}, M., {Li}, Y., {Somerville}, R.~S., {et~al.} 2021, \mnras, 503, 1940

\bibitem[{{Habouzit} {et~al.}(2022){Habouzit}, {Somerville}, {Li}, {Genel},
  {Aird}, {Angl{\'e}s-Alc{\'a}zar}, {Dav{\'e}}, {Georgiev}, {McAlpine},
  {Rosas-Guevara}, {Dubois}, {Nelson}, {Banados}, {Hernquist}, {Peirani}, \&
  {Vogelsberger}}]{Habouzit2022}
{Habouzit}, M., {Somerville}, R.~S., {Li}, Y., {et~al.} 2022, \mnras, 509, 3015

\bibitem[{{H{\"a}ring} \& {Rix}(2004)}]{HaringANDRix2004}
{H{\"a}ring}, N. \& {Rix}, H.-W. 2004, \apjl, 604, L89

\bibitem[{{Henriques} {et~al.}(2015){Henriques}, {White}, {Thomas}, {Angulo},
  {Guo}, {Lemson}, {Springel}, \& {Overzier}}]{Henriques2015}
{Henriques}, B.~M.~B., {White}, S.~D.~M., {Thomas}, P.~A., {et~al.} 2015,
  \mnras, 451, 2663

\bibitem[{{Hirschmann} {et~al.}(2014){Hirschmann}, {Dolag}, {Saro}, {Bachmann},
  {Borgani}, \& {Burkert}}]{Hirschmann2014}
{Hirschmann}, M., {Dolag}, K., {Saro}, A., {et~al.} 2014, \mnras, 442, 2304

\bibitem[{{Inayoshi} {et~al.}(2020){Inayoshi}, {Visbal}, \&
  {Haiman}}]{inayoshi_visbal_haiman2020}
{Inayoshi}, K., {Visbal}, E., \& {Haiman}, Z. 2020, \araa, 58, 27

\bibitem[{{Ivezi{\'c}} {et~al.}(2019){Ivezi{\'c}}, {Kahn}, {Tyson}, {Abel},
  {Acosta}, {Allsman}, {Alonso}, {AlSayyad}, {Anderson}, {Andrew}, {Angel},
  {Angeli}, {Ansari}, {Antilogus}, {Araujo}, {Armstrong}, {Arndt}, {Astier},
  {Aubourg}, {Auza}, {Axelrod}, {Bard}, {Barr}, {Barrau}, {Bartlett}, {Bauer},
  {Bauman}, {Baumont}, {Bechtol}, {Bechtol}, {Becker}, {Becla}, {Beldica},
  {Bellavia}, {Bianco}, {Biswas}, {Blanc}, {Blazek}, {Blandford}, {Bloom},
  {Bogart}, {Bond}, {Booth}, {Borgland}, {Borne}, {Bosch}, {Boutigny},
  {Brackett}, {Bradshaw}, {Brandt}, {Brown}, {Bullock}, {Burchat}, {Burke},
  {Cagnoli}, {Calabrese}, {Callahan}, {Callen}, {Carlin}, {Carlson},
  {Chandrasekharan}, {Charles-Emerson}, {Chesley}, {Cheu}, {Chiang}, {Chiang},
  {Chirino}, {Chow}, {Ciardi}, {Claver}, {Cohen-Tanugi}, {Cockrum}, {Coles},
  {Connolly}, {Cook}, {Cooray}, {Covey}, {Cribbs}, {Cui}, {Cutri}, {Daly},
  {Daniel}, {Daruich}, {Daubard}, {Daues}, {Dawson}, {Delgado}, {Dellapenna},
  {de Peyster}, {de Val-Borro}, {Digel}, {Doherty}, {Dubois},
  {Dubois-Felsmann}, {Durech}, {Economou}, {Eifler}, {Eracleous}, {Emmons},
  {Fausti Neto}, {Ferguson}, {Figueroa}, {Fisher-Levine}, {Focke}, {Foss},
  {Frank}, {Freemon}, {Gangler}, {Gawiser}, {Geary}, {Gee}, {Geha}, {Gessner},
  {Gibson}, {Gilmore}, {Glanzman}, {Glick}, {Goldina}, {Goldstein}, {Goodenow},
  {Graham}, {Gressler}, {Gris}, {Guy}, {Guyonnet}, {Haller}, {Harris},
  {Hascall}, {Haupt}, {Hernandez}, {Herrmann}, {Hileman}, {Hoblitt}, {Hodgson},
  {Hogan}, {Howard}, {Huang}, {Huffer}, {Ingraham}, {Innes}, {Jacoby}, {Jain},
  {Jammes}, {Jee}, {Jenness}, {Jernigan}, {Jevremovi{\'c}}, {Johns}, {Johnson},
  {Johnson}, {Jones}, {Juramy-Gilles}, {Juri{\'c}}, {Kalirai}, {Kallivayalil},
  {Kalmbach}, {Kantor}, {Karst}, {Kasliwal}, {Kelly}, {Kessler}, {Kinnison},
  {Kirkby}, {Knox}, {Kotov}, {Krabbendam}, {Krughoff}, {Kub{\'a}nek},
  {Kuczewski}, {Kulkarni}, {Ku}, {Kurita}, {Lage}, {Lambert}, {Lange},
  {Langton}, {Le Guillou}, {Levine}, {Liang}, {Lim}, {Lintott}, {Long},
  {Lopez}, {Lotz}, {Lupton}, {Lust}, {MacArthur}, {Mahabal}, {Mandelbaum},
  {Markiewicz}, {Marsh}, {Marshall}, {Marshall}, {May}, {McKercher}, {McQueen},
  {Meyers}, {Migliore}, {Miller}, {Mills}, {Miraval}, {Moeyens}, {Moolekamp},
  {Monet}, {Moniez}, {Monkewitz}, {Montgomery}, {Morrison}, {Mueller},
  {Muller}, {Mu{\~n}oz Arancibia}, {Neill}, {Newbry}, {Nief}, {Nomerotski},
  {Nordby}, {O'Connor}, {Oliver}, {Olivier}, {Olsen}, {O'Mullane}, {Ortiz},
  {Osier}, {Owen}, {Pain}, {Palecek}, {Parejko}, {Parsons}, {Pease},
  {Peterson}, {Peterson}, {Petravick}, {Libby Petrick}, {Petry},
  {Pierfederici}, {Pietrowicz}, {Pike}, {Pinto}, {Plante}, {Plate}, {Plutchak},
  {Price}, {Prouza}, {Radeka}, {Rajagopal}, {Rasmussen}, {Regnault}, {Reil},
  {Reiss}, {Reuter}, {Ridgway}, {Riot}, {Ritz}, {Robinson}, {Roby}, {Roodman},
  {Rosing}, {Roucelle}, {Rumore}, {Russo}, {Saha}, {Sassolas}, {Schalk},
  {Schellart}, {Schindler}, {Schmidt}, {Schneider}, {Schneider}, {Schoening},
  {Schumacher}, {Schwamb}, {Sebag}, {Selvy}, {Sembroski}, {Seppala}, {Serio},
  {Serrano}, {Shaw}, {Shipsey}, {Sick}, {Silvestri}, {Slater}, {Smith},
  {Smith}, {Sobhani}, {Soldahl}, {Storrie-Lombardi}, {Stover}, {Strauss},
  {Street}, {Stubbs}, {Sullivan}, {Sweeney}, {Swinbank}, {Szalay}, {Takacs},
  {Tether}, {Thaler}, {Thayer}, {Thomas}, {Thornton}, {Thukral}, {Tice},
  {Trilling}, {Turri}, {Van Berg}, {Vanden Berk}, {Vetter}, {Virieux},
  {Vucina}, {Wahl}, {Walkowicz}, {Walsh}, {Walter}, {Wang}, {Wang}, {Warner},
  {Wiecha}, {Willman}, {Winters}, {Wittman}, {Wolff}, {Wood-Vasey}, {Wu},
  {Xin}, {Yoachim}, \& {Zhan}}]{Ivezic2019}
{Ivezi{\'c}}, {\v{Z}}., {Kahn}, S.~M., {Tyson}, J.~A., {et~al.} 2019, \apj,
  873, 111

\bibitem[{{Izquierdo-Villalba}
  {et~al.}(2019{\natexlab{a}}){Izquierdo-Villalba}, {Angulo}, {Orsi}, {Hurier},
  {Vilella-Rojo}, {Bonoli}, {L{\'o}pez-Sanjuan}, {Alcaniz}, {Cenarro},
  {Crist{\'o}bal-Hornillos}, {Dupke}, {Ederoclite}, {Hern{\'a}ndez-Monteagudo},
  {Mar{\'\i}n-Franch}, {Moles}, {Mendes de Oliveira}, {Sodr{\'e}}, {Varela}, \&
  {V{\'a}zquez Rami{\'o}}}]{IzquierdoVillalba2019LC}
{Izquierdo-Villalba}, D., {Angulo}, R.~E., {Orsi}, A., {et~al.}
  2019{\natexlab{a}}, \aap, 631, A82

\bibitem[{{Izquierdo-Villalba} {et~al.}(2020){Izquierdo-Villalba}, {Bonoli},
  {Dotti}, {Sesana}, {Rosas-Guevara}, \& {Spinoso}}]{IzquierdoVillalba2020}
{Izquierdo-Villalba}, D., {Bonoli}, S., {Dotti}, M., {et~al.} 2020, \mnras,
  495, 4681

\bibitem[{{Izquierdo-Villalba}
  {et~al.}(2022{\natexlab{a}}){Izquierdo-Villalba}, {Bonoli}, {Rosas-Guevara},
  {Springel}, {White}, {Zana}, {Dotti}, {Spinoso}, {Bonetti}, \&
  {Lupi}}]{Izquierdo-Villalb_DI_2022}
{Izquierdo-Villalba}, D., {Bonoli}, S., {Rosas-Guevara}, Y., {et~al.}
  2022{\natexlab{a}}, \mnras, 514, 1006

\bibitem[{{Izquierdo-Villalba}
  {et~al.}(2019{\natexlab{b}}){Izquierdo-Villalba}, {Bonoli}, {Spinoso},
  {Rosas-Guevara}, {Henriques}, \&
  {Hern{\'a}ndez-Monteagudo}}]{IzquierdoVillalba2019}
{Izquierdo-Villalba}, D., {Bonoli}, S., {Spinoso}, D., {et~al.}
  2019{\natexlab{b}}, \mnras, 488, 609

\bibitem[{{Izquierdo-Villalba}
  {et~al.}(2022{\natexlab{b}}){Izquierdo-Villalba}, {Sesana}, {Bonoli}, \&
  {Colpi}}]{IzquierdoVillalba2021}
{Izquierdo-Villalba}, D., {Sesana}, A., {Bonoli}, S., \& {Colpi}, M.
  2022{\natexlab{b}}, \mnras, 509, 3488

\bibitem[{{Kauffmann} {et~al.}(1999){Kauffmann}, {Colberg}, {Diaferio}, \&
  {White}}]{Kauffmann1999}
{Kauffmann}, G., {Colberg}, J.~M., {Diaferio}, A., \& {White}, S. D.~M. 1999,
  \mnras, 307, 529

\bibitem[{{Kennicutt}(1998)}]{Kennicutt1998}
{Kennicutt}, Robert~C., J. 1998, \apj, 498, 541

\bibitem[{{Kitzbichler} \& {White}(2007)}]{KitzbichlerWhite2007}
{Kitzbichler}, M.~G. \& {White}, S.~D.~M. 2007, \mnras, 376, 2

\bibitem[{{Kocsis} {et~al.}(2006){Kocsis}, {Frei}, {Haiman}, \&
  {Menou}}]{Kocsis2006}
{Kocsis}, B., {Frei}, Z., {Haiman}, Z., \& {Menou}, K. 2006, \apj, 637, 27

\bibitem[{{Kormendy} \& {Ho}(2013)}]{Kormendy2013}
{Kormendy}, J. \& {Ho}, L.~C. 2013, \araa, 51, 511

\bibitem[{{Kormendy} \& {Richstone}(1995)}]{Kormendy1995}
{Kormendy}, J. \& {Richstone}, D. 1995, \araa, 33, 581

\bibitem[{{Koushiappas} {et~al.}(2004){Koushiappas}, {Bullock}, \&
  {Dekel}}]{Koushiappas2004}
{Koushiappas}, S.~M., {Bullock}, J.~S., \& {Dekel}, A. 2004, \mnras, 354, 292

\bibitem[{{Lacey} \& {Cole}(1993)}]{Lacey1993}
{Lacey}, C. \& {Cole}, S. 1993, \mnras, 262, 627

\bibitem[{{Li} {et~al.}(2022{\natexlab{a}}){Li}, {Bogdanovi{\'c}},
  {Ballantyne}, \& {Bonetti}}]{Kunyang2022}
{Li}, K., {Bogdanovi{\'c}}, T., {Ballantyne}, D.~R., \& {Bonetti}, M.
  2022{\natexlab{a}}, \apj, 933, 104

\bibitem[{{Li} {et~al.}(2022{\natexlab{b}}){Li}, {Bogdanovi{\'c}},
  {Ballantyne}, \& {Bonetti}}]{Li2022}
{Li}, K., {Bogdanovi{\'c}}, T., {Ballantyne}, D.~R., \& {Bonetti}, M.
  2022{\natexlab{b}}, \apj, 933, 104

\bibitem[{{Liao} {et~al.}(2021){Liao}, {Chen}, {Liu}, {Holgado}, {Guo},
  {Gruendl}, {Morganson}, {Shen}, {Davis}, {Kessler}, {Martini}, {McMahon},
  {Allam}, {Annis}, {Avila}, {Banerji}, {Bechtol}, {Bertin}, {Brooks},
  {Buckley-Geer}, {Carnero Rosell}, {Carrasco Kind}, {Carretero}, {Javier
  Castander}, {Cunha}, {D'Andrea}, {da Costa}, {Davis}, {De Vicente}, {Desai},
  {Thomas Diehl}, {Doel}, {Eifler}, {Evrard}, {Flaugher}, {Fosalba}, {Frieman},
  {Garcia-Bellido}, {Gaztanaga}, {Glazebrook}, {Gruen}, {Gschwend},
  {Gutierrez}, {Hartley}, {Hollowood}, {Honscheid}, {Hoyle}, {James}, {Krause},
  {Kuehn}, {Lima}, {Maia}, {Marshall}, {Menanteau}, {Miquel}, {Plazas
  Malag{\'o}n}, {Roodman}, {Sanchez}, {Scarpine}, {Schubnell}, {Serrano},
  {Smith}, {Smith}, {Soares-Santos}, {Sobreira}, {Suchyta}, {Swanson}, {Tarle},
  {Vikram}, \& {Walker}}]{Liao2021}
{Liao}, W.-T., {Chen}, Y.-C., {Liu}, X., {et~al.} 2021, \mnras, 500, 4025

\bibitem[{{Liu} {et~al.}(2019){Liu}, {Gezari}, {Ayers}, {Burgett}, {Chambers},
  {Hodapp}, {Huber}, {Kudritzki}, {Metcalfe}, {Tonry}, {Wainscoat}, \&
  {Waters}}]{Liu2019}
{Liu}, T., {Gezari}, S., {Ayers}, M., {et~al.} 2019, \apj, 884, 36

\bibitem[{{Liu} {et~al.}(2016){Liu}, {Gezari}, {Burgett}, {Chambers}, {Draper},
  {Hodapp}, {Huber}, {Kudritzki}, {Magnier}, {Metcalfe}, {Tonry}, {Wainscoat},
  \& {Waters}}]{Liu2016}
{Liu}, T., {Gezari}, S., {Burgett}, W., {et~al.} 2016, \apj, 833, 6

\bibitem[{{Loeb} \& {Rasio}(1994)}]{Loeb1994}
{Loeb}, A. \& {Rasio}, F.~A. 1994, \apj, 432, 52

\bibitem[{{Lops} {et~al.}(2023){Lops}, {Izquierdo-Villalba}, {Colpi}, {Bonoli},
  {Sesana}, \& {Mangiagli}}]{Lops2022}
{Lops}, G., {Izquierdo-Villalba}, D., {Colpi}, M., {et~al.} 2023, \mnras, 519,
  5962

\bibitem[{{Lotz} {et~al.}(2004){Lotz}, {Primack}, \& {Madau}}]{Lotz2004}
{Lotz}, J.~M., {Primack}, J., \& {Madau}, P. 2004, \aj, 128, 163

\bibitem[{{Lupi} {et~al.}(2015){Lupi}, {Haardt}, {Dotti}, \&
  {Colpi}}]{Lupi2015}
{Lupi}, A., {Haardt}, F., {Dotti}, M., \& {Colpi}, M. 2015, \mnras, 453, 3437

\bibitem[{{Lupi} {et~al.}(2016){Lupi}, {Haardt}, {Dotti}, {Fiacconi}, {Mayer},
  \& {Madau}}]{Lupi2016}
{Lupi}, A., {Haardt}, F., {Dotti}, M., {et~al.} 2016, \mnras, 456, 2993

\bibitem[{{Lupi} {et~al.}(2021){Lupi}, {Haiman}, \& {Volonteri}}]{Lupi2021}
{Lupi}, A., {Haiman}, Z., \& {Volonteri}, M. 2021, \mnras, 503, 5046

\bibitem[{{Magorrian} {et~al.}(1998){Magorrian}, {Tremaine}, {Richstone},
  {Bender}, {Bower}, {Dressler}, {Faber}, {Gebhardt}, {Green}, {Grillmair},
  {Kormendy}, \& {Lauer}}]{Magorrian1998}
{Magorrian}, J., {Tremaine}, S., {Richstone}, D., {et~al.} 1998, \aj, 115, 2285

\bibitem[{{Mancillas} {et~al.}(2019){Mancillas}, {Duc}, {Combes}, {Bournaud},
  {Emsellem}, {Martig}, \& {Michel-Dansac}}]{Mancillas2019}
{Mancillas}, B., {Duc}, P.-A., {Combes}, F., {et~al.} 2019, \aap, 632, A122

\bibitem[{{Mangiagli} {et~al.}(2022){Mangiagli}, {Caprini}, {Volonteri},
  {Marsat}, {Vergani}, {Tamanini}, \& {Inchausp{\'e}}}]{Mangiagli2022}
{Mangiagli}, A., {Caprini}, C., {Volonteri}, M., {et~al.} 2022, \prd, 106,
  103017

\bibitem[{{Mangiagli} {et~al.}(2020){Mangiagli}, {Klein}, {Bonetti}, {Katz},
  {Sesana}, {Volonteri}, {Colpi}, {Marsat}, \& {Babak}}]{Mangiagli2020}
{Mangiagli}, A., {Klein}, A., {Bonetti}, M., {et~al.} 2020, \prd, 102, 084056

\bibitem[{{Marconi} {et~al.}(2004){Marconi}, {Risaliti}, {Gilli}, {Hunt},
  {Maiolino}, \& {Salvati}}]{Marconi2004}
{Marconi}, A., {Risaliti}, G., {Gilli}, R., {et~al.} 2004, \mnras, 351, 169

\bibitem[{{Marsat} {et~al.}(2021){Marsat}, {Baker}, \& {Canton}}]{Marsat2021}
{Marsat}, S., {Baker}, J.~G., \& {Canton}, T.~D. 2021, \prd, 103, 083011

\bibitem[{{Marshall} {et~al.}(2020{\natexlab{a}}){Marshall}, {Mutch}, {Qin},
  {Poole}, \& {Wyithe}}]{Marshall2020}
{Marshall}, M.~A., {Mutch}, S.~J., {Qin}, Y., {Poole}, G.~B., \& {Wyithe}, J.
  S.~B. 2020{\natexlab{a}}, \mnras, 494, 2747

\bibitem[{{Marshall} {et~al.}(2020{\natexlab{b}}){Marshall}, {Mutch}, {Qin},
  {Poole}, \& {Wyithe}}]{Marshall2019}
{Marshall}, M.~A., {Mutch}, S.~J., {Qin}, Y., {Poole}, G.~B., \& {Wyithe}, J.
  S.~B. 2020{\natexlab{b}}, \mnras, 494, 2747

\bibitem[{{Marulli} {et~al.}(2008){Marulli}, {Bonoli}, {Branchini},
  {Moscardini}, \& {Springel}}]{Marulli2008}
{Marulli}, F., {Bonoli}, S., {Branchini}, E., {Moscardini}, L., \& {Springel},
  V. 2008, \mnras, 385, 1846

\bibitem[{{Masters} {et~al.}(2012){Masters}, {Capak}, {Salvato}, {Civano},
  {Mobasher}, {Siana}, {Hasinger}, {Impey}, {Nagao}, {Trump}, {Ikeda}, {Elvis},
  \& {Scoville}}]{Masters2012}
{Masters}, D., {Capak}, P., {Salvato}, M., {et~al.} 2012, \apj, 755, 169

\bibitem[{{Mayer} \& {Bonoli}(2019)}]{MayerBonoli2019}
{Mayer}, L. \& {Bonoli}, S. 2019, Reports on Progress in Physics, 82, 016901

\bibitem[{{Mayer} {et~al.}(2007){Mayer}, {Kazantzidis}, {Madau}, {Colpi},
  {Quinn}, \& {Wadsley}}]{Mayer2007}
{Mayer}, L., {Kazantzidis}, S., {Madau}, P., {et~al.} 2007, Science, 316, 1874

\bibitem[{{McGreer} {et~al.}(2013){McGreer}, {Jiang}, {Fan}, {Richards},
  {Strauss}, {Ross}, {White}, {Shen}, {Schneider}, {Myers}, {Brandt}, {DeGraf},
  {Glikman}, {Ge}, \& {Streblyanska}}]{McGreer2013}
{McGreer}, I.~D., {Jiang}, L., {Fan}, X., {et~al.} 2013, \apj, 768, 105

\bibitem[{{Miller} {et~al.}(2015){Miller}, {Gallo}, {Greene}, {Kelly}, {Treu},
  {Woo}, \& {Baldassare}}]{Miller2015}
{Miller}, B.~P., {Gallo}, E., {Greene}, J.~E., {et~al.} 2015, \apj, 799, 98

\bibitem[{{Milosavljevi{\'c}} \& {Merritt}(2001)}]{Milosavljevic2001}
{Milosavljevi{\'c}}, M. \& {Merritt}, D. 2001, \apj, 563, 34

\bibitem[{{Mo} {et~al.}(2010){Mo}, {van den Bosch}, \& {White}}]{MoWhite2010}
{Mo}, H., {van den Bosch}, F.~C., \& {White}, S. 2010, {Galaxy Formation and
  Evolution}

\bibitem[{{Mo} {et~al.}(1998){Mo}, {Mao}, \& {White}}]{MoMaoWhite1997}
{Mo}, H.~J., {Mao}, S., \& {White}, S.~D.~M. 1998, \mnras, 295, 319

\bibitem[{{Montuori} {et~al.}(2011){Montuori}, {Dotti}, {Colpi}, {Decarli}, \&
  {Haardt}}]{Montuori2011}
{Montuori}, C., {Dotti}, M., {Colpi}, M., {Decarli}, R., \& {Haardt}, F. 2011,
  \mnras, 412, 26

\bibitem[{{Nandra} {et~al.}(2013){Nandra}, {Barret}, {Barcons}, {Fabian}, {den
  Herder}, {Piro}, {Watson}, {Adami}, {Aird}, {Afonso}, {Alexander},
  {Argiroffi}, {Amati}, {Arnaud}, {Atteia}, {Audard}, {Badenes}, {Ballet},
  {Ballo}, {Bamba}, {Bhardwaj}, {Stefano Battistelli}, {Becker}, {De Becker},
  {Behar}, {Bianchi}, {Biffi}, {B{\^\i}rzan}, {Bocchino}, {Bogdanov}, {Boirin},
  {Boller}, {Borgani}, {Borm}, {Bouch{\'e}}, {Bourdin}, {Bower}, {Braito},
  {Branchini}, {Branduardi-Raymont}, {Bregman}, {Brenneman}, {Brightman},
  {Br{\"u}ggen}, {Buchner}, {Bulbul}, {Brusa}, {Bursa}, {Caccianiga},
  {Cackett}, {Campana}, {Cappelluti}, {Cappi}, {Carrera}, {Ceballos},
  {Christensen}, {Chu}, {Churazov}, {Clerc}, {Corbel}, {Corral}, {Comastri},
  {Costantini}, {Croston}, {Dadina}, {D'Ai}, {Decourchelle}, {Della Ceca},
  {Dennerl}, {Dolag}, {Done}, {Dovciak}, {Drake}, {Eckert}, {Edge}, {Ettori},
  {Ezoe}, {Feigelson}, {Fender}, {Feruglio}, {Finoguenov}, {Fiore}, {Galeazzi},
  {Gallagher}, {Gandhi}, {Gaspari}, {Gastaldello}, {Georgakakis},
  {Georgantopoulos}, {Gilfanov}, {Gitti}, {Gladstone}, {Goosmann}, {Gosset},
  {Grosso}, {Guedel}, {Guerrero}, {Haberl}, {Hardcastle}, {Heinz}, {Alonso
  Herrero}, {Herv{\'e}}, {Holmstrom}, {Iwasawa}, {Jonker}, {Kaastra}, {Kara},
  {Karas}, {Kastner}, {King}, {Kosenko}, {Koutroumpa}, {Kraft}, {Kreykenbohm},
  {Lallement}, {Lanzuisi}, {Lee}, {Lemoine-Goumard}, {Lobban}, {Lodato},
  {Lovisari}, {Lotti}, {McCharthy}, {McNamara}, {Maggio}, {Maiolino}, {De
  Marco}, {de Martino}, {Mateos}, {Matt}, {Maughan}, {Mazzotta}, {Mendez},
  {Merloni}, {Micela}, {Miceli}, {Mignani}, {Miller}, {Miniutti}, {Molendi},
  {Montez}, {Moretti}, {Motch}, {Naz{\'e}}, {Nevalainen}, {Nicastro}, {Nulsen},
  {Ohashi}, {O'Brien}, {Osborne}, {Oskinova}, {Pacaud}, {Paerels}, {Page},
  {Papadakis}, {Pareschi}, {Petre}, {Petrucci}, {Piconcelli}, {Pillitteri},
  {Pinto}, {de Plaa}, {Pointecouteau}, {Ponman}, {Ponti}, {Porquet}, {Pounds},
  {Pratt}, {Predehl}, {Proga}, {Psaltis}, {Rafferty}, {Ramos-Ceja}, {Ranalli},
  {Rasia}, {Rau}, {Rauw}, {Rea}, {Read}, {Reeves}, {Reiprich}, {Renaud},
  {Reynolds}, {Risaliti}, {Rodriguez}, {Rodriguez Hidalgo}, {Roncarelli},
  {Rosario}, {Rossetti}, {Rozanska}, {Rovilos}, {Salvaterra}, {Salvato}, {Di
  Salvo}, {Sanders}, {Sanz-Forcada}, {Schawinski}, {Schaye}, {Schwope},
  {Sciortino}, {Severgnini}, {Shankar}, {Sijacki}, {Sim}, {Schmid}, {Smith},
  {Steiner}, {Stelzer}, {Stewart}, {Strohmayer}, {Str{\"u}der}, {Sun}, {Takei},
  {Tatischeff}, {Tiengo}, {Tombesi}, {Trinchieri}, {Tsuru}, {Ud-Doula},
  {Ursino}, {Valencic}, {Vanzella}, {Vaughan}, {Vignali}, {Vink}, {Vito},
  {Volonteri}, {Wang}, {Webb}, {Willingale}, {Wilms}, {Wise}, {Worrall},
  {Young}, {Zampieri}, {In't Zand}, {Zane}, {Zezas}, {Zhang}, \&
  {Zhuravleva}}]{Nandra2013}
{Nandra}, K., {Barret}, D., {Barcons}, X., {et~al.} 2013, arXiv e-prints,
  arXiv:1306.2307

\bibitem[{{Nelson} {et~al.}(2019){Nelson}, {Springel}, {Pillepich},
  {Rodriguez-Gomez}, {Torrey}, {Genel}, {Vogelsberger}, {Pakmor}, {Marinacci},
  {Weinberger}, {Kelley}, {Lovell}, {Diemer}, \&
  {Hernquist}}]{NelsonTNGDataReleas2019}
{Nelson}, D., {Springel}, V., {Pillepich}, A., {et~al.} 2019, Computational
  Astrophysics and Cosmology, 6, 2

\bibitem[{{Niida} {et~al.}(2016){Niida}, {Nagao}, {Ikeda}, {Matsuoka},
  {Kobayashi}, {Toba}, \& {Taniguchi}}]{Niida2016}
{Niida}, M., {Nagao}, T., {Ikeda}, H., {et~al.} 2016, \apj, 832, 208

\bibitem[{{Petiteau} {et~al.}(2011){Petiteau}, {Babak}, \&
  {Sesana}}]{Petiteau2011}
{Petiteau}, A., {Babak}, S., \& {Sesana}, A. 2011, \apj, 732, 82

\bibitem[{{Pezzulli} {et~al.}(2017){Pezzulli}, {Valiante}, {Orofino},
  {Schneider}, {Gallerani}, \& {Sbarrato}}]{Pezzulli2017}
{Pezzulli}, E., {Valiante}, R., {Orofino}, M.~C., {et~al.} 2017, \mnras, 466,
  2131

\bibitem[{{Piro} {et~al.}(2022){Piro}, {Colpi}, {Aird}, {Mangiagli}, {Fabian},
  {Guainazzi}, {Marsat}, {Sesana}, {McNamara}, {Bonetti}, {Rossi}, {Tanvir},
  {Baker}, {Belanger}, {Dal Canton}, {Jennrich}, {Katz}, \&
  {Luetzgendorf}}]{Piro2022}
{Piro}, L., {Colpi}, M., {Aird}, J., {et~al.} 2022, arXiv e-prints,
  arXiv:2211.13759

\bibitem[{{Planck Collaboration} {et~al.}(2014){Planck Collaboration}, {Ade},
  {Aghanim}, {Armitage-Caplan}, {Arnaud}, {Ashdown}, {Atrio-Barandela},
  {Aumont}, {Baccigalupi}, {Banday}, \& et~al.}]{PlanckCollaboration2014}
{Planck Collaboration}, {Ade}, P.~A.~R., {Aghanim}, N., {et~al.} 2014, \aap,
  571, A16

\bibitem[{{Quinlan} \& {Hernquist}(1997)}]{Quinlan1997}
{Quinlan}, G.~D. \& {Hernquist}, L. 1997, \na, 2, 533

\bibitem[{{Rosas-Guevara} {et~al.}(2016){Rosas-Guevara}, {Bower}, {Schaye},
  {McAlpine}, {Dalla Vecchia}, {Frenk}, {Schaller}, \&
  {Theuns}}]{Rosas-Guevara2016}
{Rosas-Guevara}, Y., {Bower}, R.~G., {Schaye}, J., {et~al.} 2016, \mnras, 462,
  190

\bibitem[{{Sargent} {et~al.}(1978){Sargent}, {Young}, {Boksenberg},
  {Shortridge}, {Lynds}, \& {Hartwick}}]{Sargent1978}
{Sargent}, W.~L.~W., {Young}, P.~J., {Boksenberg}, A., {et~al.} 1978, \apj,
  221, 731

\bibitem[{{Sassano} {et~al.}(2023){Sassano}, {Capelo}, {Mayer}, {Schneider}, \&
  {Valiante}}]{Sassano2023}
{Sassano}, F., {Capelo}, P.~R., {Mayer}, L., {Schneider}, R., \& {Valiante}, R.
  2023, \mnras, 519, 1837

\bibitem[{{Savorgnan} {et~al.}(2016){Savorgnan}, {Graham}, {Marconi}, \&
  {Sani}}]{Savorgnan2016}
{Savorgnan}, G. A.~D., {Graham}, A.~W., {Marconi}, A.~r., \& {Sani}, E. 2016,
  \apj, 817, 21

\bibitem[{{Schneider} {et~al.}(2002){Schneider}, {Ferrara}, {Natarajan}, \&
  {Omukai}}]{Schneider2002}
{Schneider}, R., {Ferrara}, A., {Natarajan}, P., \& {Omukai}, K. 2002, \apj,
  571, 30

\bibitem[{{Sersic}(1968)}]{Sersic1968}
{Sersic}, J.~L. 1968, {Atlas de Galaxias Australes}

\bibitem[{{Sesana} {et~al.}(2006){Sesana}, {Haardt}, \& {Madau}}]{Sesana2006}
{Sesana}, A., {Haardt}, F., \& {Madau}, P. 2006, \apj, 651, 392

\bibitem[{{Sesana} \& {Khan}(2015)}]{Sesana2015}
{Sesana}, A. \& {Khan}, F.~M. 2015, \mnras, 454, L66

\bibitem[{{Shankar} {et~al.}(2004){Shankar}, {Salucci}, {Granato}, {De Zotti},
  \& {Danese}}]{Shankar2004}
{Shankar}, F., {Salucci}, P., {Granato}, G.~L., {De Zotti}, G., \& {Danese}, L.
  2004, \mnras, 354, 1020

\bibitem[{{Shankar} {et~al.}(2009){Shankar}, {Weinberg}, \&
  {Miralda-Escud{\'e}}}]{Shankar2009}
{Shankar}, F., {Weinberg}, D.~H., \& {Miralda-Escud{\'e}}, J. 2009, \apj, 690,
  20

\bibitem[{{Shankar} {et~al.}(2013){Shankar}, {Weinberg}, \&
  {Miralda-Escud{\'e}}}]{Shankar2013}
{Shankar}, F., {Weinberg}, D.~H., \& {Miralda-Escud{\'e}}, J. 2013, \mnras,
  428, 421

\bibitem[{{Shen} {et~al.}(2020){Shen}, {Hopkins}, {Faucher-Gigu{\`e}re},
  {Alexander}, {Richards}, {Ross}, \& {Hickox}}]{Shen2020}
{Shen}, X., {Hopkins}, P.~F., {Faucher-Gigu{\`e}re}, C.-A., {et~al.} 2020,
  \mnras, 495, 3252

\bibitem[{{Shen} {et~al.}(2013){Shen}, {Liu}, {Loeb}, \& {Tremaine}}]{Shen2013}
{Shen}, Y., {Liu}, X., {Loeb}, A., \& {Tremaine}, S. 2013, \apj, 775, 49

\bibitem[{{Shibuya} {et~al.}(2015){Shibuya}, {Ouchi}, \&
  {Harikane}}]{Shibuya2015}
{Shibuya}, T., {Ouchi}, M., \& {Harikane}, Y. 2015, \apjs, 219, 15

\bibitem[{{Siana} {et~al.}(2008){Siana}, {Polletta}, {Smith}, {Lonsdale},
  {Gonzalez-Solares}, {Farrah}, {Babbedge}, {Rowan-Robinson}, {Surace},
  {Shupe}, {Fang}, {Franceschini}, \& {Oliver}}]{Siana2008}
{Siana}, B., {Polletta}, M. d.~C., {Smith}, H.~E., {et~al.} 2008, \apj, 675, 49

\bibitem[{{Sijacki} {et~al.}(2015){Sijacki}, {Vogelsberger}, {Genel},
  {Springel}, {Torrey}, {Snyder}, {Nelson}, \& {Hernquist}}]{Sijacki2015}
{Sijacki}, D., {Vogelsberger}, M., {Genel}, S., {et~al.} 2015, \mnras, 452, 575

\bibitem[{{Spergel} {et~al.}(2003){Spergel}, {Verde}, {Peiris}, {Komatsu},
  {Nolta}, {Bennett}, {Halpern}, {Hinshaw}, {Jarosik}, {Kogut}, {Limon},
  {Meyer}, {Page}, {Tucker}, {Weiland}, {Wollack}, \& {Wright}}]{Spergel2003}
{Spergel}, D.~N., {Verde}, L., {Peiris}, H.~V., {et~al.} 2003, \apjs, 148, 175

\bibitem[{{Spinoso} {et~al.}(2023){Spinoso}, {Bonoli}, {Valiante}, {Schneider},
  \& {Izquierdo-Villalba}}]{Spinoso2022}
{Spinoso}, D., {Bonoli}, S., {Valiante}, R., {Schneider}, R., \&
  {Izquierdo-Villalba}, D. 2023, \mnras, 518, 4672

\bibitem[{{Springel}(2005)}]{Springel2005}
{Springel}, V. 2005, \mnras, 364, 1105

\bibitem[{{Springel} {et~al.}(2001){Springel}, {White}, {Tormen}, \&
  {Kauffmann}}]{Springel2001}
{Springel}, V., {White}, S.~D.~M., {Tormen}, G., \& {Kauffmann}, G. 2001,
  \mnras, 328, 726

\bibitem[{{Sutherland} \& {Dopita}(1993)}]{SutherlandDopita1993}
{Sutherland}, R.~S. \& {Dopita}, M.~A. 1993, \apjs, 88, 253

\bibitem[{{Tamanini} {et~al.}(2016){Tamanini}, {Caprini}, {Barausse}, {Sesana},
  {Klein}, \& {Petiteau}}]{Tamanini2016}
{Tamanini}, N., {Caprini}, C., {Barausse}, E., {et~al.} 2016, \jcap, 2016, 002

\bibitem[{{Tamburello} {et~al.}(2017){Tamburello}, {Capelo}, {Mayer},
  {Bellovary}, \& {Wadsley}}]{Tamburello2017}
{Tamburello}, V., {Capelo}, P.~R., {Mayer}, L., {Bellovary}, J.~M., \&
  {Wadsley}, J.~W. 2017, \mnras, 464, 2952

\bibitem[{{Tonry}(1984)}]{Tonry1984}
{Tonry}, J.~L. 1984, \apjl, 283, L27

\bibitem[{{Toomre} \& {Toomre}(1972)}]{Toomre1972}
{Toomre}, A. \& {Toomre}, J. 1972, \apj, 178, 623

\bibitem[{{Trinca} {et~al.}(2022){Trinca}, {Schneider}, {Valiante}, {Graziani},
  {Zappacosta}, \& {Shankar}}]{Trinca2022}
{Trinca}, A., {Schneider}, R., {Valiante}, R., {et~al.} 2022, \mnras, 511, 616

\bibitem[{{Tsalmantza} {et~al.}(2011){Tsalmantza}, {Decarli}, {Dotti}, \&
  {Hogg}}]{Tsalmantza2011}
{Tsalmantza}, P., {Decarli}, R., {Dotti}, M., \& {Hogg}, D.~W. 2011, \apj, 738,
  20

\bibitem[{{Valiante} {et~al.}(2017){Valiante}, {Agarwal}, {Habouzit}, \&
  {Pezzulli}}]{Valiante2017}
{Valiante}, R., {Agarwal}, B., {Habouzit}, M., \& {Pezzulli}, E. 2017, \pasa,
  34, e031

\bibitem[{{Valtonen} {et~al.}(2008){Valtonen}, {Lehto}, {Nilsson}, {Heidt},
  {Takalo}, {Sillanp{\"a}{\"a}}, {Villforth}, {Kidger}, {Poyner}, {Pursimo},
  {Zola}, {Wu}, {Zhou}, {Sadakane}, {Drozdz}, {Koziel}, {Marchev}, {Ogloza},
  {Porowski}, {Siwak}, {Stachowski}, {Winiarski}, {Hentunen}, {Nissinen},
  {Liakos}, \& {Dogru}}]{Valtonen2008}
{Valtonen}, M.~J., {Lehto}, H.~J., {Nilsson}, K., {et~al.} 2008, \nat, 452, 851

\bibitem[{{Vasiliev} {et~al.}(2014){Vasiliev}, {Antonini}, \&
  {Merritt}}]{Vasiliev2014}
{Vasiliev}, E., {Antonini}, F., \& {Merritt}, D. 2014, \apj, 785, 163

\bibitem[{{Visbal} \& {Haiman}(2018)}]{Visbal2018}
{Visbal}, E. \& {Haiman}, Z. 2018, \apjl, 865, L9

\bibitem[{{Volonteri} {et~al.}(2003){Volonteri}, {Haardt}, \&
  {Madau}}]{Volonteri2003}
{Volonteri}, M., {Haardt}, F., \& {Madau}, P. 2003, \apj, 582, 559

\bibitem[{{Volonteri} {et~al.}(2021){Volonteri}, {Habouzit}, \&
  {Colpi}}]{Volonteri2021}
{Volonteri}, M., {Habouzit}, M., \& {Colpi}, M. 2021, Nature Reviews Physics,
  3, 732

\bibitem[{{Volonteri} {et~al.}(2020){Volonteri}, {Pfister}, {Beckmann},
  {Dubois}, {Colpi}, {Conselice}, {Dotti}, {Martin}, {Jackson}, {Kraljic},
  {Pichon}, {Trebitsch}, {Yi}, {Devriendt}, \& {Peirani}}]{Volonteri2020}
{Volonteri}, M., {Pfister}, H., {Beckmann}, R.~S., {et~al.} 2020, \mnras, 498,
  2219

\bibitem[{{Weinberger} {et~al.}(2018){Weinberger}, {Springel}, {Pakmor},
  {Nelson}, {Genel}, {Pillepich}, {Vogelsberger}, {Marinacci}, {Naiman},
  {Torrey}, \& {Hernquist}}]{Weinberger2018}
{Weinberger}, R., {Springel}, V., {Pakmor}, R., {et~al.} 2018, \mnras, 479,
  4056

\bibitem[{{White} \& {Frenk}(1991)}]{WhiteFrenk1991}
{White}, S.~D.~M. \& {Frenk}, C.~S. 1991, \apj, 379, 52

\bibitem[{{White} \& {Rees}(1978)}]{WhiteandRees1978}
{White}, S.~D.~M. \& {Rees}, M.~J. 1978, \mnras, 183, 341

\bibitem[{{Witt} {et~al.}(2021){Witt}, {Charisi}, {Taylor}, \&
  {Burke-Spolaor}}]{Witt2021}
{Witt}, C.~A., {Charisi}, M., {Taylor}, S.~R., \& {Burke-Spolaor}, S. 2021,
  arXiv e-prints, arXiv:2110.07465

\bibitem[{{Yoshida} {et~al.}(2003){Yoshida}, {Abel}, {Hernquist}, \&
  {Sugiyama}}]{Yoshida2003}
{Yoshida}, N., {Abel}, T., {Hernquist}, L., \& {Sugiyama}, N. 2003, \apj, 592,
  645

\bibitem[{{Yu}(2002)}]{Yu2002}
{Yu}, Q. 2002, \mnras, 331, 935

\bibitem[{{Yuan} {et~al.}(2021){Yuan}, {Murase}, {Zhang}, {Kimura}, \&
  {M{\'e}sz{\'a}ros}}]{Yuan2021}
{Yuan}, C., {Murase}, K., {Zhang}, B.~T., {Kimura}, S.~S., \&
  {M{\'e}sz{\'a}ros}, P. 2021, \apjl, 911, L15

\end{thebibliography}
\end{document}